\documentclass[11pt,a4paper]{article}
\pdfoutput=1

\usepackage{jheppub}
\usepackage{latexsym}
\usepackage{multirow}
\usepackage{color}
\usepackage[usenames,dvipsnames,table]{xcolor}
\usepackage{graphicx}
\usepackage{epsfig} 
\usepackage{epsf}   
\usepackage{dcolumn}
\usepackage{bm}
\usepackage{dcolumn}
\usepackage{textcomp}
\usepackage{float}
\usepackage{subfig}
\usepackage{hypcap}
\usepackage[]{hyperref}
\usepackage{makecell}
\usepackage{color}
\usepackage{pifont}
\usepackage{appendix}
\usepackage{amsmath}
\usepackage{multirow,bigdelim}  
\usepackage{lineno}
\usepackage[normalem]{ulem}
\graphicspath{{fig/}}

\newcommand{\comment}[1]{}
\usepackage{amssymb}
\hypersetup{
	bookmarks=true,         
	unicode=false,          
	pdftoolbar=true,        
	pdfmenubar=true,        
	pdffitwindow=true,   
	pdfstartview={FitH},    
	pdfsubject={scalar NSI},   
	pdfnewwindow=true,      
	pdfcreator={RevTeX},
	colorlinks=true,     
	linkcolor=red,         
	citecolor=blue,        
	filecolor=black,     
	urlcolor=blue,           
}

\preprint{}

\title{Imprints of scalar NSI on the CP-violation sensitivity using synergy among DUNE, T2HK and T2HKK}

\author[a]{Abinash Medhi,}
\author[a]{Moon Moon Devi}
\author[b]{and Debajyoti Dutta}

\affiliation[a]{Department of Physics, Tezpur University, Napaam, Sonitpur, Assam-784028, India}
\affiliation[b]{Department of Physics, Assam Don Bosco University, Kamarkuchi, Sonapur, Assam-782402, India}

\emailAdd{amedhi@tezu.ernet.in}
\emailAdd{devimm@tezu.ernet.in}
\emailAdd{debajyoti.dutta@dbuniversity.ac.in}

\date{\today}

	\abstract{
The Non--Standard Interactions (NSIs) are subdominant effects, often appearing in various extensions of SM, which may impact the neutrino oscillations through matter. It is important and interesting to explore the impact of NSIs in the ongoing and upcoming precise neutrino oscillations experiments. In this work, we have studied the imprints of a scalar--mediated NSI in three upcoming long--baseline (LBL) experiments (DUNE, T2HK and T2HKK). The effects of scalar NSI appears as a medium--dependent correction to the neutrino mass term. Its contribution scales linearly with matter density, making LBL experiments a suitable candidate to probe its effects. We show that the scalar NSI may significantly impact the oscillation probabilities, event rates at the detectors and the $\chi^2$--sensitivities of $\delta_{CP}$ measurements. We present the results of a combined analysis involving the LBL experiments (DUNE+T2HK and DUNE+T2HKK) which offer a better capability of constraining the scalar NSI parameters as well as an improved sensitivity towards CP-violation.  
}  
\arxivnumber{2209.05287}

\begin{document}
	\maketitle
\section{Introduction} \label{sec:introduction}
The discovery of neutrino oscillations jointly by Super--Kamiokande (SK)~\cite{Super-Kamiokande:1998kpq} and Sudbury Neutrino Observatory (SNO)~\cite{SNO:2002tuh} have given a new insight to probe new--physics beyond the Standard Model (BSM). Neutrino oscillations essentially confirms that neutrinos are massive and provide the first clear experimental hint of BSM--physics. The parameters associated with the neutrino oscillations are being widely probed in different neutrino experiments \cite{Super-Kamiokande:2004orf,KamLAND:2004mhv,MINOS:2008kxu,MINOS:2011neo}. The neutrinos are one of the promising portals to explore new--physics in the leptonic sector. The BSM models, which describe the neutrino masses and mixing, often explore new unknown couplings of neutrinos termed as non--standard interactions (NSIs). Looking at the unprecedented accuracy and the precision provided by the current and upcoming neutrino experiments, these subdominant effects on neutrino oscillations may have a significant impact on the physics reach of these experiments. In this work, we have primarily explored the impacts of a scalar mediated NSI on the measurements of the leptonic phase $\delta_{CP}$ in the three long baseline (LBL) neutrino experiments DUNE \cite{Abi:2020loh}, T2HK \cite{Hyper-KamiokandeProto-:2015xww}, and T2HKK \cite{Hyper-Kamiokande:2016srs}. We have performed a synergy analysis combining these LBL experiments to probe the impact of scalar NSI in a model--independent way.

The ongoing and future neutrino experiments aim at measuring the neutrino oscillation parameters with utmost accuracy. However, the presence of parameter degeneracy \cite{Burguet-Castell:2001ppm, Minakata:2001qm, Fogli:1996pv, Barger:2001yr} among the mixing parameters creates difficulty in measurement of these parameters. Due to the underlying degeneracy, various sets of mixing parameters can produce same oscillation probabilities and can bring ambiguity in pin pointing the values of oscillation parameters. To overcome these effects, data from various experiments may be combined. The degenerate parameter space is different for different neutrino experiments and the combination of such experiments may help in determining the oscillation parameters unambiguously. The combination of different experiments often provides better sensitivity and also highlights various possible synergies among the experiments. For unambiguous determination of neutrino oscillation parameters, combination of various neutrino experiments are needed as the degenerate parameter space is different for different experiments \cite{Barger:2001yr,Burguet-Castell:2002ald,Minakata:2001qm,Fogli:1996pv}. In~\cite{Choubey:2017cba}, the authors showed that in presence of a light sterile neutrino, a combination of three different LBL experiments (DUNE, T2HK, T2HKK) give better sensitivity (more that 5$\sigma$) towards the CP--violation measurement as compared to individual sensitivity. In the same work, the authors also pointed that the combination of the experiments significantly improved the mass hierarchy as well as octant discovery potential sensitivities. It has also been explored \cite{Prakash:2012az} that the mass hierarchy--$\delta_{CP}$ degeneracy can be resolved using the synergy between two LBL experiments T2K \cite{T2K:2011qtm} and NO$\nu$A \cite{Prakash:2012az}. In~\cite{Masud:2016bvp}, the authors combined DUNE, T2K and No$\nu$A to explore possible synergy among these experiments towards a vector NSI. It is found that a combined sensitivity study from these experiments can be crucial to pin--point the CP--violation and CP--measurement in leptonic sector. In \cite{Choubey:2022gzv}, the authors have shown that the synergy between T2HK \cite{Hyper-KamiokandeProto-:2015xww} and JUNO \cite{JUNO:2015zny} experiments can provide an improved sensitivity up--to 9$\sigma$ towards mass ordering of neutrinos. In \cite{Agarwalla:2013ju}, the authors pointed out that the $\theta_{23}$ octant ambiguity can be resolved by combining the sensitivities of T2K and NO$\nu$A, irrespective of the hierarchy and $\delta_{CP}$. The physics potential can be significantly enhanced by combining a number of experiments, as it boosts up the sensitive energy range as well as the event distributions. The synergy between various neutrino experiments are often used for better understanding as well as for optimizing the fundamental knowledge of neutrino oscillations \cite{Cao:2020ans,Ghosh:2017ged,Ghosh:2015ena,Ghosh:2012px,Ghosh:2014dba,Bharti:2016hfb,Ballett:2016daj,Fukasawa:2016yue,Minakata:2003wq,Ghosh:2014zea}.

In this precision era of neutrino physics all the ongoing and upcoming neutrino experiments focus on measuring the neutrino mixing parameters with utmost accuracy. The primary goal of these experiments are to address the three main unknowns in the neutrino sector, i.e., the hierarchy of neutrino masses~\cite{Capozzi:2017ipn}, the octant of mixing angle $\theta_{23}$~\cite{Agarwalla:2013ju} and the determination of CP phase ($\delta_{CP}$) in leptonic sector~\cite{Kobayashi:1973fv}. The robust nature of the ongoing and future neutrino experiments make them sensitive to the subdominant effects of neutrinos. One such subdominant effect is NSI, which may have a significant impact on the measurement of oscillation parameters in various neutrino experiments. Initially the idea of NSI~\cite{Wolfenstein:1977ue} was introduced with a coupling of neutrinos with the environmental fermions by a vector mediator. These kind of vector mediated NSIs appear as a matter potential term in the neutrino oscillation Hamiltonian. The vector mediated NSI has been widely explored \cite{Miranda:2015dra, Farzan:2017xzy, Biggio:2009nt, Babu:2019mfe, Ohlsson:2012kf}, and it is an excellent candidate to probe physics beyond the Standard Model. It can have a significant effect on the physics reach of various neutrino experiments~\cite{Liao:2016orc,Friedland:2012tq,Coelho:2012bp,Rahman:2015vqa,Coloma:2015kiu,deGouvea:2015ndi,Liao:2016hsa,Forero:2016cmb,Huitu:2016bmb,Bakhti:2016prn,Kumar:2021lrn,Agarwalla:2015cta,Agarwalla:2014bsa,Agarwalla:2012wf,Blennow:2016etl,Blennow:2015nxa,Deepthi:2016erc,Masud:2021ves,Soumya:2019kto,Masud:2018pig,Masud:2017kdi,Masud:2015xva,Ge:2016dlx,Fukasawa:2016lew,Chatterjee:2021wac} and these effects are being widely probed~\cite{Khatun:2019tad,Chatterjee:2014gxa,Super-Kamiokande:2011dam, Davidson:2003ha,Choubey:2014iia,Denton:2018xmq,Farzan:2015hkd,Farzan:2015doa,Esmaili:2013fva,Khan:2021wzy,Liu:2020emq,Chatterjee:2020kkm,Denton:2020uda,Babu:2020nna,Flores:2020lji,Farzan:2019xor,Pandey:2019apj}. A global status on the bounds of the vector NSI parameters can be found in~\cite{Esteban:2019lfo,Coloma:2019mbs}. 

We have explored here, the non-standard coupling of neutrinos with a scalar~\cite{Ge:2018uhz, Yang:2018yvk, Khan:2019jvr, Medhi:2021wxj}. The scalar mediated NSI affects the neutrino mass in the neutrino Hamiltonian and can provide unique phenomenology in neutrino oscillations. Unlike the vector NSI, the effects of scalar NSI linearly scale with the environmental matter density and this makes long-baseline neutrino experiments one of the most suitable candidates to probe scalar NSI. In~\cite{Ge:2018uhz}, the authors initiate the idea of scalar NSI to fit the recent data from Borexino experiment. Although there are currently not any stringent bounds on the scalar NSI parameters, a few studies have tried putting some constraints under astrophysical and cosmological limits \cite{Babu:2019iml,Venzor:2020ova}. In our work~\cite{Medhi:2021wxj}, we have explored the possible impacts of scalar NSI on the CP-violation sensitivities at LBL experiments taking DUNE as a case study. It is found that the presence of scalar NSI significantly impacts the CP--sensitivities of DUNE. These results are interesting and acts as a motivation to explore the scalar NSI in LBL experiments further. Combining various LBL experiments also become crucial as the synergy study would provide a more precise sensitivity scenario.  

In this paper we have performed, for the first time, a synergy study on the effects of scalar NSI on three LBL experiments viz. DUNE, T2HK and T2HKK in a model independent way. We have probed the effects of scalar NSI, one element at a time, and have found notable impacts of scalar NSI on the physics sensitivities of the chosen neutrino experiments. We have primarily explored the possible impacts of scalar NSI parameters on the CP--violation (CPV) sensitivities. We have then performed a combined analysis of DUNE with T2HK as well as DUNE with T2HKK for testing possible synergies among these experiments. We show that for some chosen values of NSI parameters the CPV sensitivities get enhanced and give improved precision in $\delta_{CP}$ measurements. It is found that, for all the chosen negative values of the NSI parameters the CPV sensitivities always get suppressed. We also see that a positive NSI parameter can fake the CP effects and mimic the standard CPV sensitivity at DUNE and T2HKK. The joint study of the LBL experiments (DUNE+T2HK and DUNE+T2HKK) improves the overall sensitivities and can help in lifting the underlying degeneracy in CPV measurements. It is highly crucial to put constraints on these NSI parameters for accurate measurements and better understandings of the data coming from various neutrino experiments.

The paper is organized as follows: In section \ref{sec:framework} we discuss the detailed formalism of scalar NSI. In section \ref{sec:methdology}, we describe the simulation methodology used in our analysis. The technical details of the three neutrino experiments used in our simulations are presented in section \ref{sec:experiment}. The impacts of NSI on oscillation probabilities and CP--asymmetry are shown in section \ref{sec:oscillation_probabilities} and section \ref{sec:CP_asymmetry} respectively. We discuss the results of the $\chi^2$ analyses on NSI parameter sensitivity and CP-violation sensitivity in section \ref{sec:results}. We conclude our findings in section \ref{sec:summary}.
\section{Scalar NSI Formalism}  \label{sec:framework}
The elusive neutrinos interact with matter through weak interaction and gravity. The neutrino interactions take place through mediating a W$^{\pm}$ boson (Charge Current -- CC) or a Z boson (Neutral Current -- NC) \cite{Linder:2005fc}. Both of the interactions appear as matter potentials in the neutrino Hamiltonian, however, only the CC--interactions contribute to the oscillation probabilities. The NC--interactions do not contribute to the oscillations as they appear as a common term in the Hamiltonian. The Lagrangian for neutrino--matter coupling via CC interactions may be written as \cite{Wolfenstein:1977ue, Nieves:2003in, Nishi:2004st, Maki:1962mu},
\begin{equation}
\mathcal L^{\rm eff}_{\rm cc}
=
- \frac {4 G_F}{\sqrt 2}
\left[
\overline{\nu_e}(p_3) \gamma_\mu P_L \nu_e(p_2)
\right]
\left[
\overline e(p_1) \gamma^\mu P_L e(p_4)
\right],
\label{eq:Leff}
\end{equation}

\noindent where, $G_F$ is the Fermi coupling constant, $p_{i}$'s are momenta of incoming and outgoing states and $P_L = (1 - \gamma_5)/2$, $P_R = (1+ \gamma_5)/2$) are left and right chiral projection operators.

 The effective Hamiltonian, $\mathcal{H_{\rm eff}}$, for neutrino oscillations in matter is framed as \cite{Bilenky:1987ty},
\begin{equation}
\mathcal{H_{\rm eff}} = E_\nu + \frac{1}{2E_\nu} \, \mathcal{U} {\rm diag}(0, \Delta m^2_{21}, \Delta m^2_{31}) \mathcal{U}^\dag + {\rm diag} (V_{\rm CC}, 0 , 0)\,,
\label{eq:matter_H2}
\end{equation}

\noindent where, 
\begin{itemize}
    \item $\mathcal{U}$ = Pontecorvo-Maki-Nakagawa-Sakata (PMNS) matrix \cite{Pontecorvo:1957cp,Pontecorvo:1957qd,Pontecorvo:1967fh,ParticleDataGroup:2020ssz},
\item $E_\nu$ = neutrino energy, 
\item $\Delta m^2_{ij} = m_i^2 - m_j^2$, are the neutrino mass-squared differences, and 
\item $V_{\rm SI} = \pm \sqrt 2 G_F n_e$, comes due to CC neutrino matter interactions. 
\end{itemize}
The non-standard coupling of neutrinos with a scalar~\cite{Ge:2018uhz,Babu:2019iml} is also an interesting sector to probe new--physics beyond SM. The effective Lagrangian for neutrinos coupling via a scalar, $\phi$ may be framed as, 
\begin{align}
{\cal L}_{\rm eff}^{\rm S} \ = \ \frac{y_f y_{\alpha\beta}}{m_\phi^2}(\bar{\nu}_\alpha(p_3) \nu_\beta(p_2))(\bar{f}(p_1)f(p_4)) \,, 
\label{eq:nsi_L}
\end{align}\\
where, 
\begin{itemize}
    \item 
\noindent $\alpha$, $\beta$ refer to the neutrino flavours e, $\mu$, $\tau$,
\item  \noindent $f$ = e, u, d indicate the matter fermions, (e: electron, u:
up-quark, d: down-quark),
\item  \noindent $\bar{f}$ is for corresponding anti fermions, 
 \item  \noindent $y_{\alpha\beta}$ is the Yukawa couplings of the neutrinos with the scalar mediator $\phi$, 
  \item  \noindent$y_f$ is the Yukawa coupling of $\phi$ with $f$, and, 
  \item \noindent $m_\phi$ is the mass of the scalar mediator $\phi$.
  \end{itemize}
  
  The Lagrangian is composed of Yukawa terms and hence it is not possible to convert it into vector currents. So, the effect of scalar NSI appears as an addition to the neutrino mass term. The corresponding Dirac equation taking into account the effect of scalar NSI gets the following form,
  \begin{equation}
  \bar \nu_\beta
\left[
  i \partial_\mu \gamma^\mu
+
\left(
  M_{\beta \alpha}
+ \frac {\sum_f n_f y_f y_{\alpha \beta}}{m^2_\phi}
\right)
\right] \nu_\alpha
=
  0 \,,
\end{equation}

\noindent where, $n_f$ is the number density of the environmental fermions.
  
  Hence we see that the effect of scalar NSI appears as a perturbation to the neutrino mass term. So, the effective Hamiltonian in presence of scalar NSI takes the form,
  
  \begin{equation}
\mathcal H_{\rm SNSI}
\approx
E_\nu
+ \frac { M_{\rm eff}  M_{\rm eff}^\dagger}{2 E_\nu}
\pm V_{\rm SI} \,,
\label{eq:Hs}
\end{equation}
  where, $M_{\rm eff}$ = $M + M_{\rm SNSI}$, is the effective mass matrix that includes both the regular mass matrix $M$ and the contribution from the scalar NSI, $M_{SNSI}  \equiv \sum_f n_f y_f y_{\alpha\beta} / m^2_\phi$. The active neutrino mass ($\equiv$ $\mathcal{U^{'}} D_{\nu} \mathcal{U^{'}}^{\dagger}$) may be diagonalized by the mixing matrix $\mathcal{U^{'}} \equiv P \mathcal{U} Q^{\dagger}$. Here $D_\nu$ is the diagonal mass matrix of neutrinos represented by $D_\nu$ $\equiv$ diag($m_1, m_2, m_3$). The matrix $\mathcal{U^{'}}$ is a combination of Majorana rephasing matrix, Q and a diagonal rephasing matrix P. The Majorana rephasing matrix can be absorbed by $Q D_\nu Q^{\dagger} = D_\nu$, however the unphysical rephasing matrix P cannot be rotated away. The effective neutrino mass term, after rotating the unphysical rephasing matrix P into the scalar NSI contribution, can therefore be written as , 
  
  \begin{equation}
 M_{eff} \equiv \mathcal{U} D_\nu \mathcal{U}^{\dagger} + P^{\dagger} M_{SNSI} P \equiv M + \delta M 
 \label{effectiveM}
\end{equation}
  
  The scalar NSI contribution $\delta M$ includes the unphysical rephasing matrix P after proper rotation. We have used the following parametrization of $\delta M$ to probe the effects of scalar NSI in neutrino oscillations,
  \begin{equation}
\delta M
\equiv
\sqrt{|\Delta m^2_{31}|}
\left\lgroup
\begin{matrix}
\eta_{ee}     & \eta_{e \mu}    & \eta_{e \tau}   \\
\eta_{\mu e}  & \eta_{\mu \mu}  & \eta_{\mu \tau} \\
\eta_{\tau e} & \eta_{\tau \mu} & \eta_{\tau \tau}
\end{matrix}
\right\rgroup \,.
\label{eq:dM}
\end{equation}\\
The dimensionless elements $\eta_{\alpha \beta}$ quantify the size of scalar NSI. The Hermicity of the Hamiltonian requires the diagonal elements to be real and off-diagonal elements to be complex. In this work we have explored the diagonal elements of the scalar NSI matrix, one at a time. For the three cases that we have used, the non-zero diagonal elements the effective modified Hamiltonian take the forms as shown below,
\begin{equation}
~~~~~~~{\rm Case~I:}~ M_{\rm eff} = \mathcal{U} {\rm diag}\left(m_1, m_2, m_3
 \right)\mathcal{U}^\dag + \sqrt{|\Delta m^2_{31}|}~ \rm diag \left( \eta_{ee}, 0, 0
 \right).
 \label{MeffCase1}
  \end{equation}
   
\begin{equation}
~~~~~~~~~{\rm Case~II:}~ M_{\rm eff} = \mathcal{U} {\rm diag}\left(m_1, m_2, m_3
 \right)\mathcal{U}^\dag + \sqrt{|\Delta m^2_{31}|}~ \rm diag \left( 0, \eta_{\mu\mu}, 0
 \right).
  \label{MeffCase2}
  \end{equation}
  
  \begin{equation}
~~~~~~~~~~~{\rm Case~III:}~ M_{\rm eff} = \mathcal{U} {\rm diag}\left(m_1, m_2, m_3
 \right)\mathcal{U}^\dag + \sqrt{|\Delta m^2_{31}|}~ \rm diag \left( 0, 0, \eta_{\tau\tau}
 \right).
  \label{MeffCase3}
  \end{equation} \\
Interestingly, $\mathcal H_{\rm SNSI}$ has a direct dependence on the absolute masses of neutrinos. We have taken the value of $m_1$ to be $10^{-5}$ eV in this work.  The values of $m_2$ and $m_3$ have been accordingly calculated from $\Delta m_{21}^2$ and $\Delta m_{31}^2$.

\section{Methodology}
 \label{sec:methdology}
To explore the impact of NSI on various neutrino experiments we have used  GLoBES (Global Long Baseline Experiment Simulator) \cite{Huber:2004ka, Kopp:2006wp, Huber:2007ji}. GLoBES is a widely used sophisticated neutrino experiment simulator for long baseline experiments. The values of mixing parameters used in our simulation studies are listed in table~\ref{tab:mixing_parameters}. Throughout the analysis, we have considered normal hierarchy to be the true hierarchy and higher octant to be the true octant. We have considered three proposed super-beam experiments DUNE, T2HK and T2HKK to explore the impact of scalar NSI. The systematics and background information are incorporated from the corresponding Technical Design Reports (TDR) of the experiments. The uncertainties on signal and background are summarized in table \ref{tab:norm-uncertain-exp}. In this study we have considered the diagonal scalar NSI elements one at a time. We have, at first, explored the impact of scalar NSI at the probability level as well as the event level at the detector. We have then studied the effects of scalar NSI on the CP asymmetry parameter.
In the following subsections we describe the technical details of the three experiments and the impact of scalar NSI on the oscillation probabilities as well as on the CP-asymmetry parameter.
\begin{table}[h]
	\centering
	
	\begin{tabular}{|c|c|c|}
		\hline
		Parameters & True Values\\
		\hline
		$\theta_{12}$ & 34.51$^\circ$    \\
		$\theta_{13}$ & 8.44$^\circ$   \\
		$\theta_{23}$ & 47$^\circ$   \\
		$\delta_{CP}$ & -$\pi$/2 \\
		$\Delta m_{21}^2$ & 7.56 $\times$ 10$^{-5}$ $eV^2$  \\
		$\Delta m_{31}^2$ & 2.43 $\times$ 10$^{-3}$ $eV^2$  \\
		\hline
	\end{tabular}
	\caption{The benchmark values of oscillation parameters used \cite{NuFIT5.0}.}
	\label{tab:mixing_parameters}
\end{table} 

\subsection{Experimental setup}
 \label{sec:experiment}
The technical details of DUNE, T2HK and T2HKK have been illustrated below and a
comparison of their baselines, L/E and fiducial volumes has been included in table \ref{tab:expt-details}
\subsubsection{DUNE} The Deep Underground Neutrino Experiment (DUNE) \cite{DUNE1,DUNE2,DUNE3,DUNE4,DUNE5} is a proposed long baseline neutrino experiments which will be located in the USA. The Near Detector for the experiment will be located at Long-Baseline Neutrino Facility (LBNF) at a distance of 574 meters and 60 meters underground from the source of neutrino beam site at Fermilab. The neutrinos will be detected after travelling a distance of 1300 km at the Far detector (FD) which will be located in Homestake Mine in South Dakota. The FD is made of four modules of liquid argon time projection chamber (LArTPC), each having a fiducial mass of 10kt. The TPC which is used to detect the charge ionization from neutrino interactions has good spatial resolution, energy resolution, 3D tract reconstruction and identify particle track using the energy loss information along the track. The neutrino beam used for the DUNE will be produced at Fermilab having a power of 1.2 MW-120 GeV and will deliver $10^{21}$ proton-on-target (POT) per year. The experiment is expected to start the operation in 2026.

\subsubsection{T2HK} T2HK (Tokai to Hyper-Kamiokande) \cite{Hyper-KamiokandeProto-:2015xww} is one of the proposed promising long baseline experiment which is planned to have a baseline of 295 km. In the proposed set up the intense neutrino beam will be produced at J-PARC facility and will be detected in Hyper-Kamiokande (HK) detector. The neutrino beam from J-PARC will have a power of 1.3 MW which will generate 27 $\times$ $10^{21}$ POT (proton on target) per year. The HK detector in Japan is an upgradation of the Super-Kamiokande (SK) detector and is expected to have about twenty times the fiducial mass of Super-Kamiokande. The detector will have two cylindrical water Cherenkov module each having a fiducial mass of 187 kt. It will be located 2.5$^\circ$ off-axis from the J-PARC neutrino beam in Japan. For our simulation studies we have taken a baseline of 295 km and fiducial volume to be 374 kt (two cylindrical detector each having fiducial volume of 187 kt). The total run time of 10 years has been divided into 2.5 years in neutrino mode and 7.5 years in antineutrino mode (1:3 ratio) to have an equal contribution from the neutrino and the antineutrino signal events.

\subsubsection{T2HKK} T2HKK \cite{Hyper-Kamiokande:2016srs} is another proposed detector set--up involving T2HK where there is a plan to put the second cylindrical detector of HK in Korea. The second detector will be located at a distance of 1100 km from the J-PARC proton synchrotron facility. So, the T2HKK experiment will have two far detector set--ups, one at a distance of 295 km at HK site and another in Korea at a distance of 1100 km. Both the detector module will have fiducial volumes of 187 kt and the detection principle will be based on the water Cherenkov technique. The detector will be placed at an angle of 2.5$^\circ$ off axis from the neutrino beam and the peak of the second oscillation maximum will be at 0.66 GeV. In this work we have considered the background and systematic uncertainties of T2HKK to be identical to that of T2HK.

\begin{table}[h!]
\begin{center}
\begin{tabular}{|c|c|c|c|}
\hline
Experiment   & Baseline (L in km)  & L/E (in km/GeV) & Fiducial Volume (in kton) \\       
\hline
\hline
T2HK         & 295 km  & 527  & 187 $\times$ 2  \\
\hline
T2HKK         & 295 km; 1100 km  & 527(295 km); 1964(1100 km)  & 187(295 km) + 187(1100 km) \\
\hline
DUNE        & 1300 km & 1543    & 40   \\
\hline
\end{tabular}
\end{center}
\caption{{\footnotesize The baselines, L/E and fiducial volumes of each detector for T2HK, T2HKK, and DUNE. }}
\label{tab:expt-details}
\end{table}

\begin{table}[h!]
\begin{center}
\begin{tabular}{|c|c|c|c|}
\hline
Channel   & T2HK (295 km) & T2HK (1100 km)  & DUNE (1300 km)  \\       
\hline
\hline
$\nu_e$ appearance         & 3.2\%(5\%)  & 3.8\%(5\%)   & 3.2\%(5\%)   \\
\hline
$\nu_{\bar{e}}$ appearance           & 3.9\%(5\%)  & 4.1\%(5\%)    & 3.9\%(5\%)  \\
\hline
$\nu_\mu$ disappearance         & 3.6\%(5\%)  & 3.8\%(5\%)    & 3.6\%(5\%)    \\
\hline
$\nu_{\bar{\mu}}$ disappearance         & 3.6\%(5\%)  & 3.8\%(5\%)   & 3.6\%(5\%)   \\
\hline
\end{tabular}
\end{center}
\caption{{\footnotesize The signal (background) normalization uncertainties of the experiments for the various channels for T2HK, T2HKK and DUNE.}}
\label{tab:norm-uncertain-exp}
\end{table}

 \subsection{Effects on oscillation probabilities}
 \label{sec:oscillation_probabilities}
 In this section we discuss the effects of scalar NSI (the three diagonal cases as mentioned in eq. \ref{MeffCase1}, eq. \ref{MeffCase2} and eq. \ref{MeffCase3}) on the neutrino oscillation probabilities. To perform this analysis we have used NuOscProbExact package \cite{Bustamante:2019ggq}. NuOscProbExact is a flexible python based numerical oscillation probability calculator for both the two and three flavour cases. It employs SU(2) and SU(3) expansions of the evolution operators to compute the numerical probabilities for time-independent Hamiltonian. We have modified the neutrino Hamiltonian accordingly as in eq. \ref{eq:Hs} and have incorporated the three scalar NSI cases. We have used the oscillation parameter values as listed in table \ref{tab:mixing_parameters}. Unless otherwise mentioned, we considered NH to be the true mass hierarchy and HO to be the true octant. 
 
  \begin{figure}[h]
 	\centering
 	\includegraphics[width=0.32\linewidth, height = 5cm]{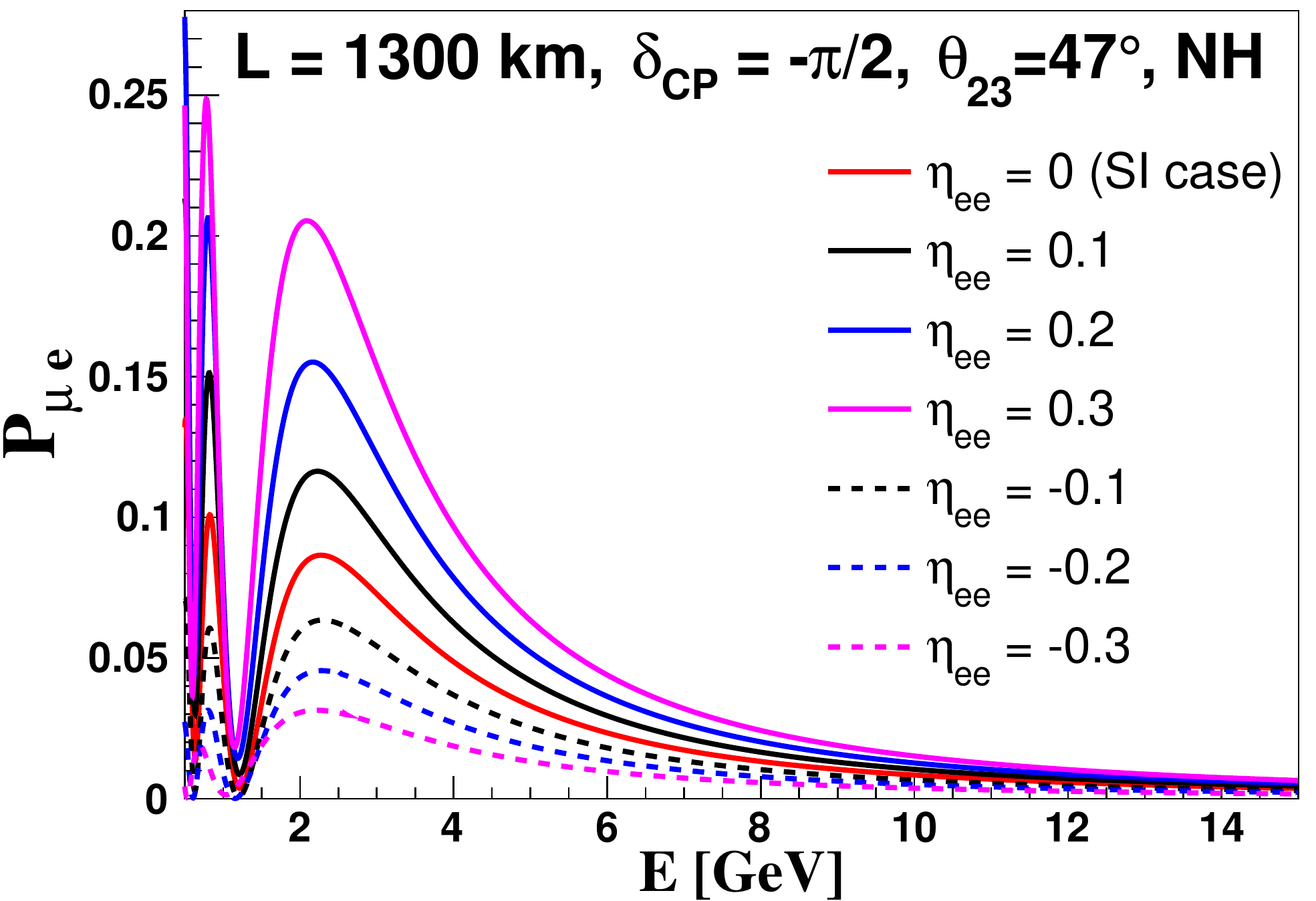} 
 	\includegraphics[width=0.32\linewidth, height = 5cm]{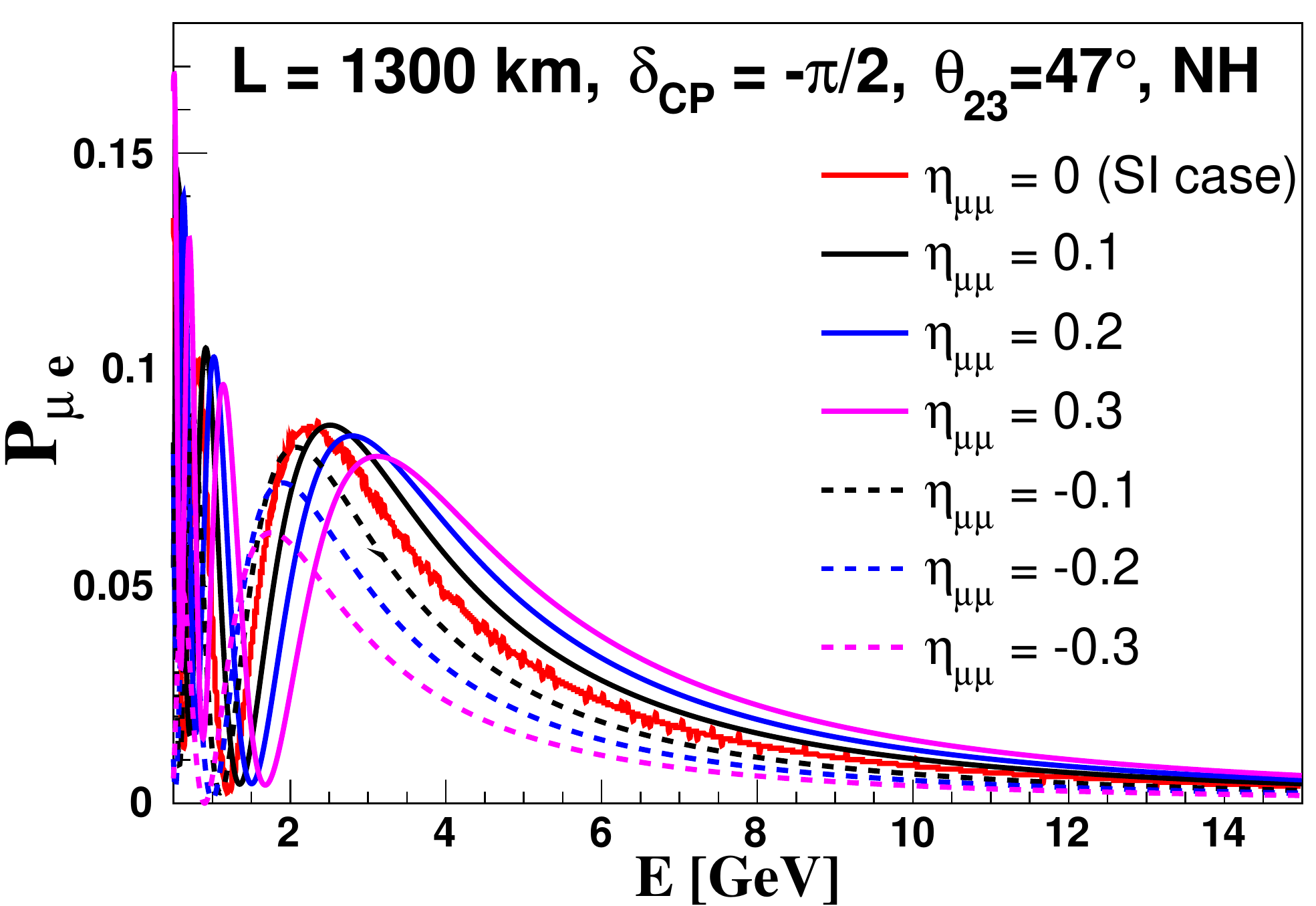} 
 	\includegraphics[width=0.32\linewidth, height = 5cm]{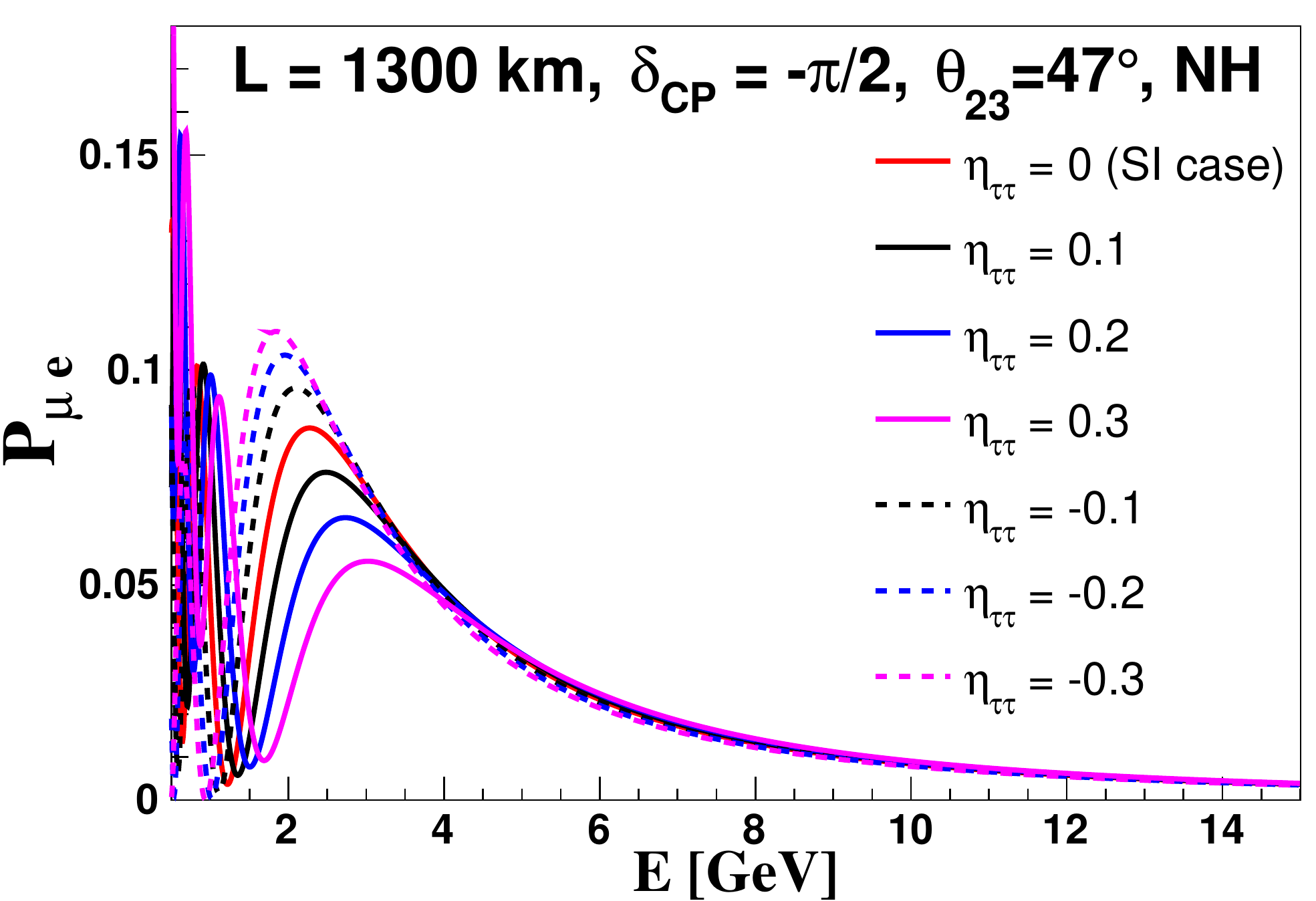}

 	\includegraphics[width=0.32\linewidth, height = 5cm]{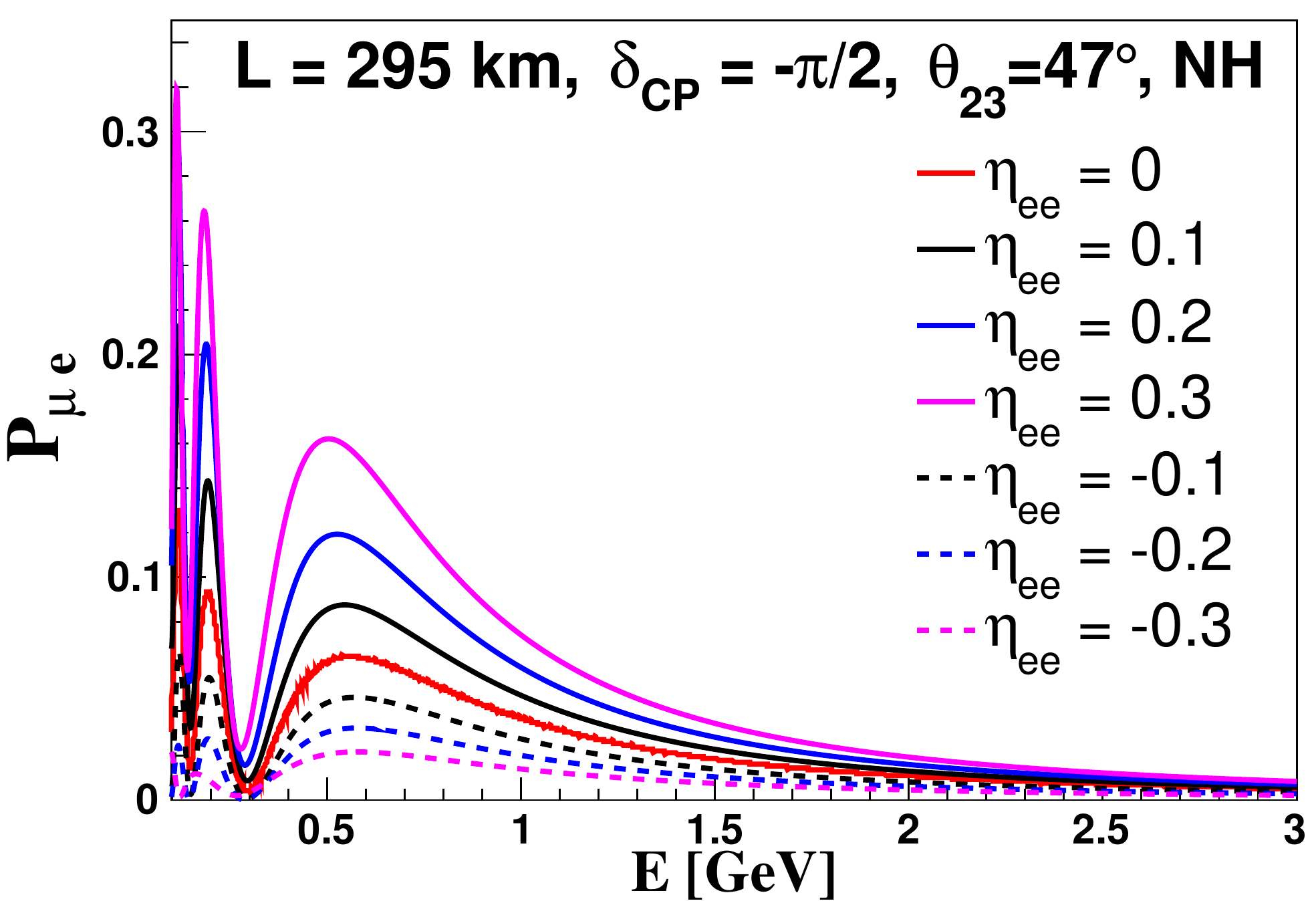} 
 	\includegraphics[width=0.32\linewidth, height = 5cm]{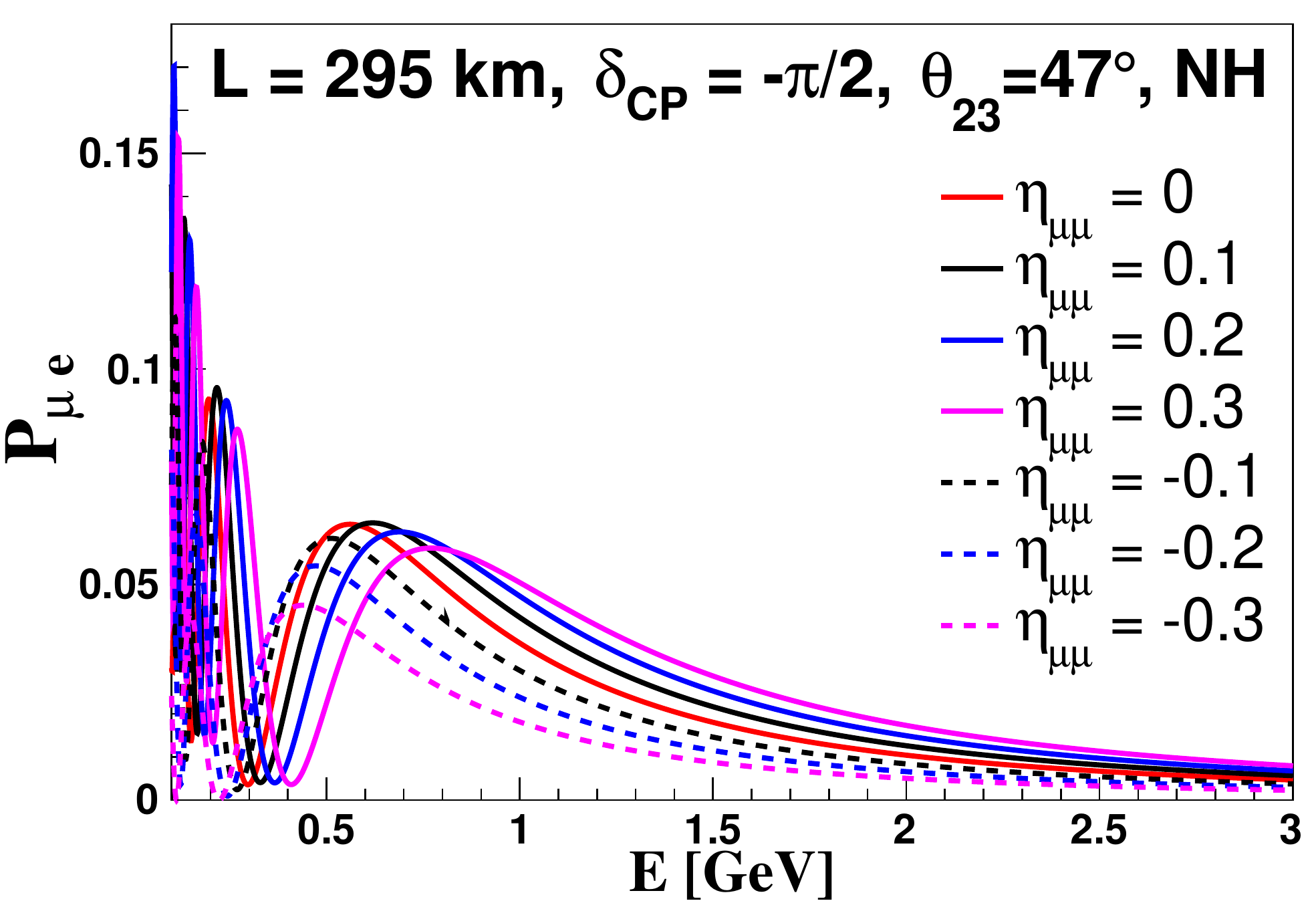} 
 	\includegraphics[width=0.32\linewidth, height = 5cm]{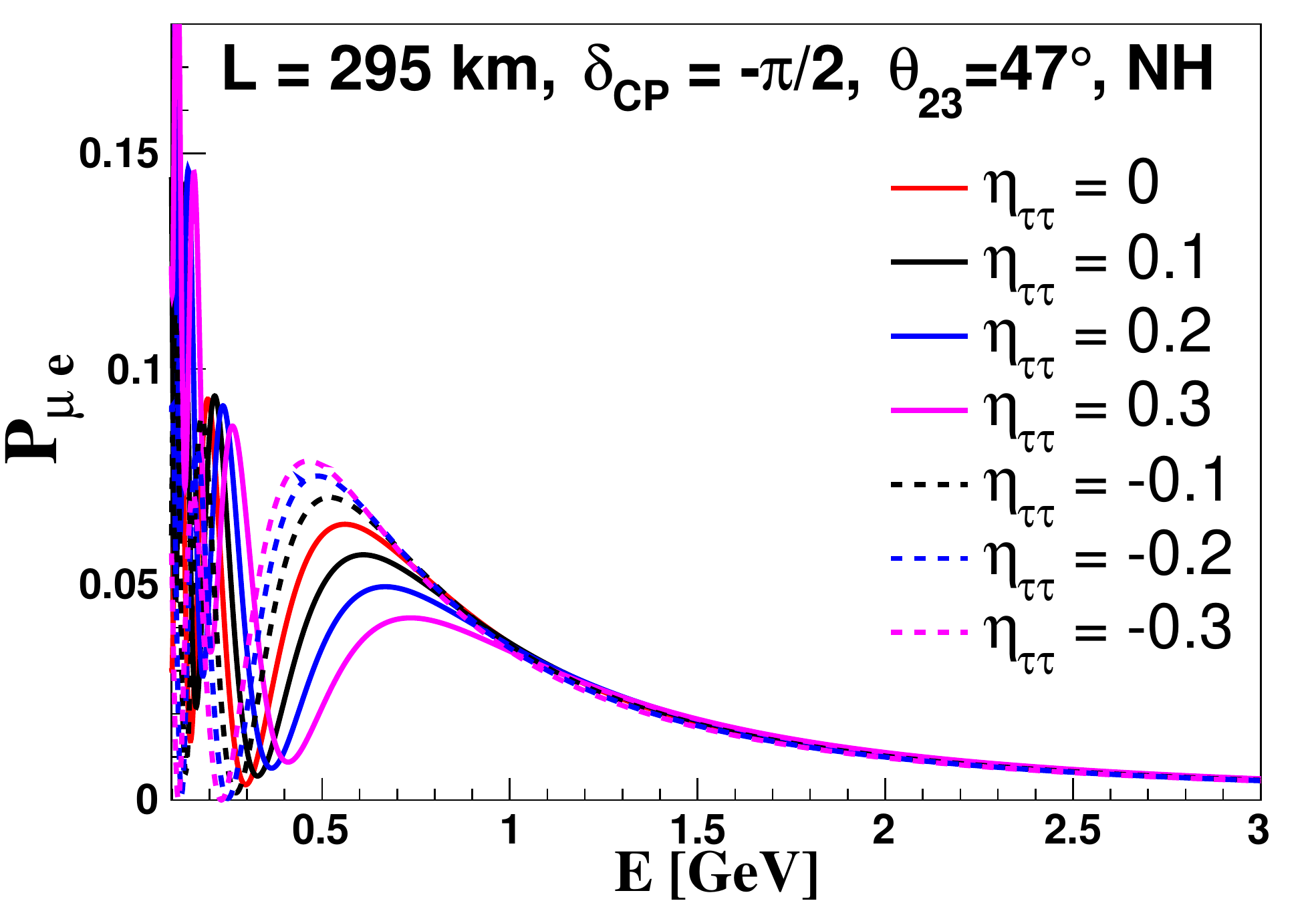} 
 	
 	\includegraphics[width=0.32\linewidth, height = 5cm]{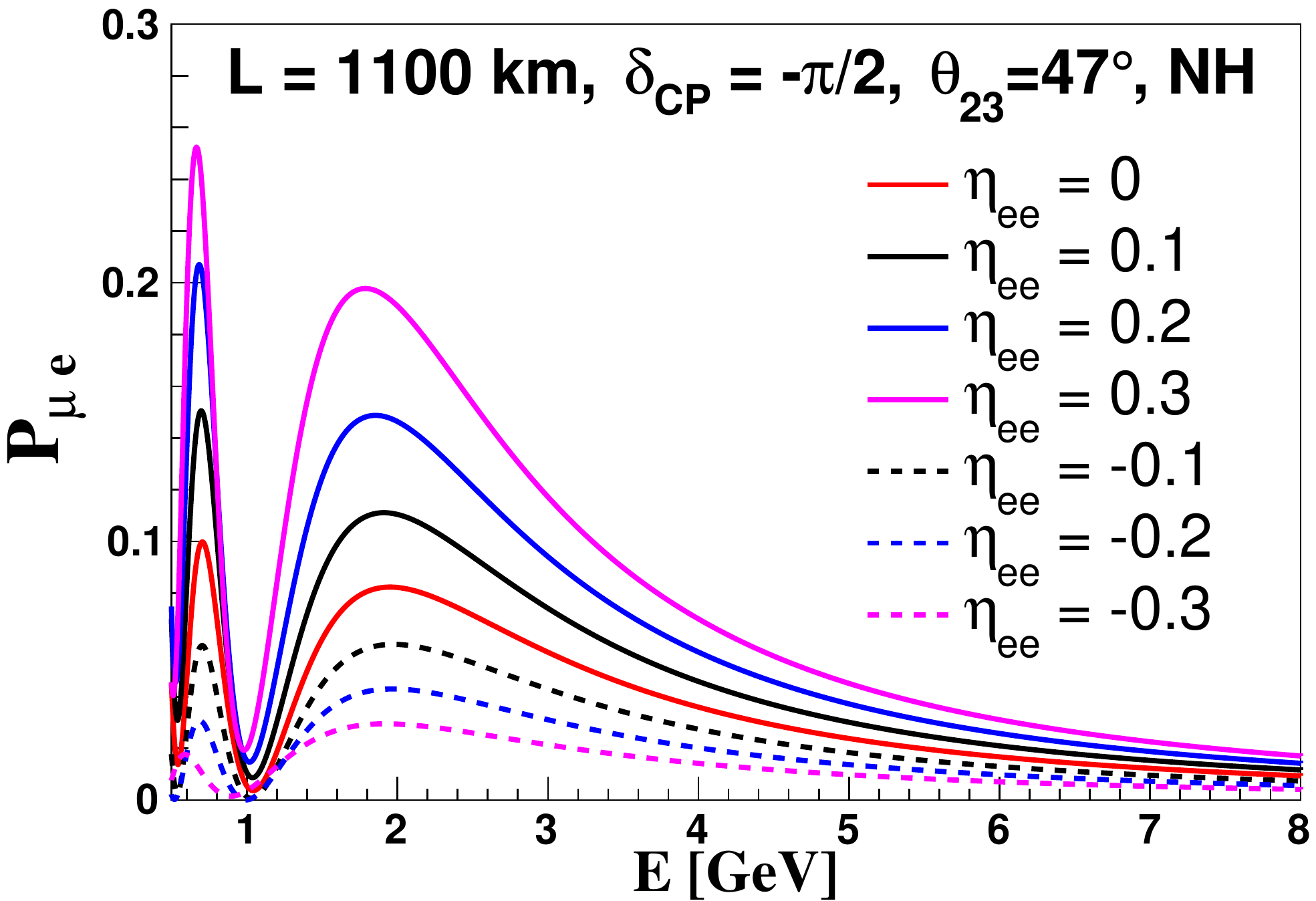} 
 	\includegraphics[width=0.32\linewidth, height = 5cm]{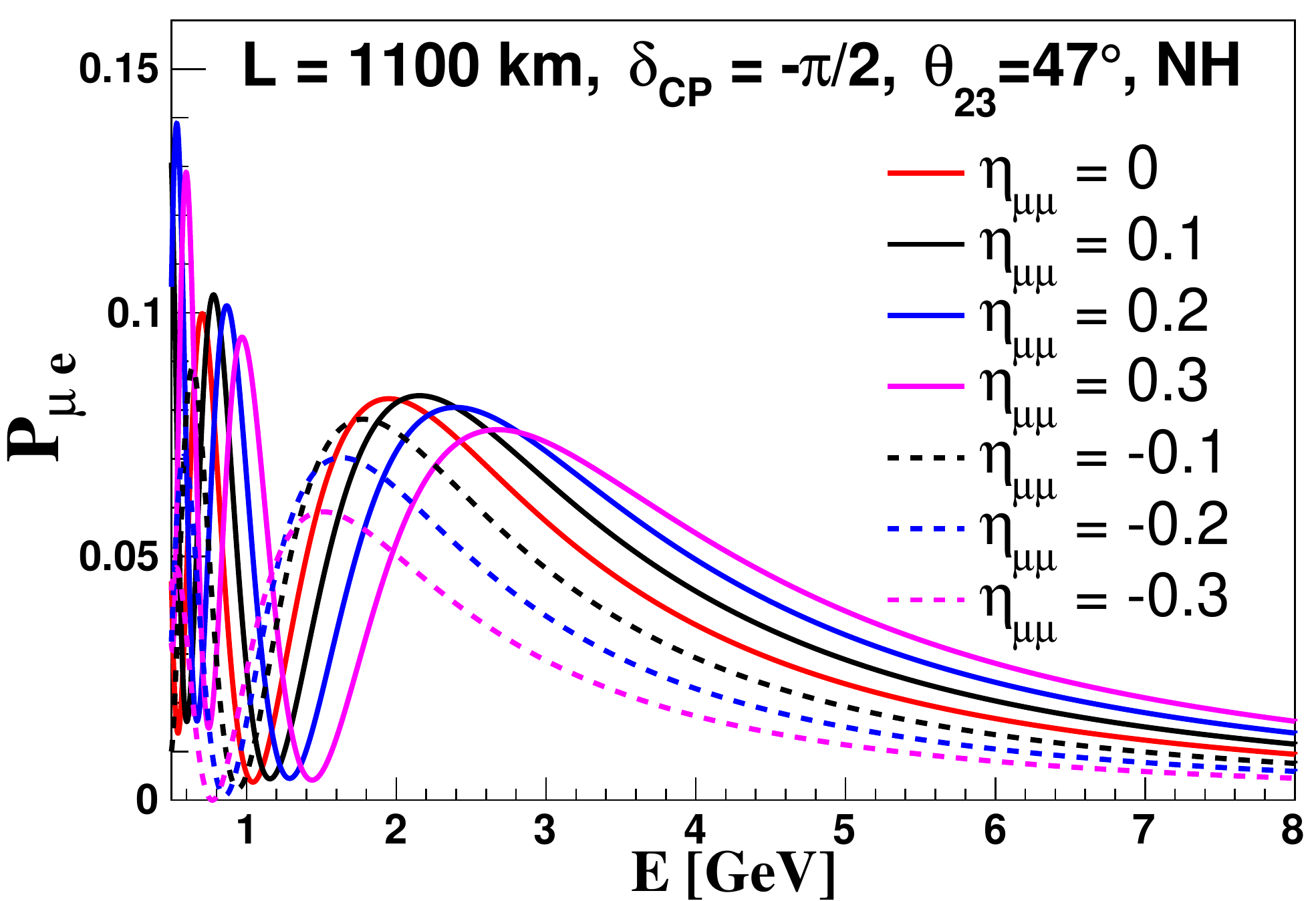} 
 	\includegraphics[width=0.32\linewidth, height = 5cm]{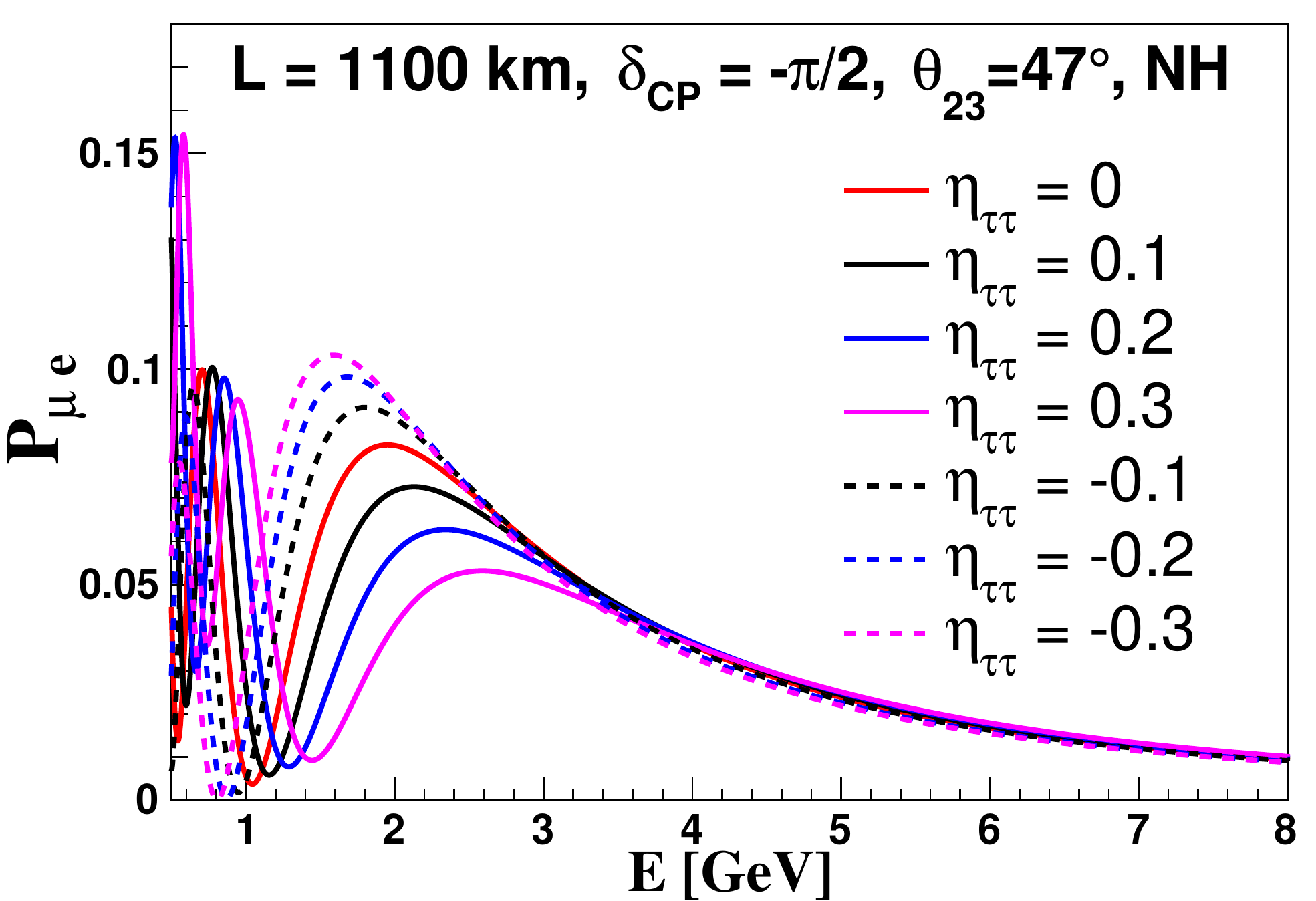} 
 	\caption{The effects of $\eta_{ee}$ (left--column), $\eta_{\mu\mu}$ (middle--column) and $\eta_{\tau\tau}$ (right--column) on $P_{\mu e}$ at the baselines corresponding to DUNE (top--row), T2HK (middle--row) and T2HKK (bottom--row). Here, $\delta_{CP}$ = -$\pi$/2, $\theta_{23}$ = 47$^\circ$ and true mass Hierarchy = NH. In all the plots, the red solid--curve is for no--NSI case while other solid (dashed) curves are for positive (negative) NSI parameters.}
 	\label{fig:probability3}
 \end{figure}
 
 The effects of the diagonal scalar NSI elements $\eta_{ee}$ (left--column), $\eta_{\mu\mu}$ (middle--column) and $\eta_{\tau\tau}$ (right--column) on $P_{\mu e}$ as a function of neutrino energy are shown in figure \ref{fig:probability3}. The plots corresponding to the baselines of DUNE (top--row), T2HK (middle--row) and T2HKK (bottom--row) are shown here. The probabilities are calculated for $\delta_{CP}$ = -- $90^\circ$ and $\theta_{23}$ = $47^\circ$. In all the plots, the solid--red line represents the case without scalar NSI i.e. $\eta_{\alpha\beta}$ = 0. The solid (dashed) lines in black, blue and magenta are for the chosen positive (negative) $\eta_{ee}$, $\eta_{\mu\mu}$ and $\eta_{\tau\tau}$ respectively. We observe that,

 \begin{itemize}
     \item The presence of Scalar NSI parameters show significant effects on the oscillation probabilities at all the three baselines, especially around the oscillation maxima.
     \item A positive (negative) $\eta_{ee}$ enhances (suppresses) the probabilities around the oscillation maxima while a positive (negative) $\eta_{\tau\tau}$ exhibits complementary variations.
     \item A positive (negative) $\eta_{\mu\mu}$ shifts the oscillation maxima towards the higher (lower) energies with minor suppression on the amplitude.
 \end{itemize}
 
 The visible effects of scalar NSI on neutrino oscillations are interesting and we explore it further by constructing a CP-asymmetry parameter at the probability level.

 \subsection{Effects on CP asymmetry}
 \label{sec:CP_asymmetry}
In this work, we are primarily exploring the possible impact of scalar NSI on the CP-measurement potential of the three chosen long--baseline experiments. We construct the CP--asymmetry parameter at the probability level as,

\begin{equation}\label{def_asym}
A_{CP} = \frac{P_{\mu e}-\bar{P}_{\mu e}}{P_{\mu e}+\bar{P}_{\mu e}}\,,
\end{equation}
where, $P_{\mu e}$ and $\bar{P}_{\mu e}$ are the appearance probabilities of $\nu_e$ and $\bar{\nu_e}$ respectively. The CP asymmetry parameter ($A_{CP}$) can be an estimate of CP violation as it quantifies the change in oscillation probabilities when CP phase changes its sign. The shape and size of the CP--asymmetry curve largely depends on the baseline and energy.

\begin{figure}[!h]
	\centering
	\includegraphics[width=0.32\linewidth, height = 5.5cm]{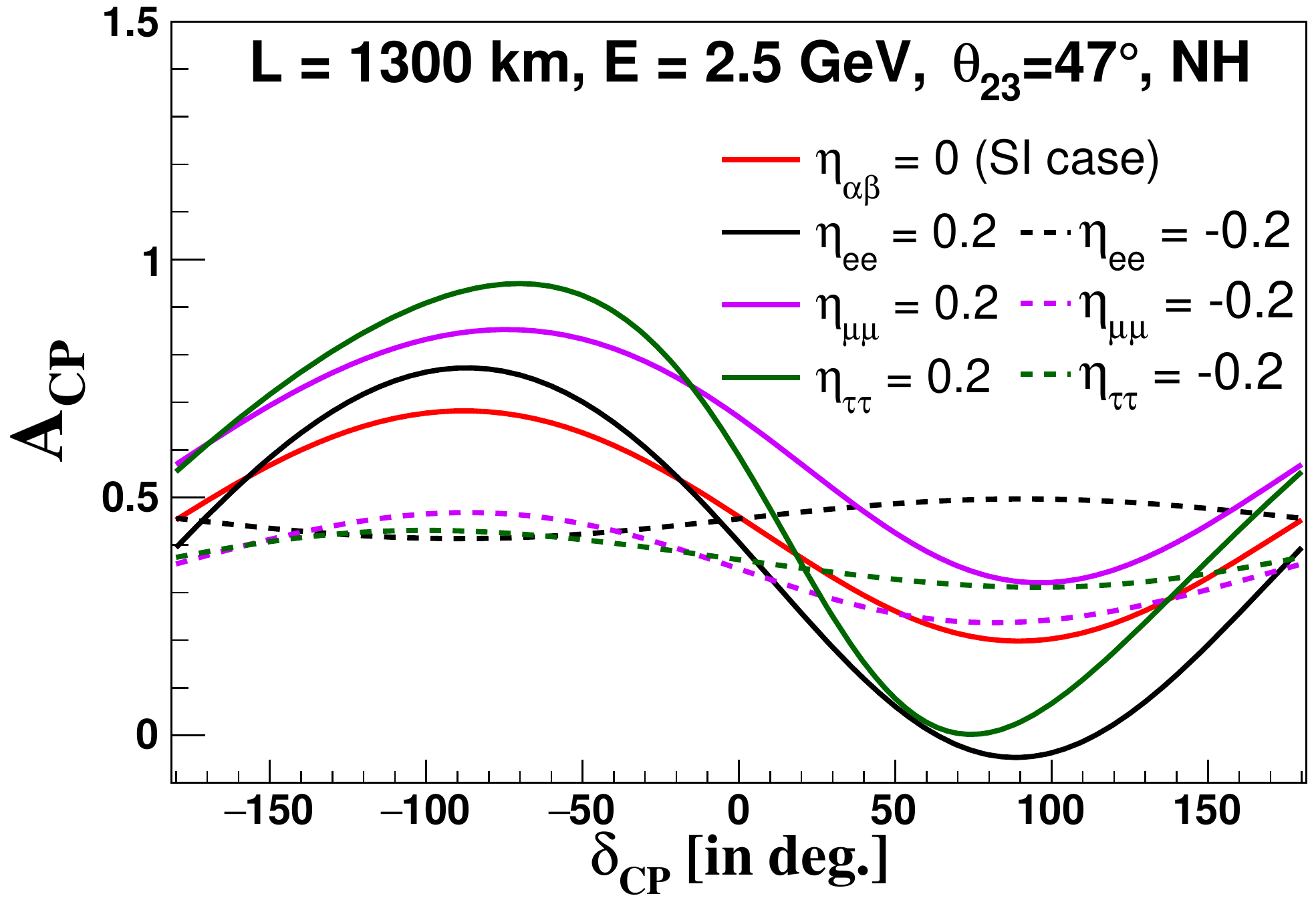} 
	\includegraphics[width=0.32\linewidth, height = 5.5cm]{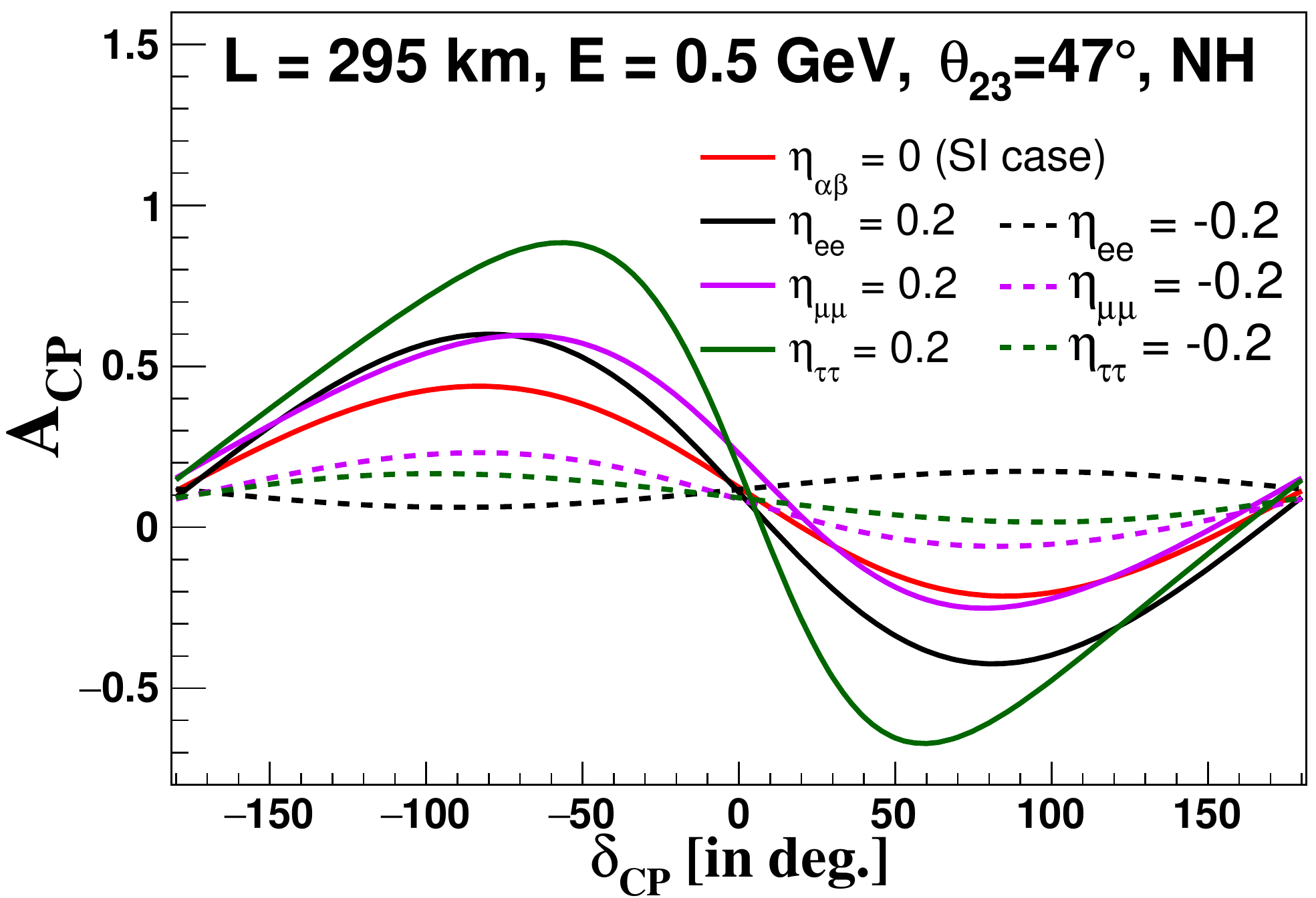} 
	\includegraphics[width=0.32\linewidth, height = 5.5cm]{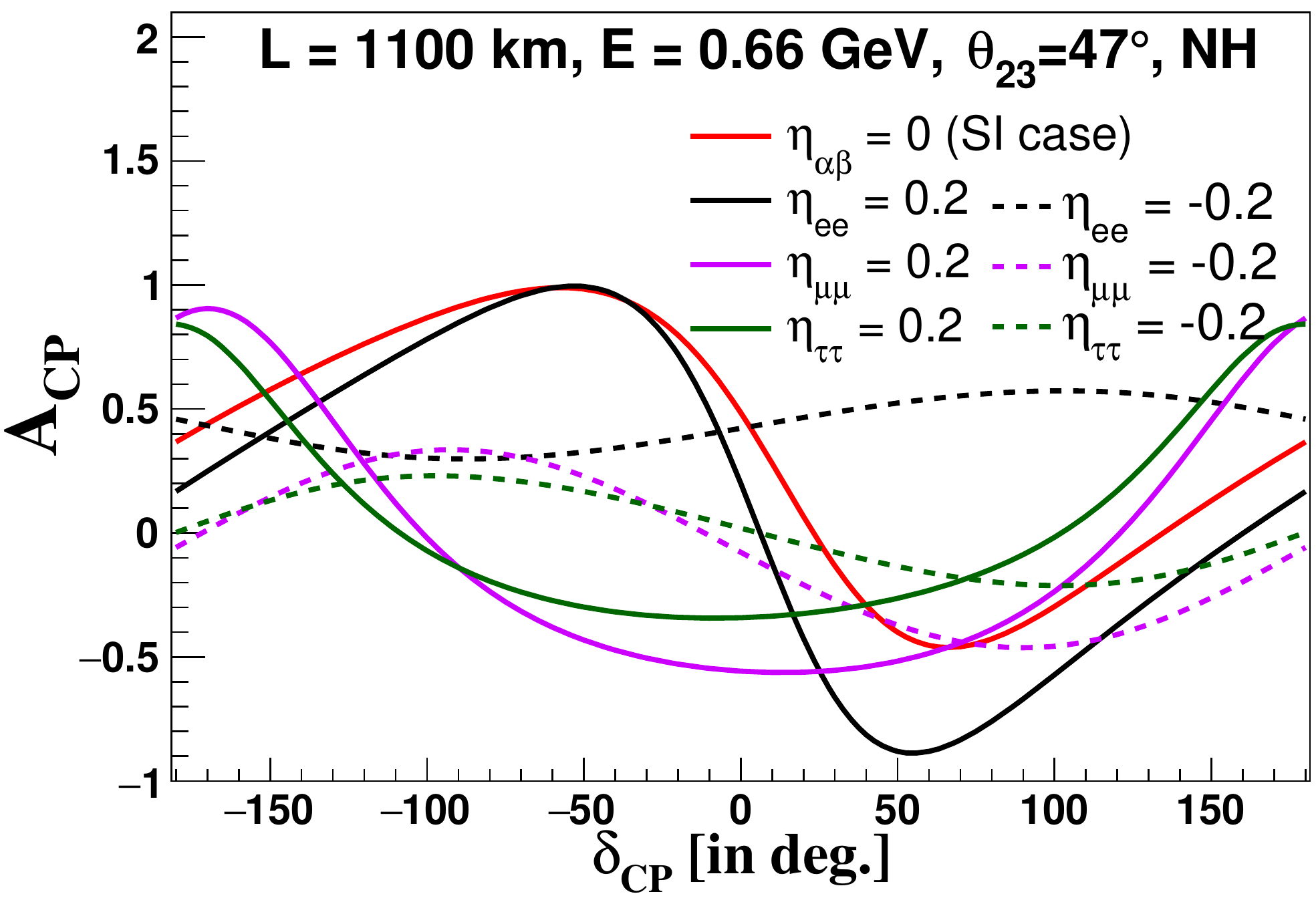} 
	\caption{The CP--asymmetry vs $\delta_{CP}$ plot for DUNE (left--panel), T2HK (middle--panel) and T2HKK (right--panel) in presence of $\eta_{\alpha\beta}$ at corresponding peak energies. Here, $\theta_{23}$ = 47$^\circ$ and true mass hierarchy = NH. In all the three plots, the solid--red curve is for no scalar NSI case and other coloured solid (dashed) curves are for chosen positive (negative) $\eta_{\alpha\beta}$. }
	\label{fig:CP_assymetry}
\end{figure}

We show the CP--asymmetry in presence of scalar NSI as a function of $\delta_{CP}$ at the baselines and peak energies of DUNE (left--panel), T2HK (middle--panel) and T2HKK (right--panel) in figure \ref{fig:CP_assymetry}. Note that, the peak energies for DUNE, T2HK and T2HKK have been considered as 2.5 GeV, 0.5 GeV and 0.66 GeV respectively. The solid--red curve in all the plots represent the no-scalar NSI case, i.e. $\eta_{\alpha\beta}$ = 0. The solid (dashed) curves in black, magenta and green are for positive (negative) values of scalar NSI elements. The observations from figure \ref{fig:CP_assymetry} are listed below.

\begin{itemize}
    \item The presence of scalar NSI results in degeneracy for different sets of ($\eta_{\alpha\beta}$, $\delta_{CP}$), which would impact the expected CP asymmetry at DUNE, T2HK and T2HKK.
    
    \item At DUNE, a positive $\eta_{ee}$ enhances $A_{CP}$ in the range $\delta_{CP}$ $\in$ [-150$^\circ$, 10$^\circ$], while a negative $\eta_{ee}$ enhances $A_{CP}$ in the range $\delta_{CP}$ $\in$ [0, 180$^\circ$]. At T2HK, a positive (negative) $\eta_{ee}$ enhances (suppresses) the $A_{CP}$ values in $\delta_{CP}$ $\in$ [-180$^\circ$, 0]. For T2HKK, however, the chosen positive $\eta_{ee}$ suppresses the $A_{CP}$ parameter throughout the entire $\delta_{CP}$ range.
    
    \item At DUNE, a positive $\eta_{\mu\mu}$ enhances the $A_{CP}$ values in the whole $\delta_{CP}$ range. For a negative $\eta_{\mu\mu}$, we see an enhancement in $A_{CP}$ in the range $\delta_{CP}$ $\in$ [60$^\circ$, 140$^\circ$] while for other values of $\delta_{CP}$ we observe a suppression. At T2HK, a positive (negative) $\eta_{\mu\mu}$ enhances (suppresses) $A_{CP}$ in the range $\delta_{CP}$ $\in$ [-180$^\circ$, 40$^\circ$]. At T2HKK, we mostly see a suppression in $A_{CP}$ for a negative $\eta_{\mu\mu}$. However, for a positive $\eta_{\mu\mu}$ at T2HKK we observe a fluctuation in the variation pattern.
    
    \item At DUNE, a positive $\eta_{\tau\tau}$ enhances $A_{CP}$ for $\delta_{CP}$ <0. We note a crossover and suppression through the $delta_{CP}$ range [30$^\circ$, 140$^\circ$]. We observe a similar trend at T2HK as well. For a negative $\eta_{\tau\tau}$, at both DUNE and T2HK, $A_{CP}$ appear to be very mild dependent on $\delta_{CP}$. At T2HKK, we note a strong fluctuation with $\eta_{\tau\tau}$ of either polarity.
    \end{itemize}

\section{Results and Discussion}
 \label{sec:results}
 Motivated by the significant impact on the oscillation probabilities and on $A_{CP}$, we focus into the effects of the scalar NSI on the event rates at the three detectors. We then perform a statistical analysis by constructing various $\chi^2$ parameters to probe the scalar NSI effects on $\delta_{CP}$. 
 
\begin{figure}[h]
	\centering
	\includegraphics[width=0.32\linewidth, height = 5cm]{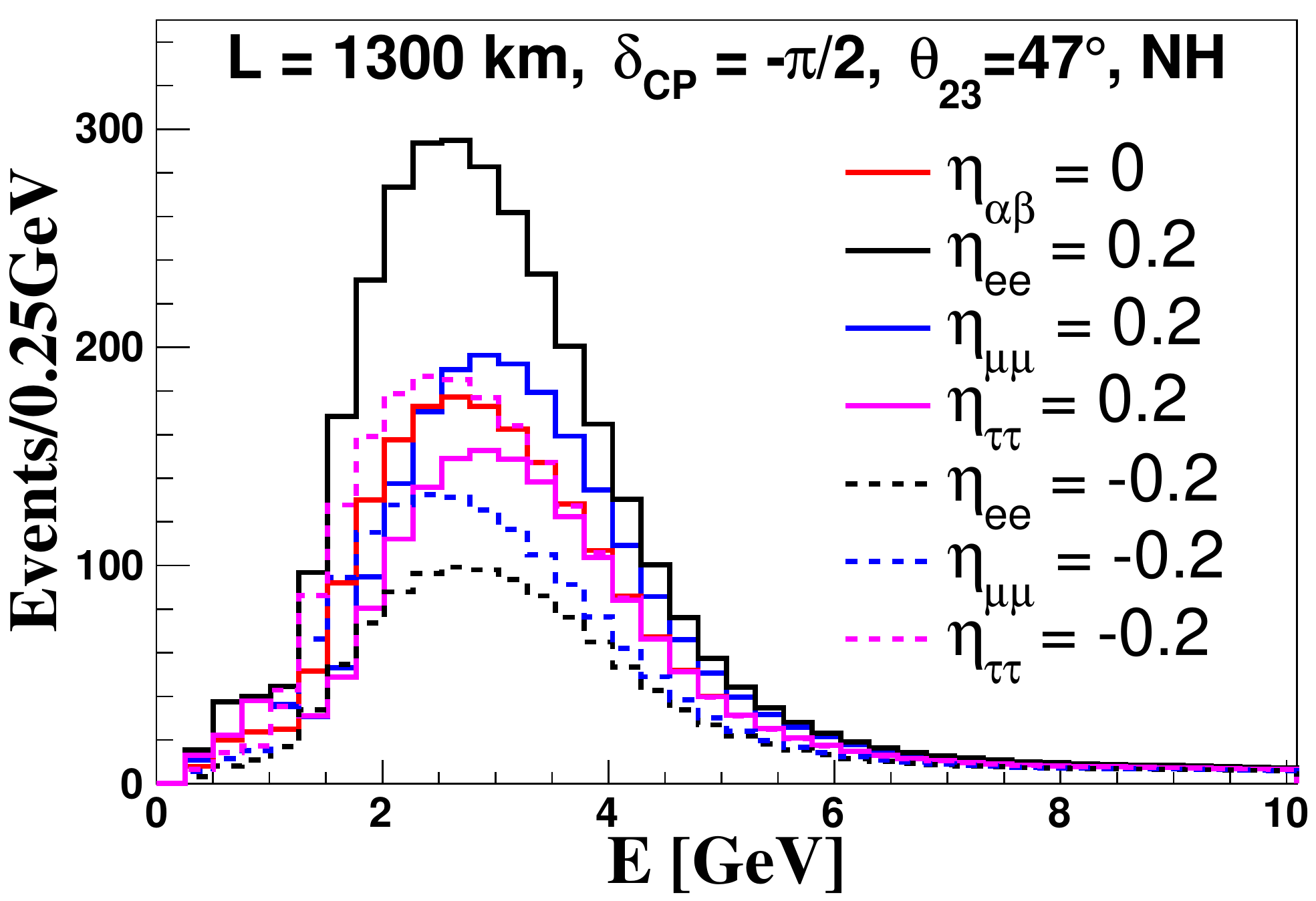} 
	\includegraphics[width=0.32\linewidth, height = 5cm]{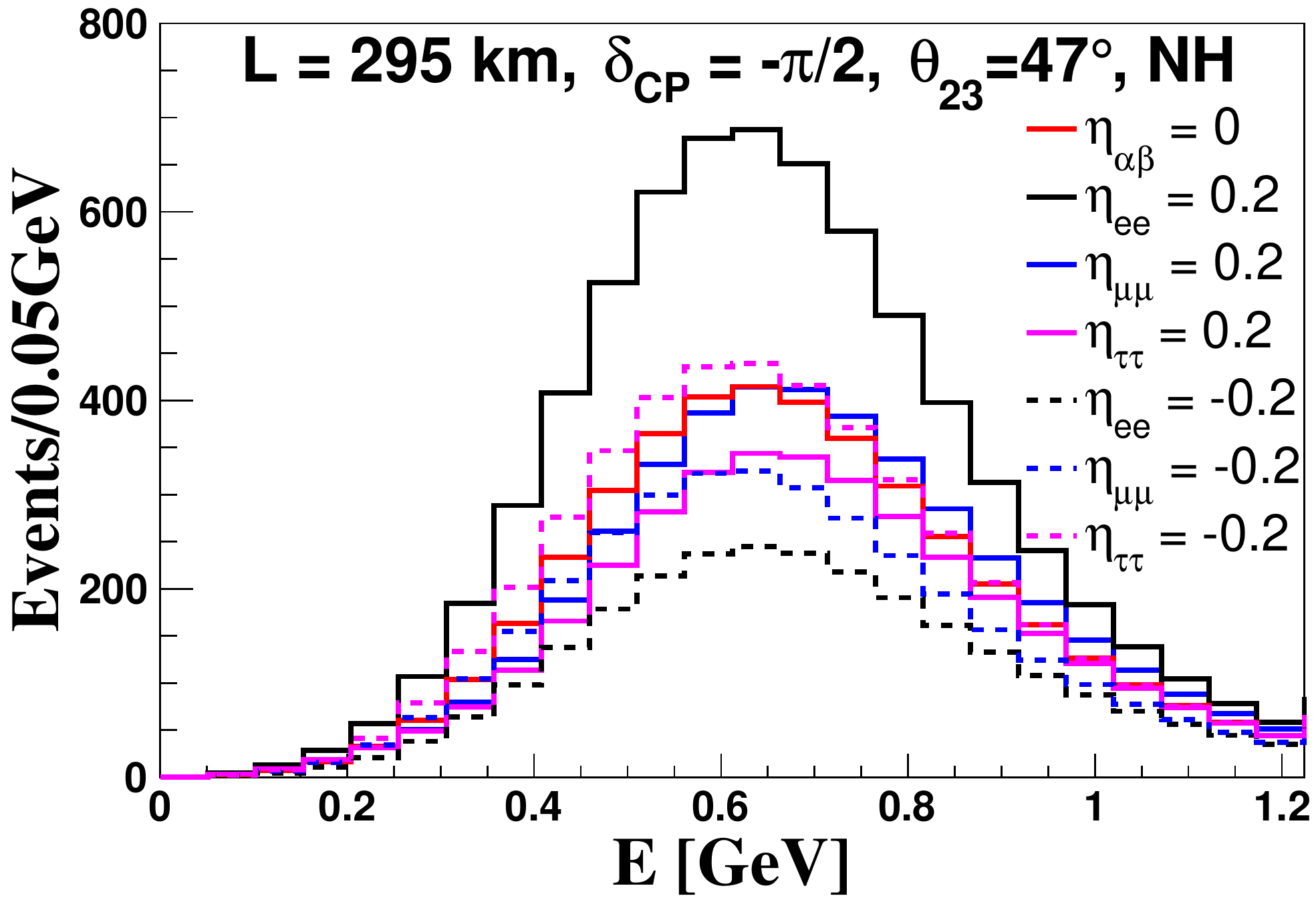} 
	\includegraphics[width=0.32\linewidth, height = 5cm]{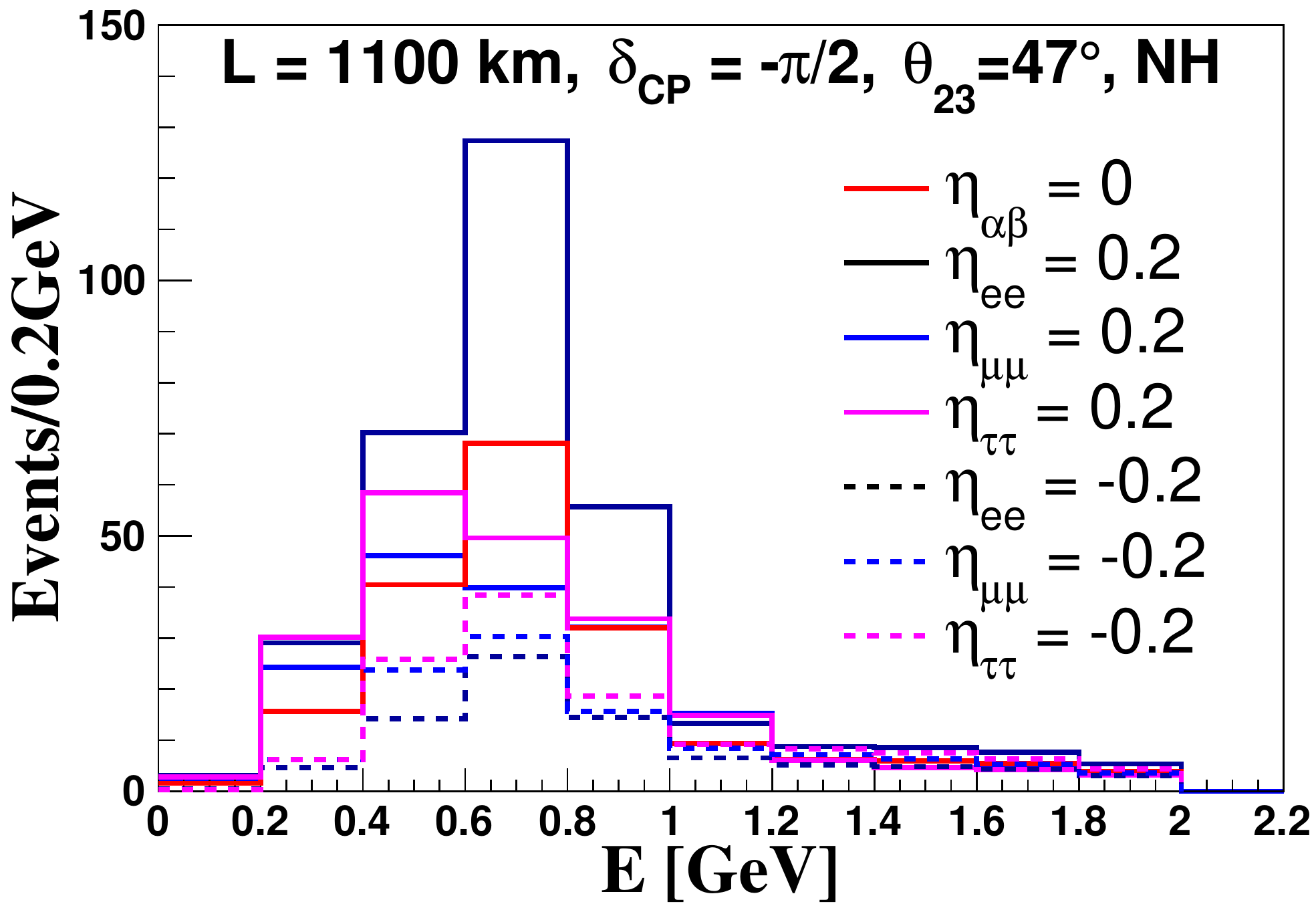} 
	\caption{The binned event rates of DUNE (left), T2HK (middle) and T2HKK (right) as a function of neutrino energy at $\delta_{CP}$ = -$\pi$/2, $\theta_{23}$ = 47$^\circ$ and NH for different choices of $\eta_{\alpha\beta}$.}
	\label{fig:event_rate_1}
\end{figure}
 
 \subsection{Effects on event rates}
\begin{figure}[h]
	\centering
	\includegraphics[width=0.32\linewidth, height = 5cm]{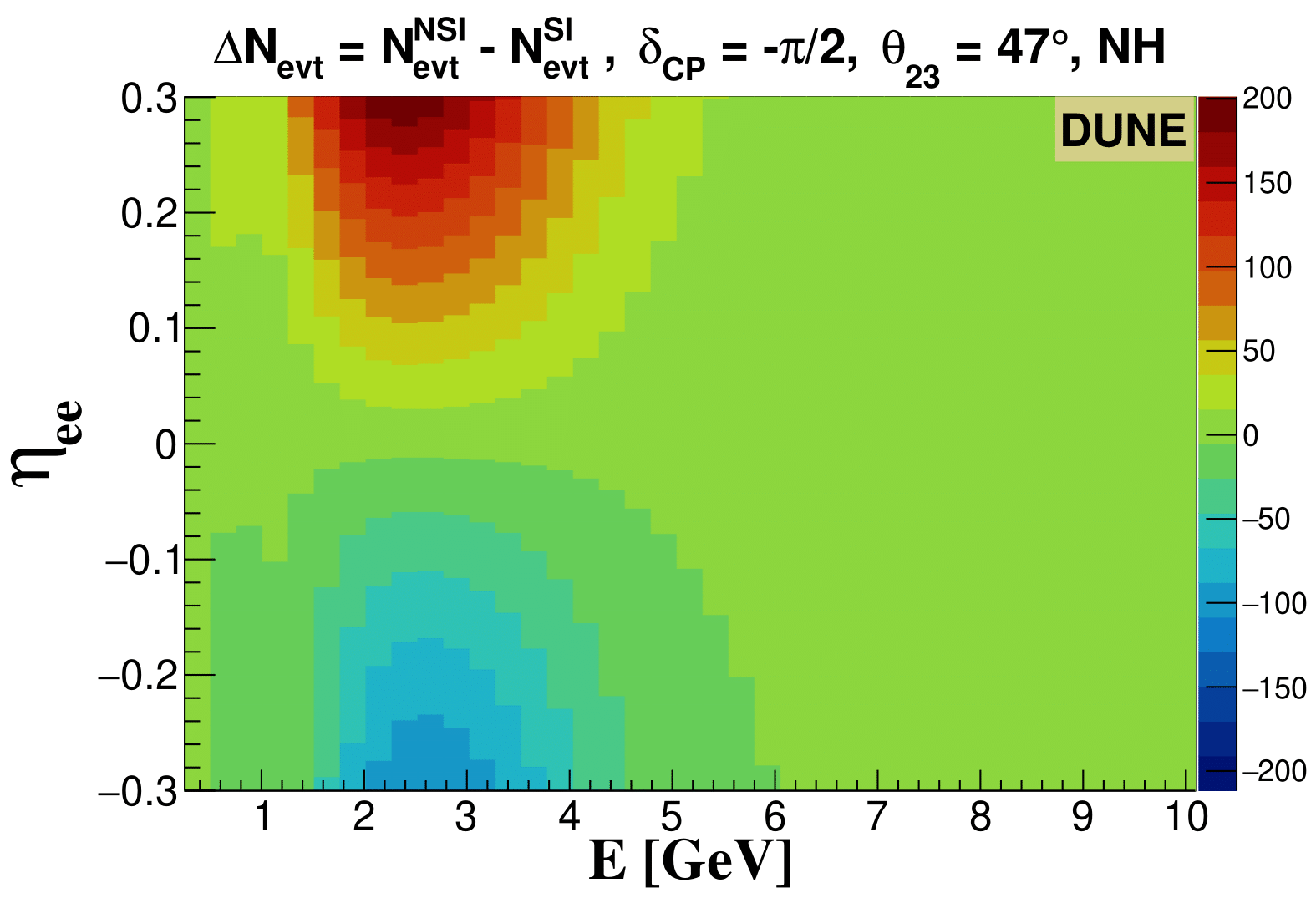} 
	\includegraphics[width=0.32\linewidth, height = 5cm]{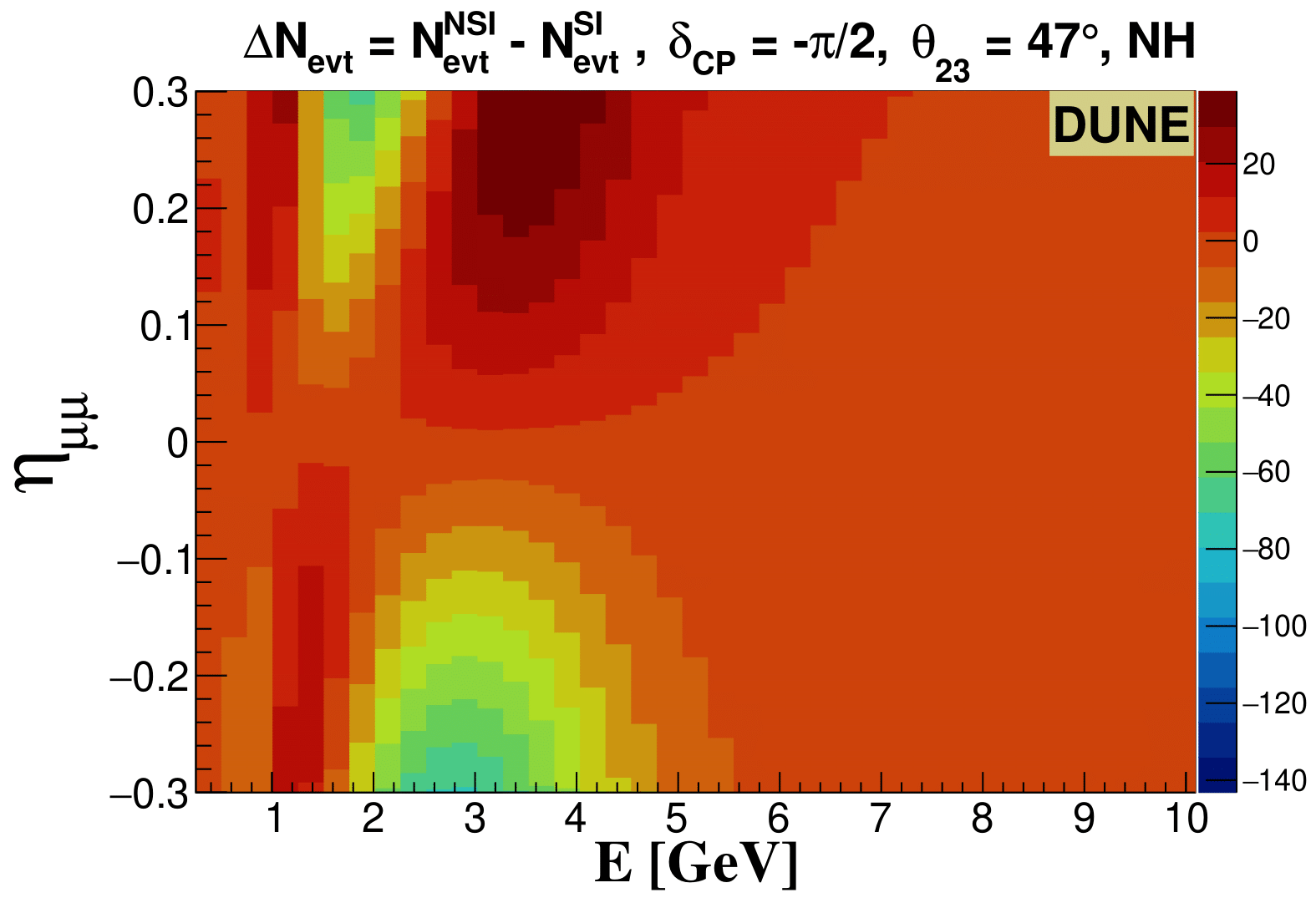} 
	\includegraphics[width=0.32\linewidth, height = 5cm]{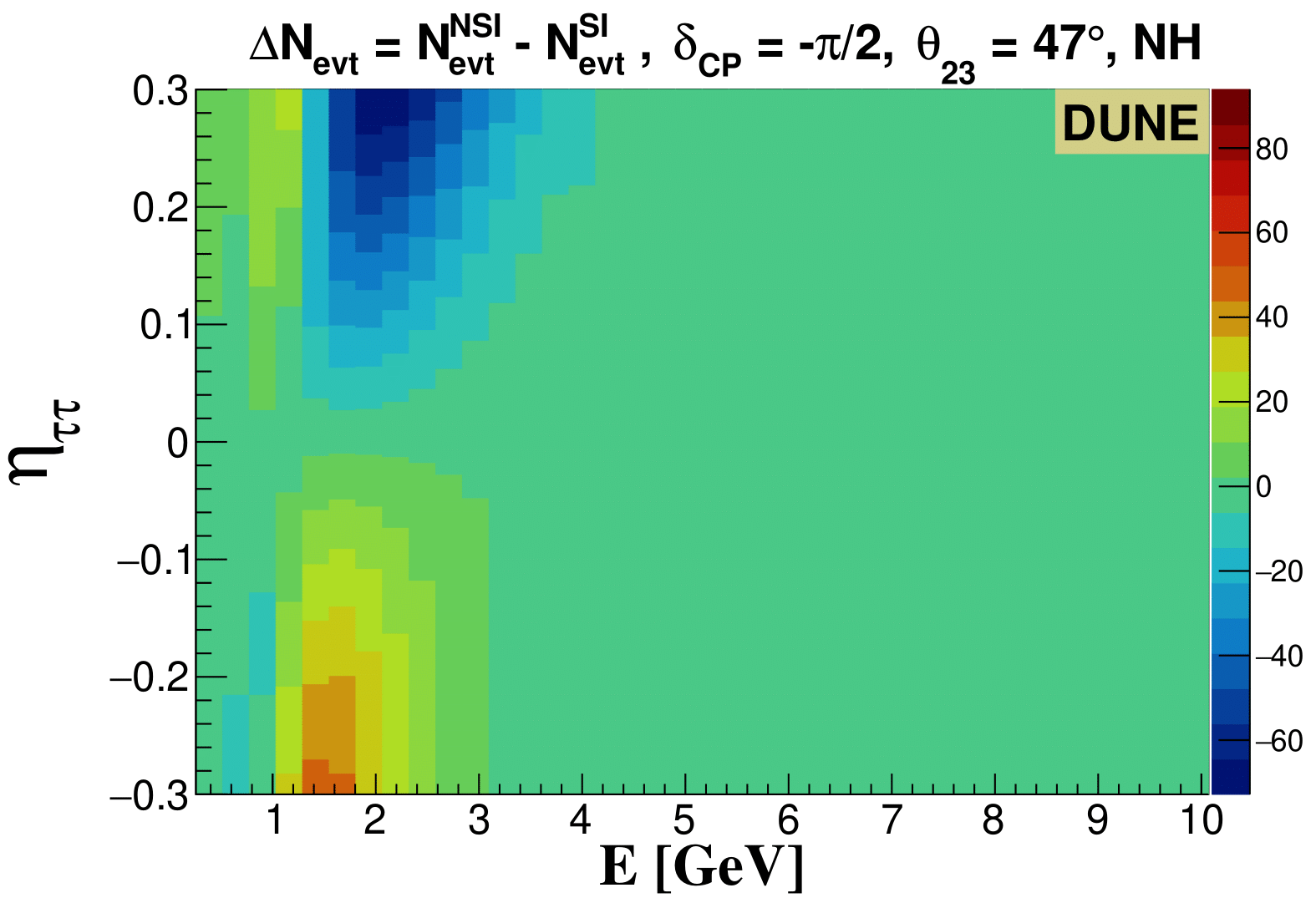} 

	\includegraphics[width=0.32\linewidth, height = 5cm]{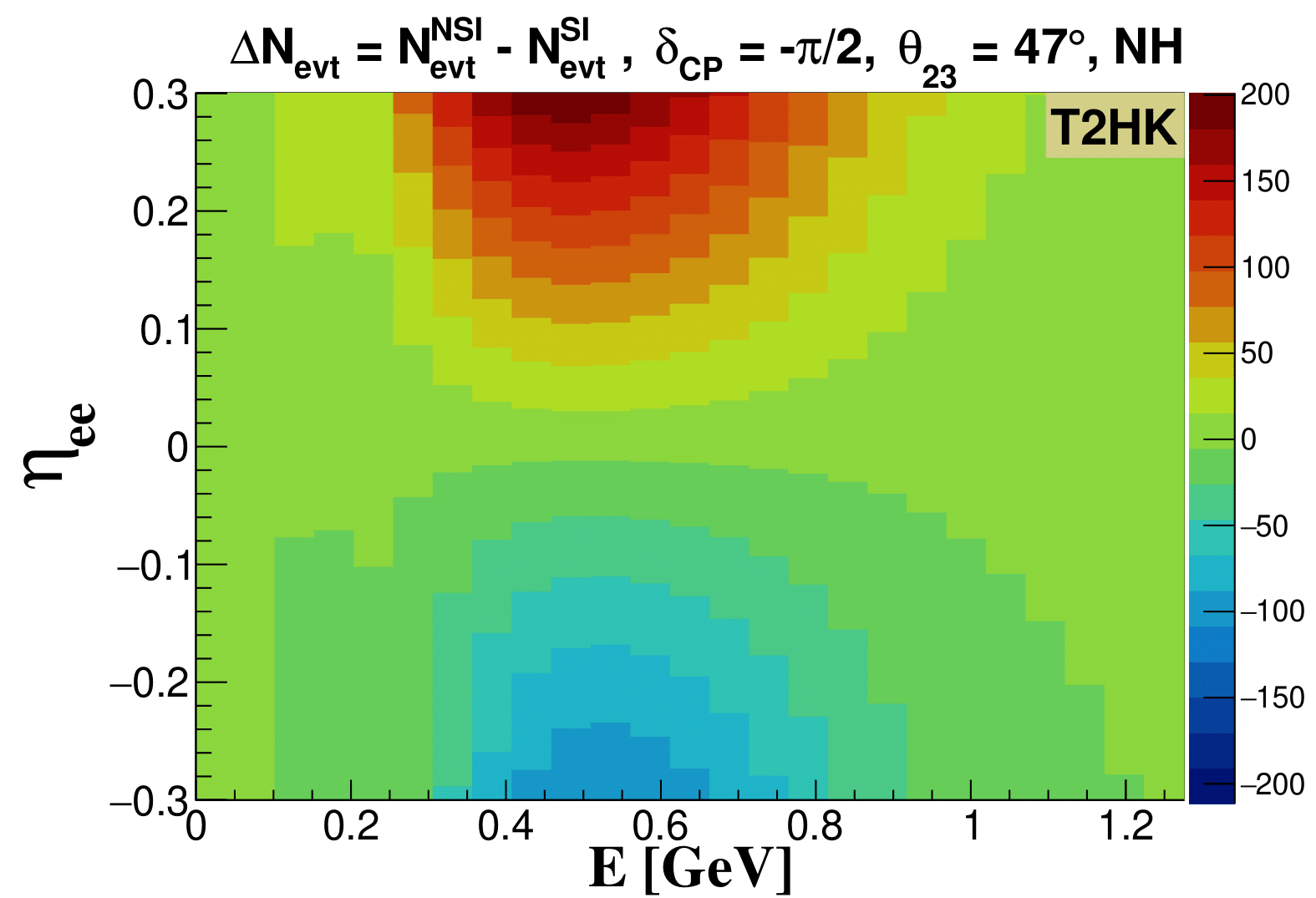} 
	\includegraphics[width=0.32\linewidth, height = 5cm]{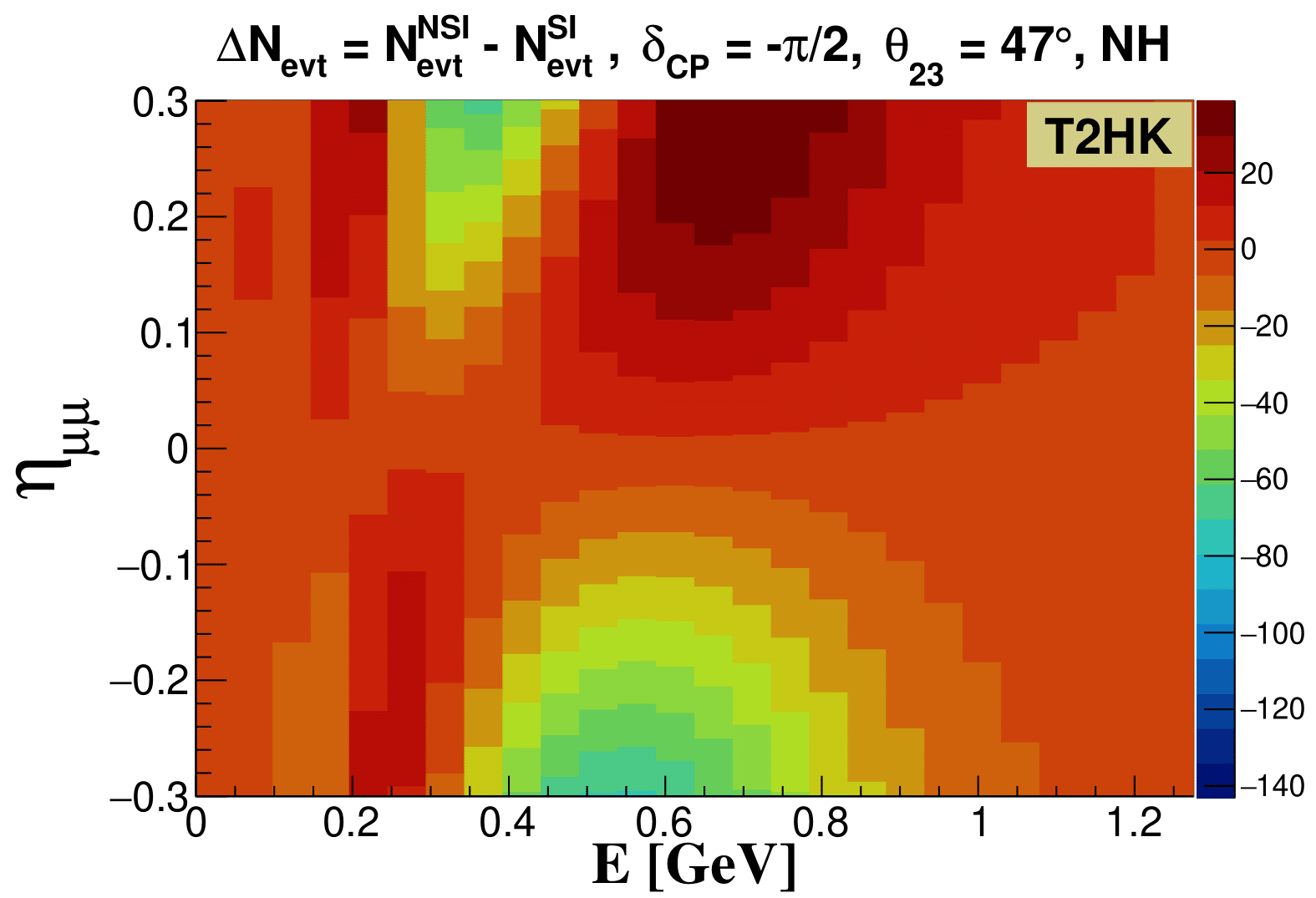} 
	\includegraphics[width=0.32\linewidth, height = 5cm]{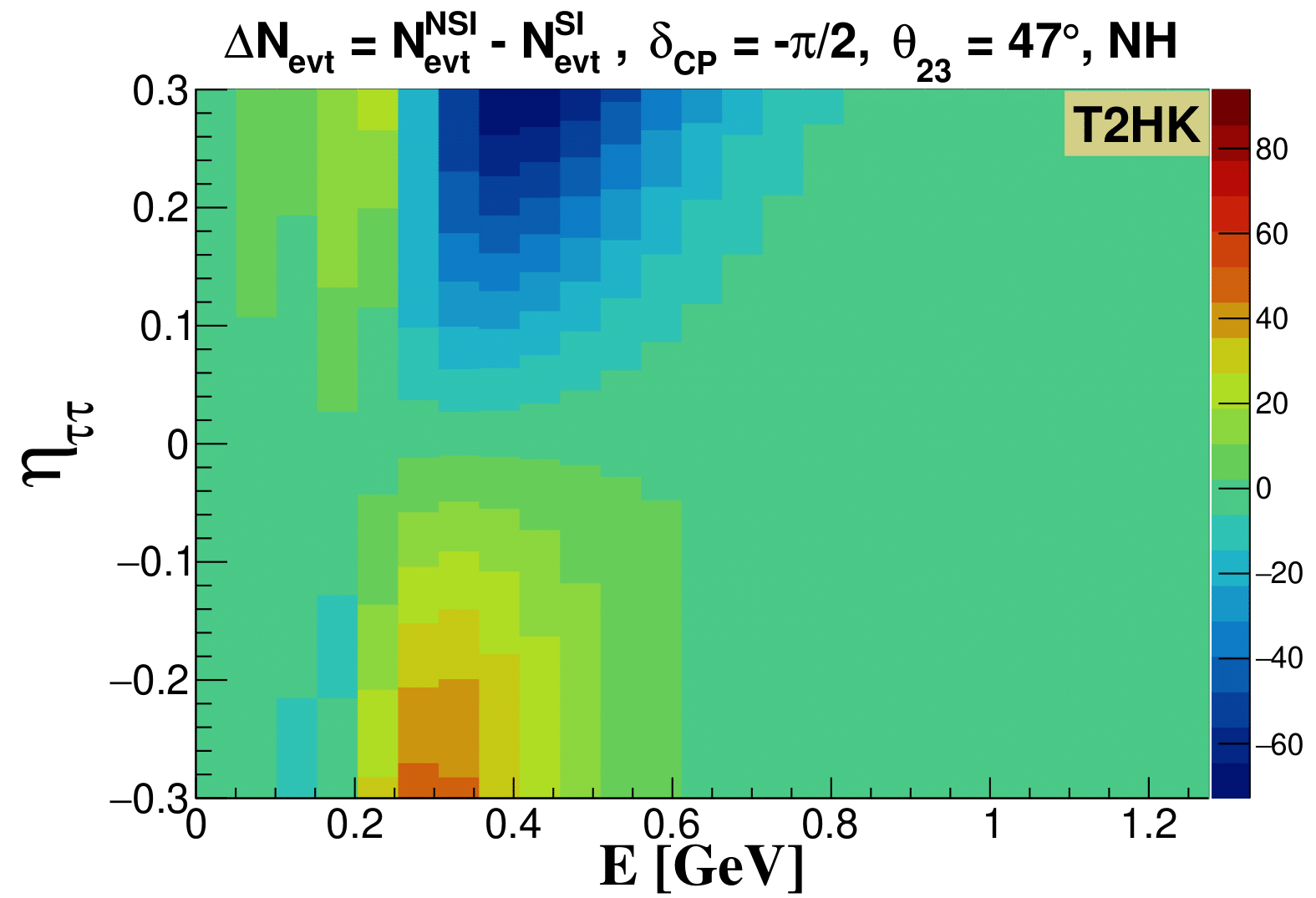} 

	\includegraphics[width=0.32\linewidth, height = 5cm]{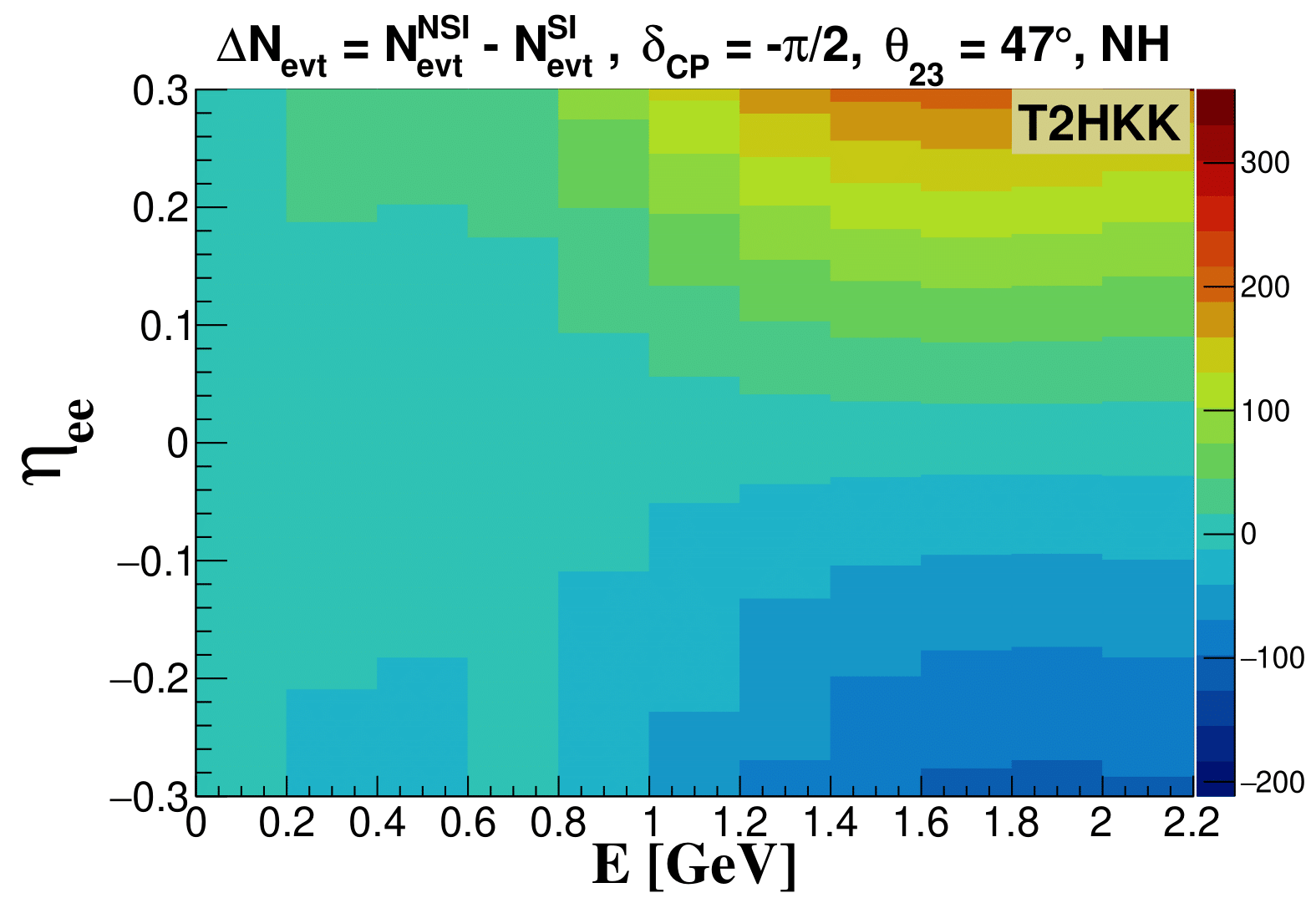} 
	\includegraphics[width=0.32\linewidth, height = 5cm]{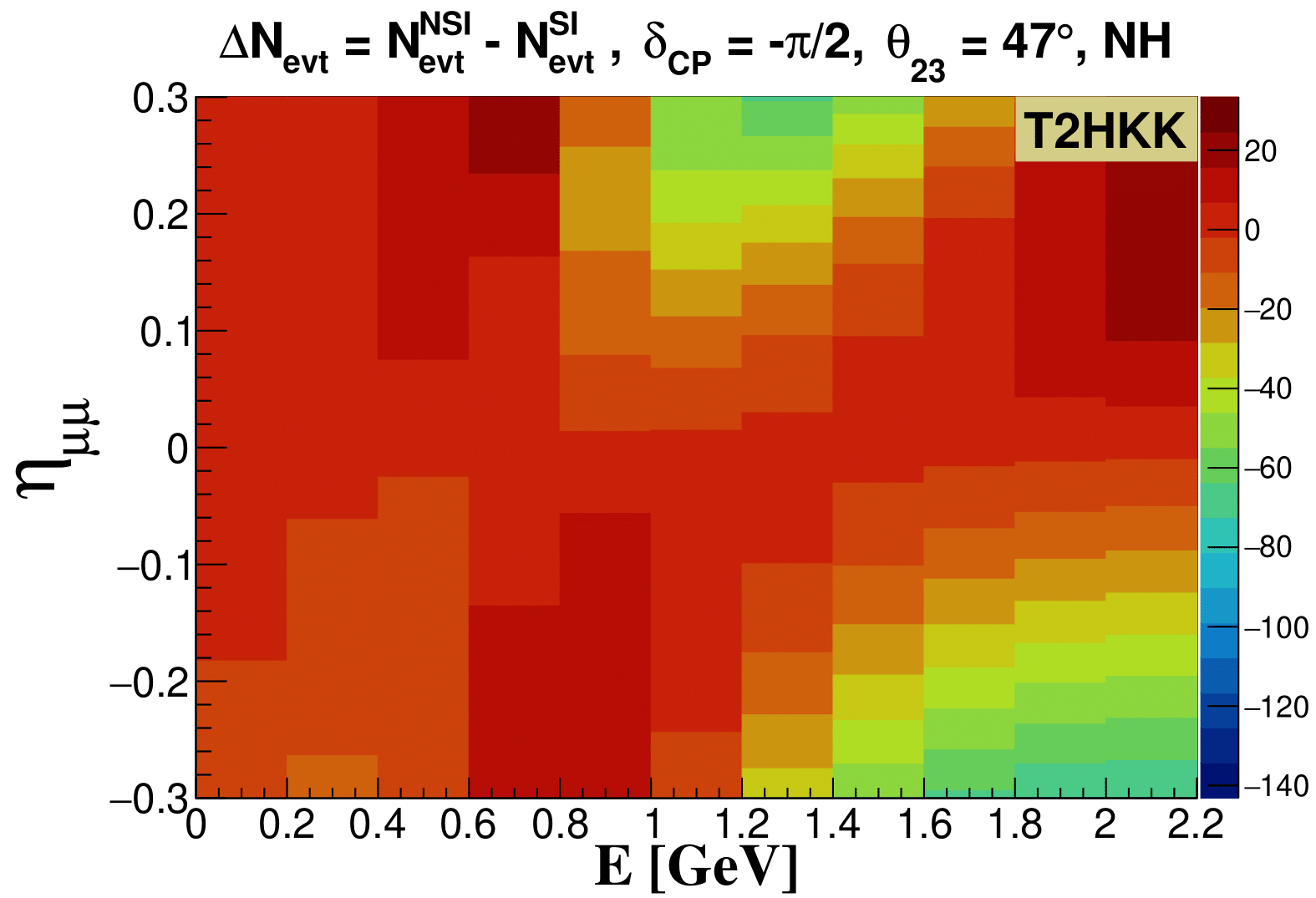} 
	\includegraphics[width=0.32\linewidth, height = 5cm]{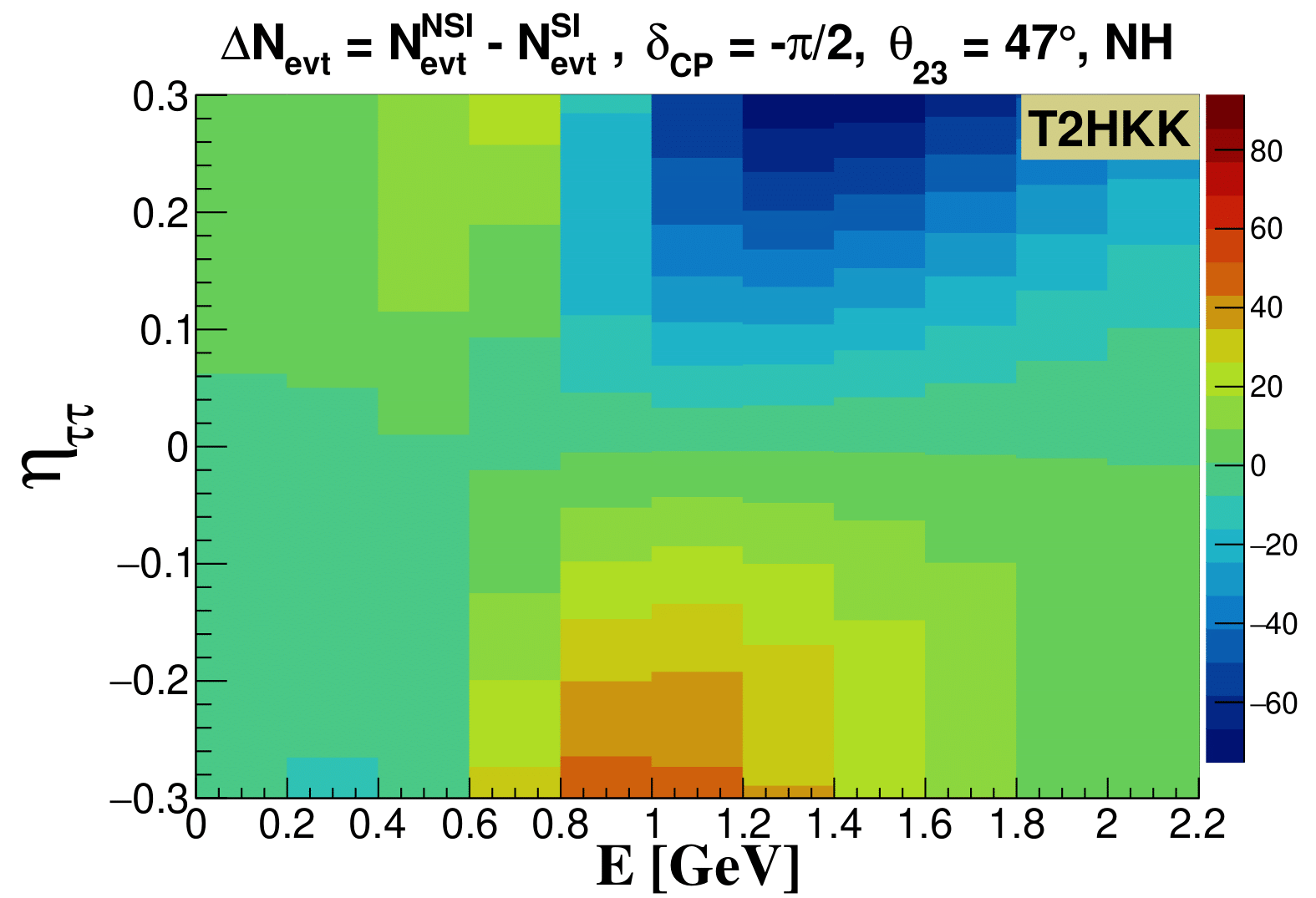} 
	\caption{The variation of $\Delta N_{evt}$ of DUNE (top-row), T2HK (middle--row) and T2HKK (bottom--row) as a function of neutrino energy at fixed $\delta_{CP}$ = -$\pi$/2, $\theta_{23}$ = 47$^\circ$ and NH for different choices of $\eta_{\alpha\beta}$ $\in$ [-0.3, 0.3]. In the figure the left--column is for non-zero $\eta_{ee}$, the middle--column is for non-zero $\eta_{\mu\mu}$ and the right--column is for non-zero $\eta_{\tau\tau}$.}
	\label{fig:event_rate_4}
\end{figure}

We discuss here in detail, the effects of scalar NSI parameters on the binned event rates at the three LBL experiments. In figure \ref{fig:event_rate_1}, we show the raw binned event rates of DUNE (left--panel), T2HK (middle--panel) and T2HKK (right--panel) as a function of the true neutrino energy. We have varied the $\eta_{\alpha\beta}$ parameters, one at a time, in the range [-0.3, 0.3] while keeping $\delta_{CP}$ (true) = -$\pi/2$ and $\theta_{23}$ (true) = $47^\circ$. We quantify the effects of scalar NSI on the event rates in terms of the parameter $\Delta N_{evt}$ which is defined as 
\begin{equation}
\Delta N_{evt} = N_{evt}^{NSI} - N_{evt}^{SI},    
\end{equation}

\noindent where, $N_{evt}^{NSI}$ ($N_{evt}^{SI}$) are the binned event rates at the far detector of the experiment in presence of scalar NSI (in absence of scalar NSI). The values of $\Delta N_{evt}$ quantify the impact of scalar NSI on the event rates. In figure \ref{fig:event_rate_4}, we have shown $\Delta N_{evt}$ as a function of neutrino energy and $\eta_{\alpha\beta}$ for DUNE (top--row), T2HK (middle--row), and T2HKK (bottom--row). We see from figure \ref{fig:event_rate_1} and figure~\ref{fig:event_rate_4} that,

\begin{itemize}
    \item A positive (negative) $\eta_{ee}$ increases (decreases) the binned events around the first oscillation maxima for DUNE and T2HK and around the second oscillation maxima for T2HKK.
    \item For $\eta_{\mu\mu}$, however, we observe a varied scenario with energy. For both positive and negative $\eta_{\mu\mu}$, there are certain increments and decrements of the event rates at various energy ranges. For example, if we look into the effects of a positive $\eta_{\mu\mu}$, we see that, at DUNE the event rates gets enhanced in E $\in$ [2.5 GeV, 5 GeV] and gets reduced in E $\in$ [1.5 GeV, 2.5 GeV]. At T2HK, we observe enhanced (suppressed) event rates in E $\in$ [2.0 GeV, 2.2 GeV] (E $\in$ [1 GeV, 1.8 GeV]) and at T2HKK, we find enhanced (suppressed) event rates in E $\in$ [0.6 GeV, 1.2 GeV] (E $\in$ [0.3 GeV, 0.5 GeV]). For a negative $\eta_{\mu\mu}$, we come across an opposite variation with energy at all the three experiments. 
   
     \item A positive (negative) $\eta_{\tau\tau}$ mostly decreases (increases) the event rates around the oscillation maxima prominently for DUNE and T2HK. However, at some lower energies ([1 GeV, 2GeV] for DUNE, [0.1 GeV, 0.3GeV] for T2HK and [0.2 GeV, 0.8 GeV] for T2HKK) we see a nominal increase in the event rates with a positive $\eta_{\tau\tau}$. 
     
    \item The behaviour of the binned event rates of the experiments as shown in figure \ref{fig:event_rate_1}, are in good agreement with the neutrino oscillation probabilities as observed in figure~\ref{fig:probability3}
\end{itemize}

\subsection{Exploring the sensitivities using a $\chi^2$ analysis}
We now focus on exploring the possible impact of $\eta_{\alpha\beta}$ on the $\delta_{CP}$ measurement potential of the three experiments. We probe the three experiments' sensitivity towards the CP--conserving and CP--violating phases of $\delta_{CP}$ through the statistical $\chi^2$ as defined below.

\begin{equation}
\label{eq:chisq}
\chi^2 \equiv  \min_{\eta}  \sum_{i} \sum_{j}
\frac{\left[N_{true}^{i,j} - N_{test}^{i,j} \right]^2 }{N_{true}^{i,j}},
\end{equation}
where, $N_{true}^{i,j}$ and $N_{test}^{i,j}$ are the number of true and test events in the $\{i,j\}$-th bin respectively. 

We have performed a sensitivity analysis of the experiments' capability towards constraining $\eta_{\alpha\beta}$. We have also explored the effects of $\eta_{\alpha\beta}$ on the CP--violation measurements of these experiments. The CP--violation sensitivity may be defined as the experiments' ability to differentiate between CP--conserving and CP--violating values. We have marginalized over the systematic uncertainties. The sensitivities are first obtained for the individual experiments. We then consider DUNE+T2HK and DUNE+T2HKK combinations to explore the synergy. We discuss the results in the following.

\subsection{Sensitivity to scalar NSI parameters}
\begin{figure}[h]
	\centering
	\includegraphics[width=0.32\linewidth, height = 5cm]{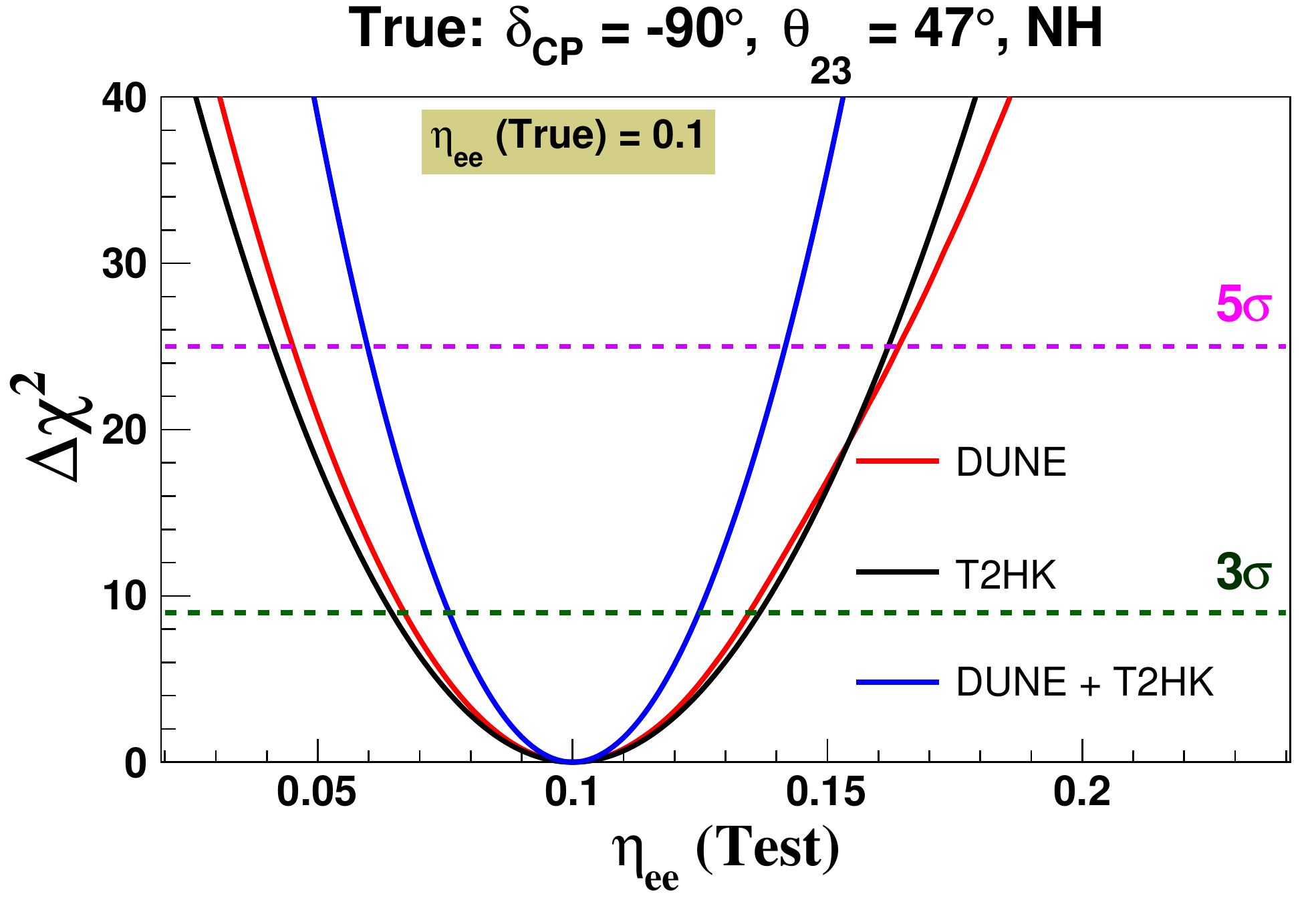} 
	\includegraphics[width=0.32\linewidth, height = 5cm]{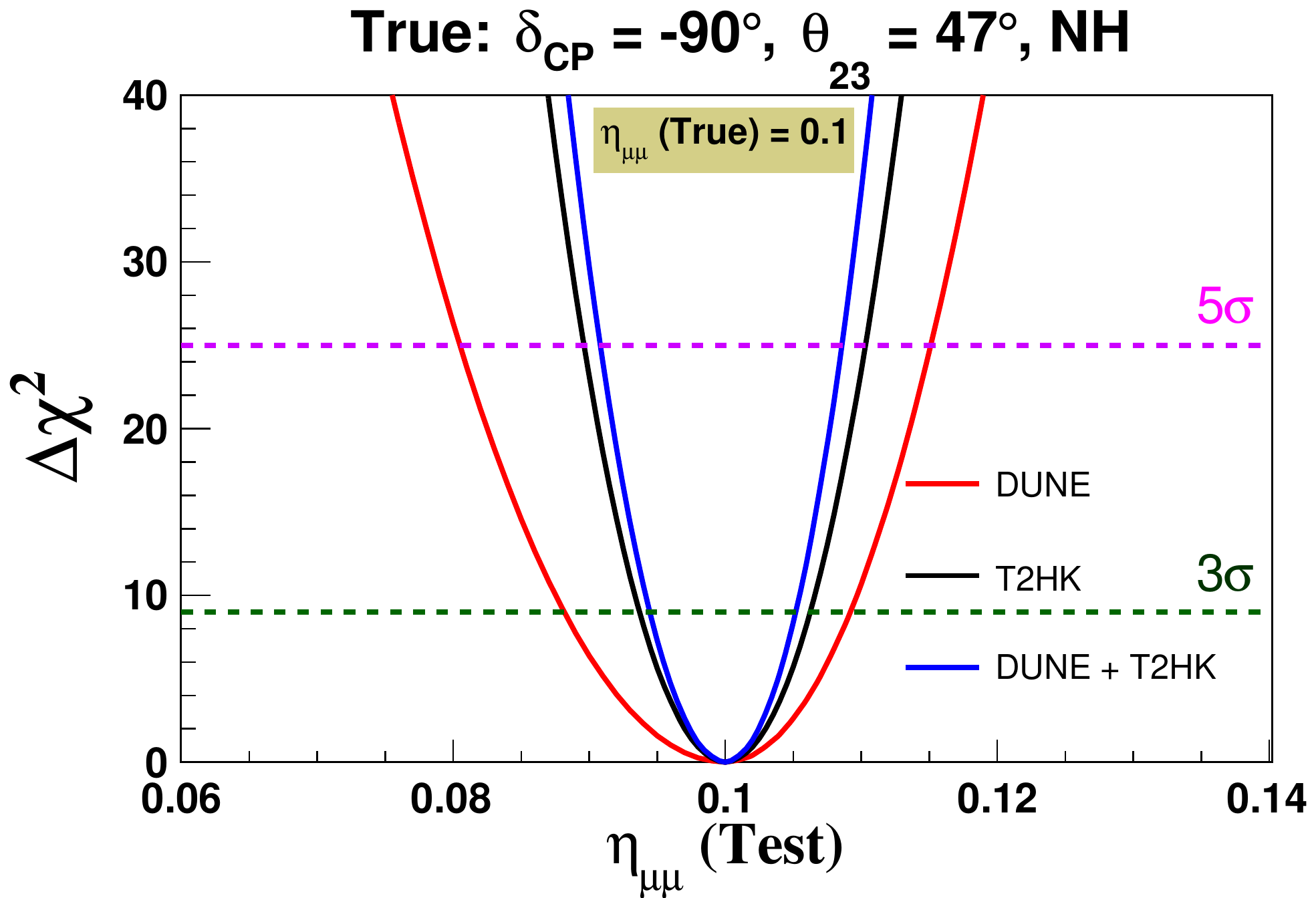} 
	\includegraphics[width=0.32\linewidth, height = 5cm]{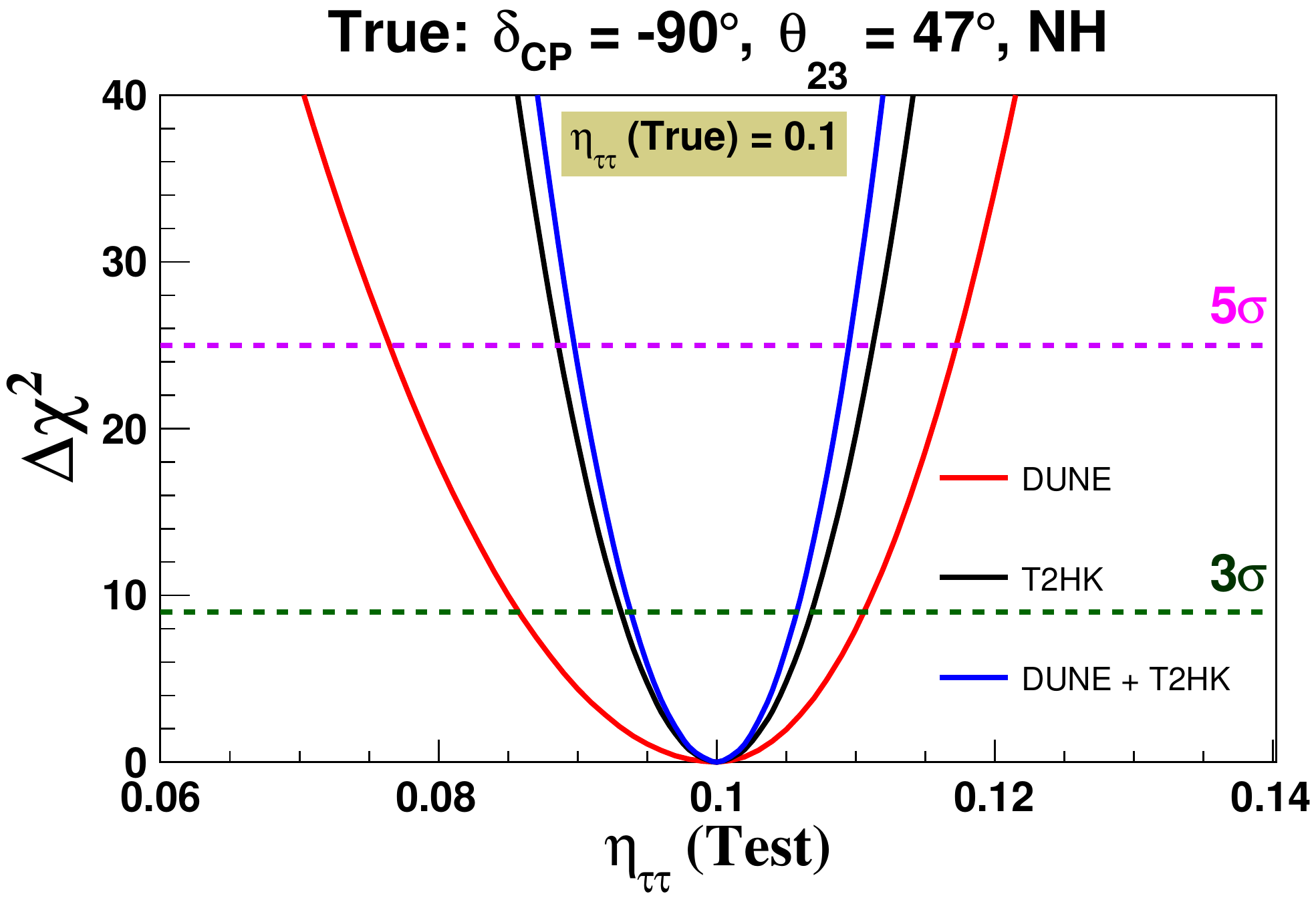} 
	\caption{The sensitivity of DUNE, T2HK and DUNE + T2HK towards constraining non--zero $\eta_{ee}$ (left--panel), $\eta_{\mu\mu}$ (middle--panel), and $\eta_{\tau\tau}$ (right--panel) at true $\delta_{CP}$ = -$\pi$/2 and true $\theta_{23}$ = 47$^\circ$. In all the plots, the sensitivities for DUNE, T2HK and DUNE+T2HK are shown in red, black and blue respectively.}
	\label{fig:fixed_chi2_1}
\end{figure}

In figure \ref{fig:fixed_chi2_1}, we show the experiments' sensitivity towards constraining the scalar NSI parameters, $\eta_{\alpha\beta}$ for DUNE, T2HK and for the combination DUNE+T2HK. We see that, both DUNE and T2HK may constrain the NSI parameters significantly. The constraining capability may further improve in the combined analysis. The plots for $\eta_{ee}$, $\eta_{\mu\mu}$ and $\eta_{\tau\tau}$ are shown in left--panel, middle--panel and right--panel respectively. We have kept the true values of $\eta_{\alpha\beta}$ fixed at 0.1 and marginalized the test $\eta_{\alpha\beta}$ in the range [-0.5, 0.5]. We considered normal hierarchy (NH) to be true neutrino mass hierarchy and higher octant (HO) to be true octant. Throughout the analysis, we have taken true $\delta_{CP}$ = -90$^\circ$ and true $\theta_{23}$ =  47$^\circ$  unless otherwise mentioned. Then we plotted $\Delta \chi^2$ as a function of test $\eta_{\alpha\beta}$ parameters. The dashed green and the dashed magenta line represent the 3$\sigma$ and 5$\sigma$ CL respectively. We observe that,

\begin{itemize}
    \item The sensitivity of DUNE towards constraining $\eta_{ee}$ (for a true $\eta_{ee}$ = 0.1) is nominally better at 3$\sigma$ as compared to that of T2HK. On the other hand T2HK shows better constraining capability towards $\eta_{\mu\mu}$ and $\eta_{\tau\tau}$ (for true $\eta_{\alpha\beta}$ = 0.1), as compared to DUNE. This is due to the large detector size of T2HK ($\sim$ 374kt) as compared to DUNE which leads to improved statistics at T2HK.
    \item The combined study with DUNE+T2HK improves the sensitivity towards constraining the $\eta_{\alpha\beta}$ parameters and is capable to putting stringer bounds on $\eta_{\alpha\beta}$. Combining DUNE with T2HK gives always improved sensitivity due to huge combined data from both the detectors.
\end{itemize}

\begin{figure}[h]
	\centering
	\includegraphics[width=0.32\linewidth, height = 5cm]{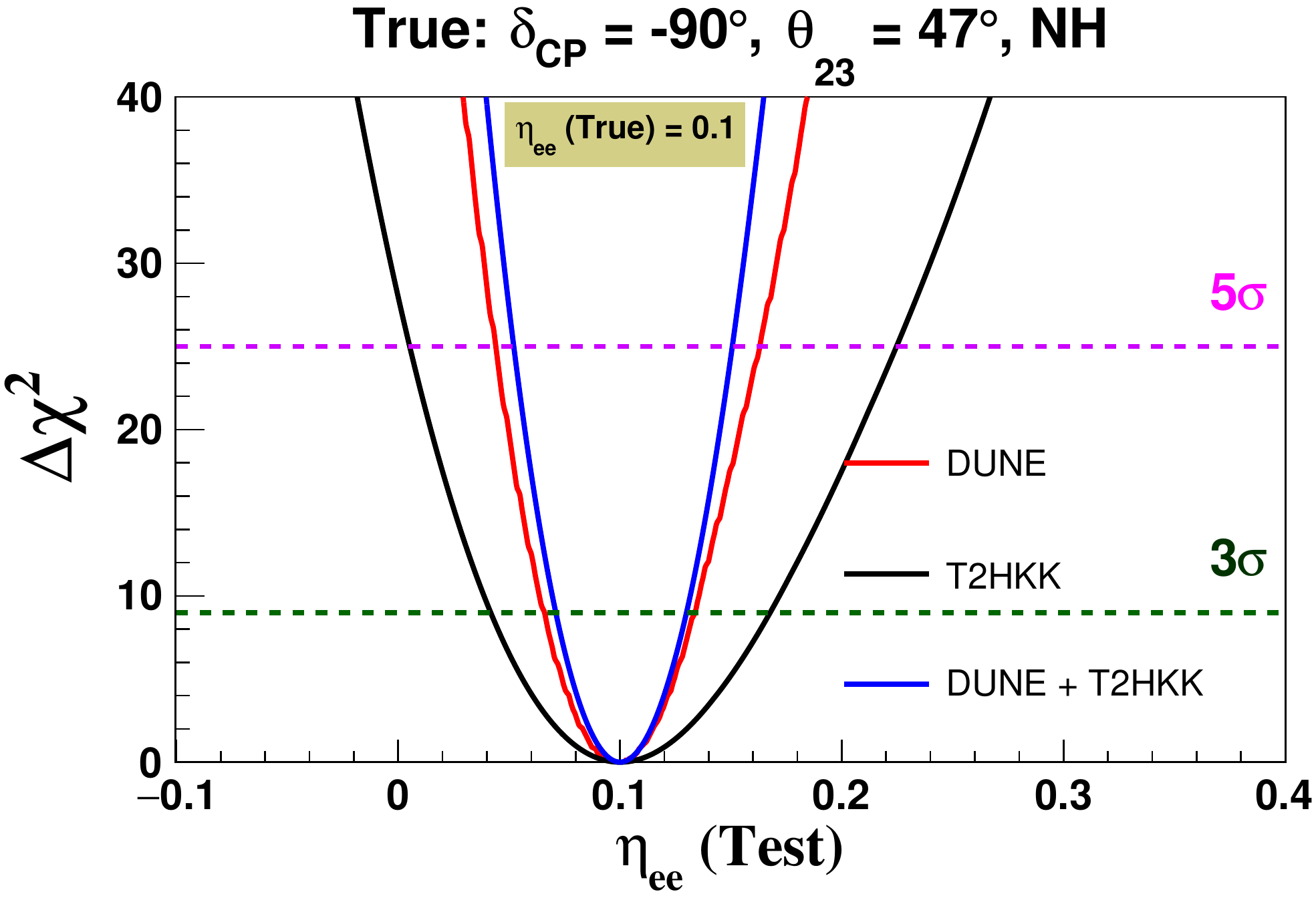} 
	\includegraphics[width=0.32\linewidth, height = 5cm]{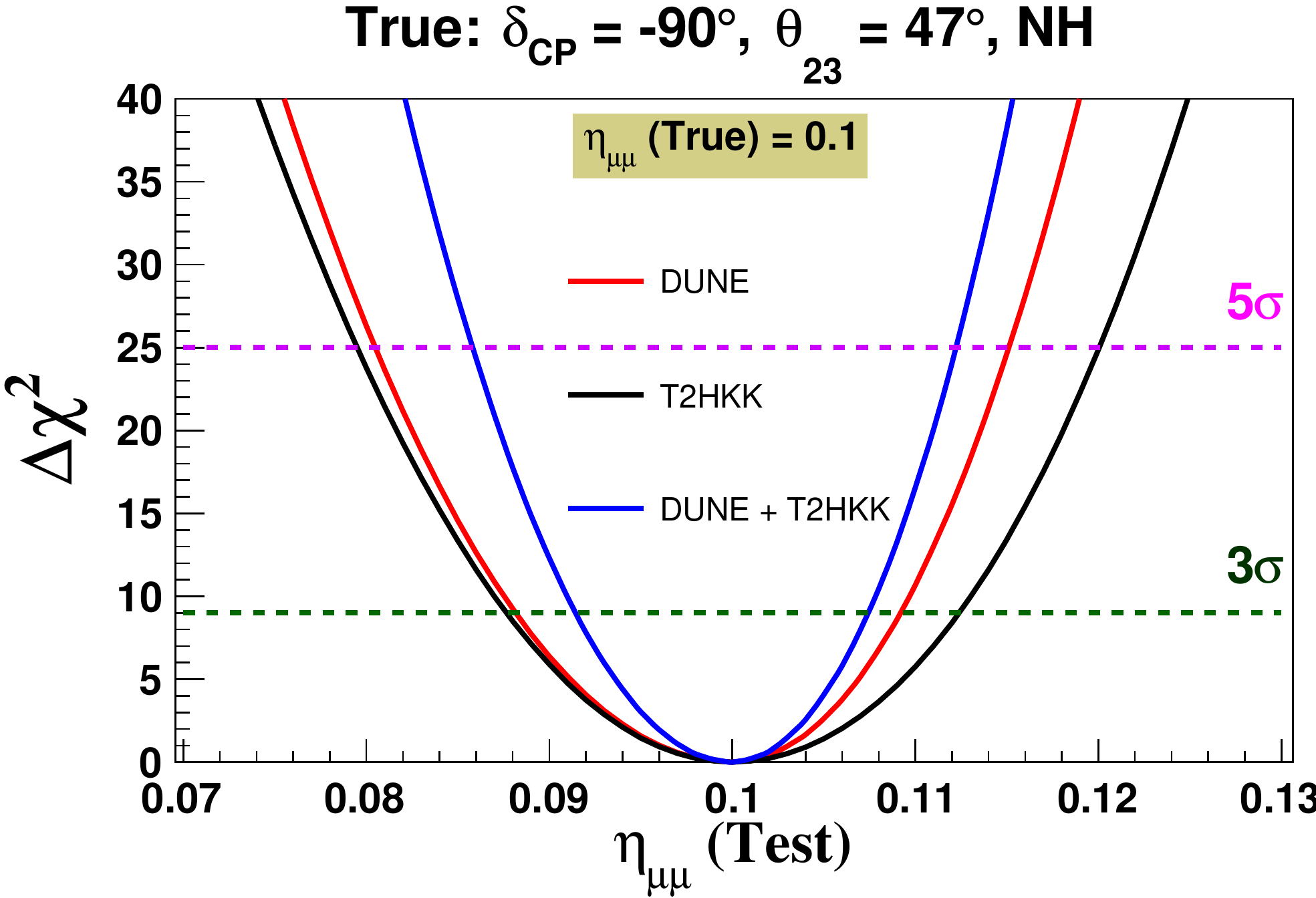} 
	\includegraphics[width=0.32\linewidth, height = 5cm]{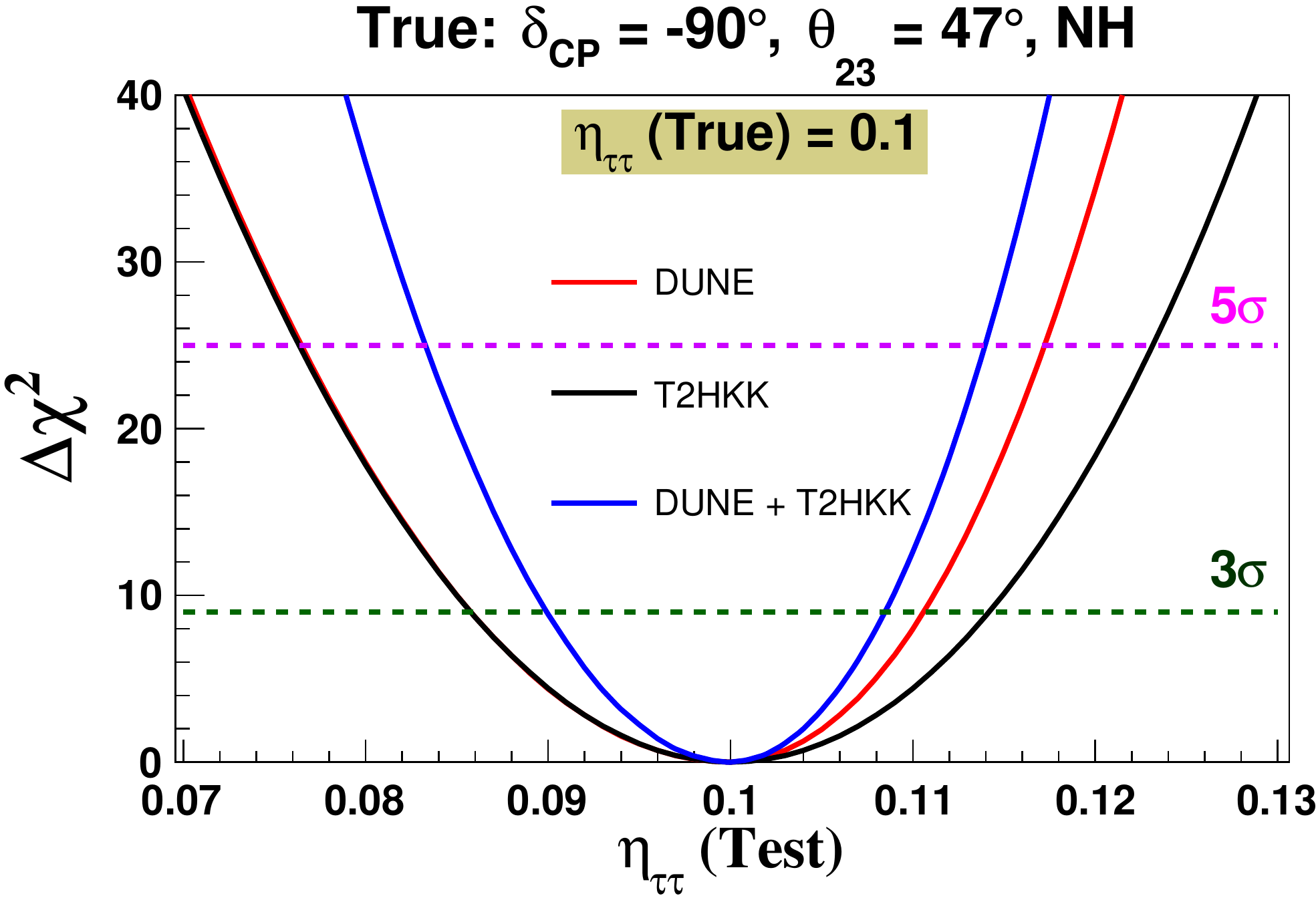} 
	\caption{The sensitivity of DUNE, T2HKK and DUNE + T2HKK towards constraining $\eta_{ee}$ (left--panel), $\eta_{\mu\mu}$ (middle--panel), and $\eta_{\tau\tau}$ (right--panel) at true $\delta_{CP}$ = -$\pi$/2 and true $\theta_{23}$ = 47$^\circ$. In all the three plots the results for DUNE, T2HKK and DUNE+T2HKK are shown in red, black and blue respectively.}
	\label{fig:fixed_chi2_2}
\end{figure}

In figure \ref{fig:fixed_chi2_2}, the sensitivity of DUNE, T2HKK and DUNE+T2HKK towards constraining $\eta_{\alpha\beta}$ are shown. The results for $\eta_{ee}$, $\eta_{\mu\mu}$ and $\eta_{\tau\tau}$ are shown in left--panel, middle--panel and right--panel respectively. We have plotted $\Delta \chi^2$ as a function of test $\eta_{\alpha\beta}$. Our observations are listed below.

\begin{itemize}
    \item  The constraining capability of T2HKK towards $\eta_{ee}$ and $\eta_{\mu\mu}$ are weaker than that of DUNE. For $\eta_{\tau\tau} (test) $ $\leq$ $\eta_{\tau\tau}$ (true), we see an overlap of the DUNE and T2HKK capabilities. For the rest of the $\eta_{\tau\tau}$ range, DUNE comes with a better sensitivity.
    \item Combining DUNE and T2HKK constrains $\eta_{\alpha\beta}$ with a stronger bound than those of DUNE and T2HKK individually. This improved sensitivity is due to substantial statistics and wider parameter space obtained by combining DUNE and T2HKK. 
\end{itemize}

\subsection{CP Violation sensitivity}
The measurement of $\delta_{CP}$ in the leptonic sector is one of the prime goal of various ongoing and upcoming neutrino experiments. The detection of CP--violation may be crucial in explaining the baryon asymmetry of the Universe i.e. the dominance of matter over antimatter \cite{Steigman:1976ev, Cohen:1997ac, Fong:2012buy}. It is interesting to explore the subdominant effects of scalar NSI on $\delta_{CP}$ related measurements at the neutrino sector \cite{Medhi:2021wxj}. We discuss here the effects of $\eta_{\alpha\beta}$ on the CPV sensitivities at DUNE, T2HK and T2HKK. We have obtained the sensitivities by varying the true values of $\delta_{CP}$ in the allowed range [-$\pi$, $\pi$]. The true values of other mixing parameters used in this analysis are as listed in table~\ref{tab:mixing_parameters}. In the test spectrum of $\delta_{CP}$, we have only considered the CP-conserving values i.e. 0 and $\pm$ $\pi$. We have marginalized $\theta_{23}$ and $\Delta m_{31}^2$ over the allowed 3$\sigma$ ranges~\cite{NuFIT5.0} and have minimized the $\chi^2$ over all the marginalization ranges. The CPV sensitivity is calculated as ,
\begin{equation}
{\Delta \chi}^{2}_{\rm CPV}~(\delta^{\rm true}_{\rm CP}) = {\rm min}~\left[\chi^2~(\delta^\text{true}_{CP},\delta^\text{test}_{CP}=0),~\chi^2 (\delta^\text{true}_{CP},\delta^\text{test}_{CP}=\pm \pi)\right ].
\end{equation}

\begin{figure}[h]
	\centering
	\includegraphics[width=0.32\linewidth, height = 5cm]{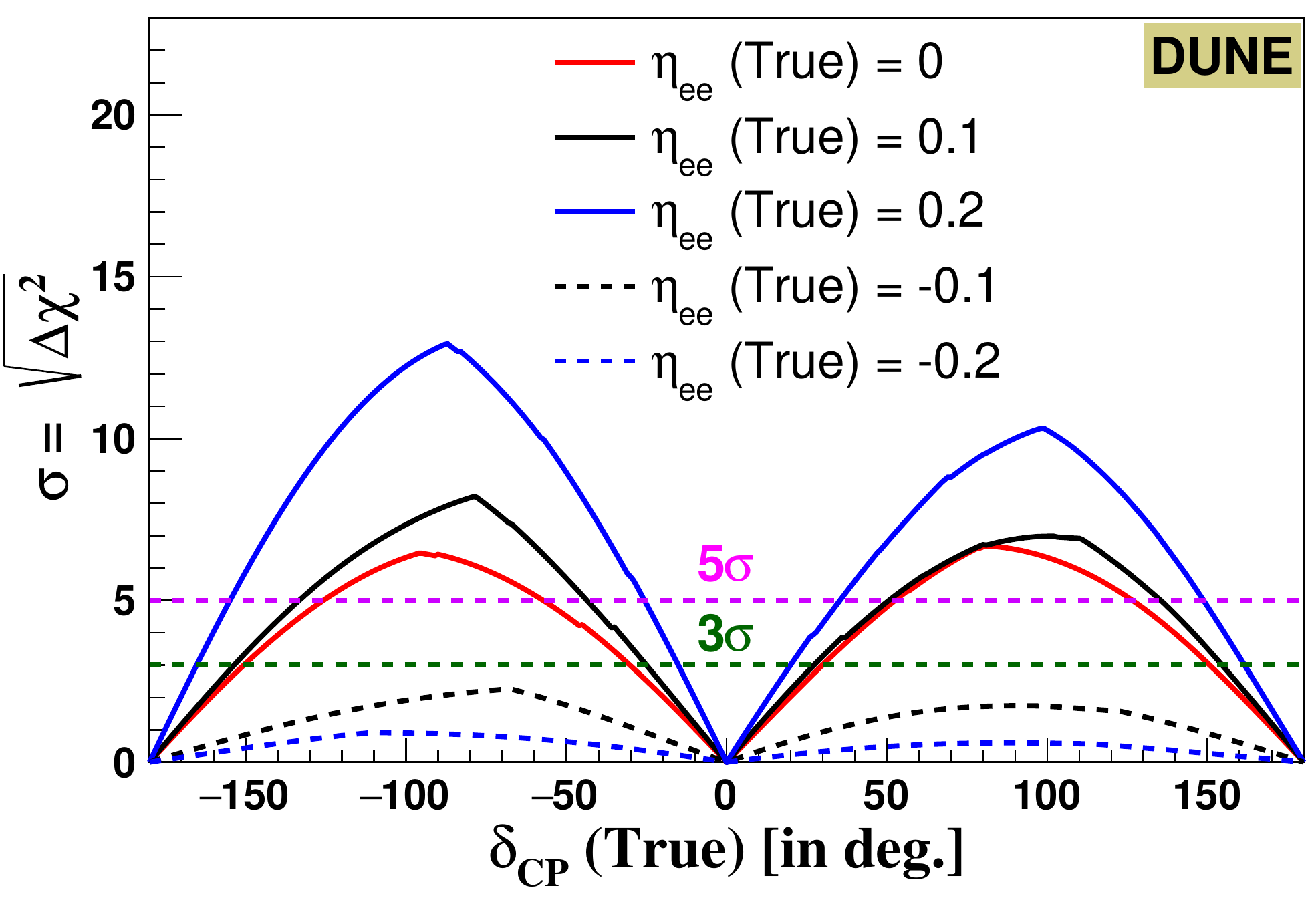} 
	\includegraphics[width=0.32\linewidth, height = 5cm]{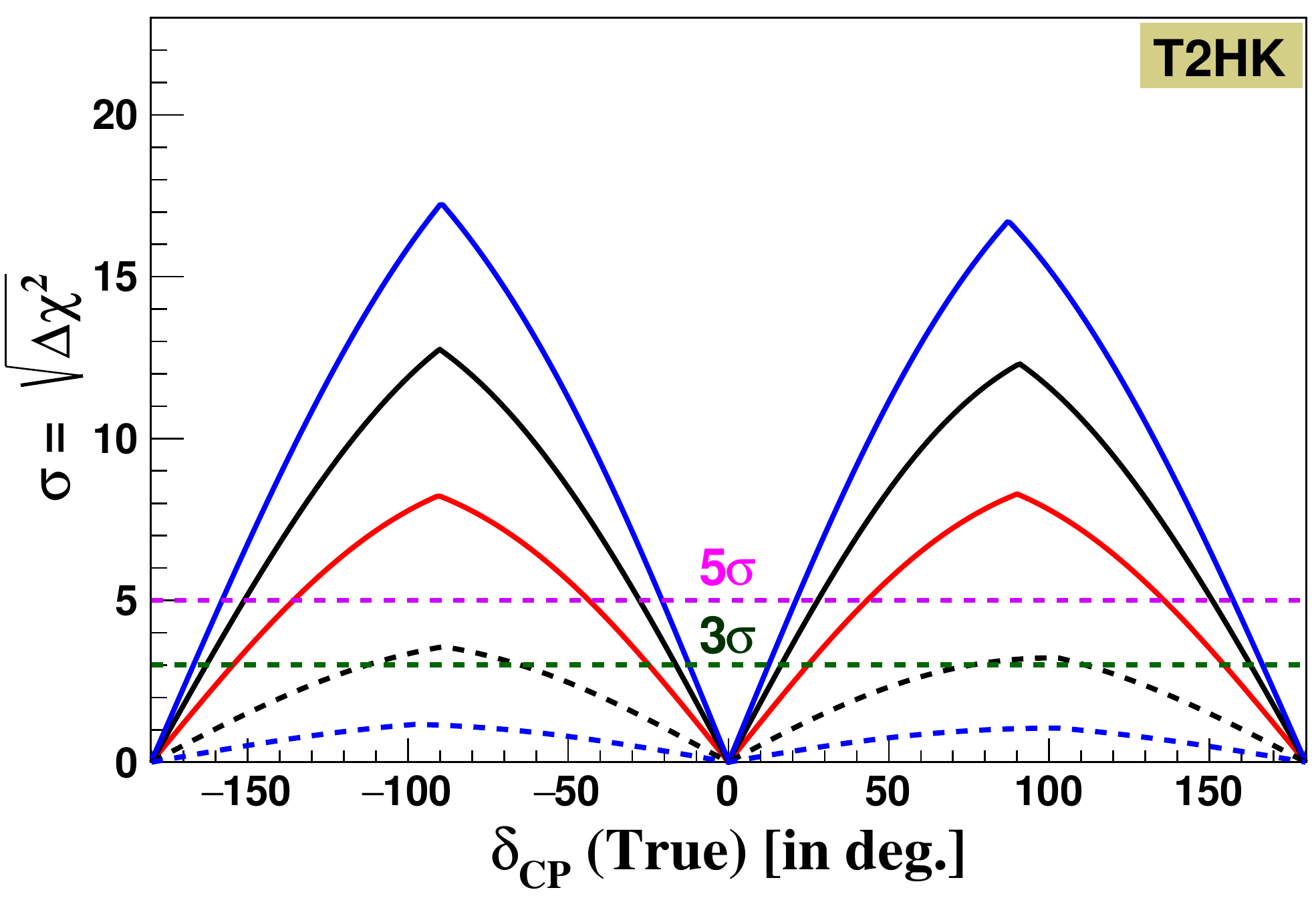} 
	\includegraphics[width=0.32\linewidth, height = 5cm]{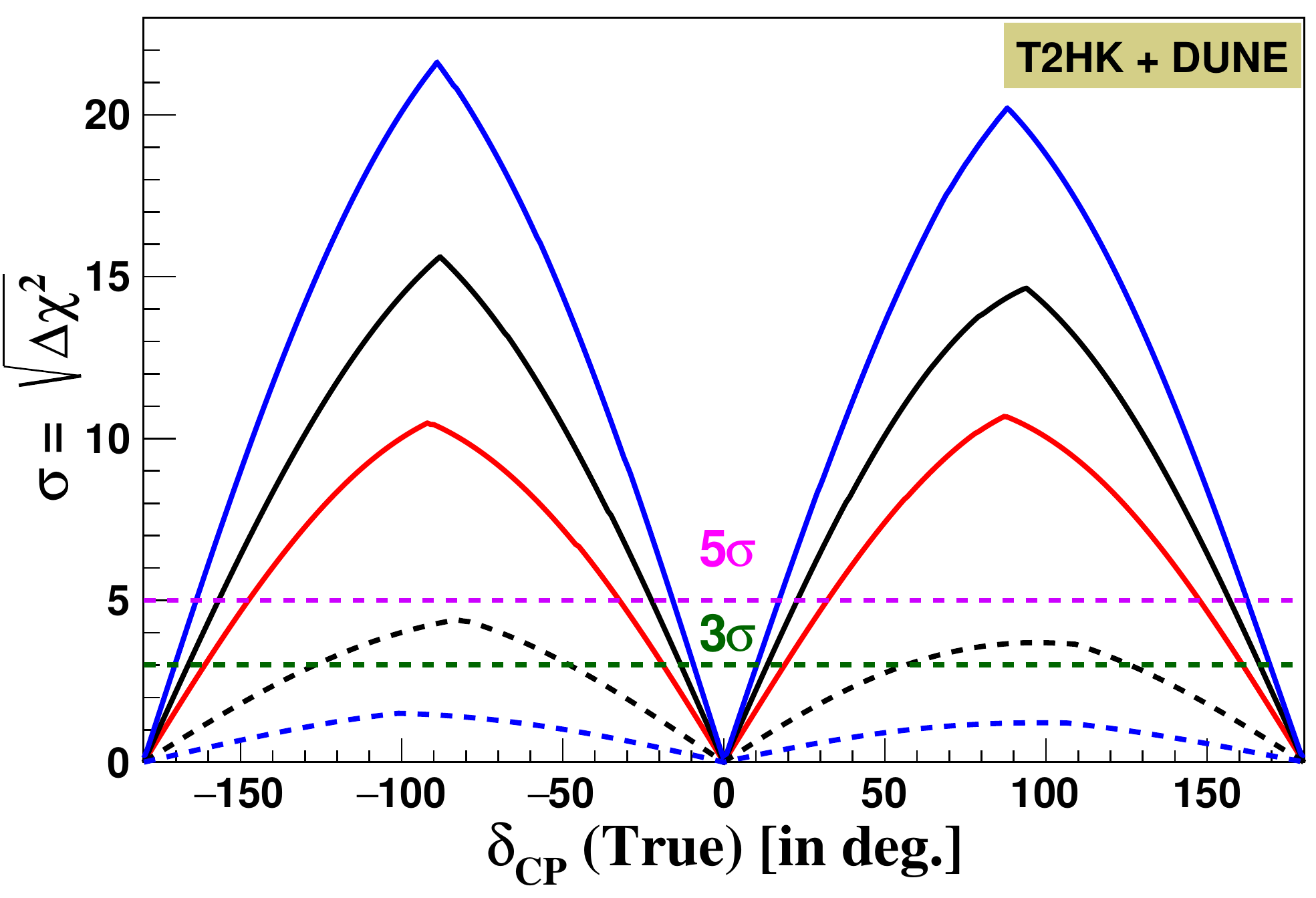} 

	\includegraphics[width=0.32\linewidth, height = 5cm]{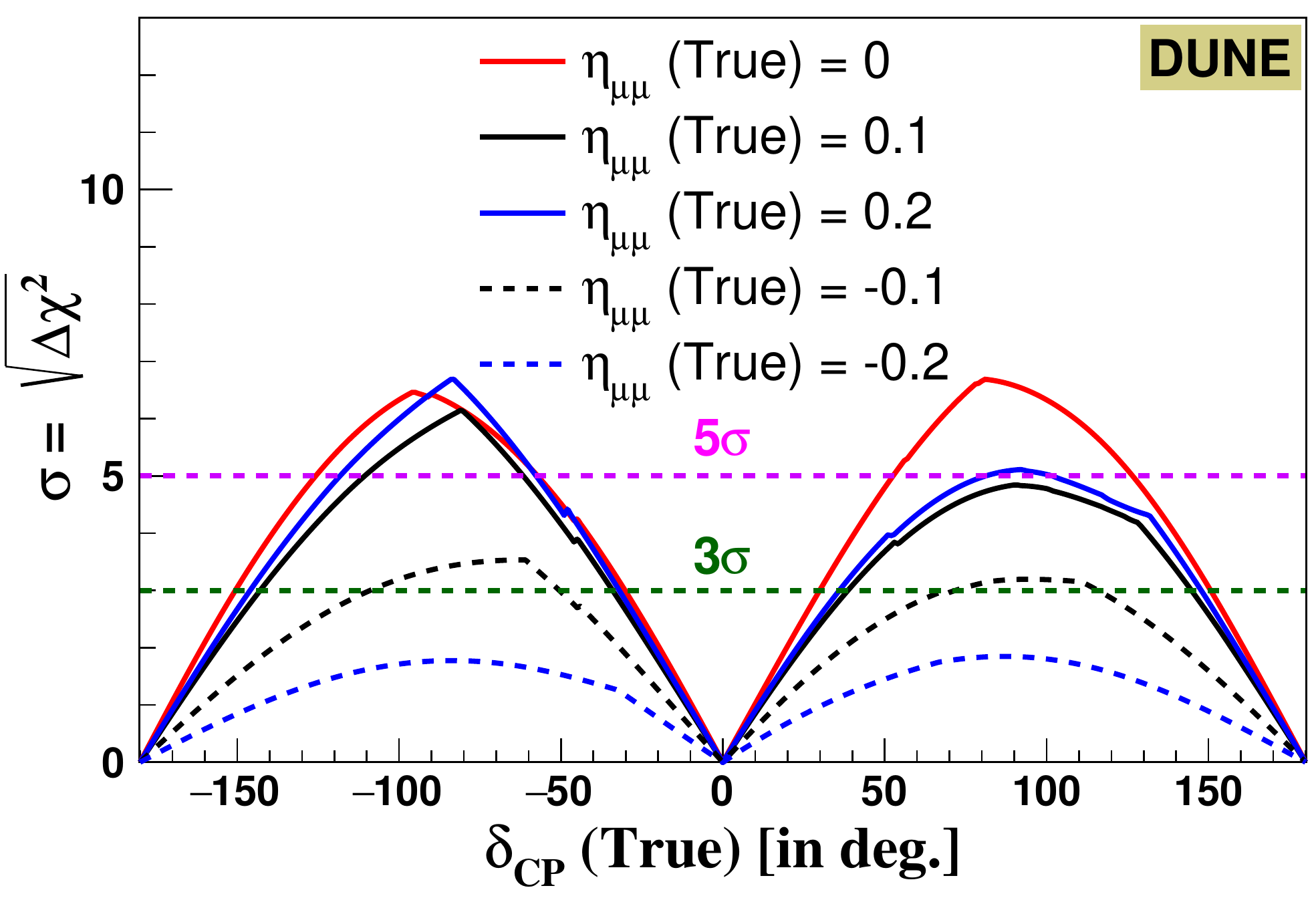} 
	\includegraphics[width=0.32\linewidth, height = 5cm]{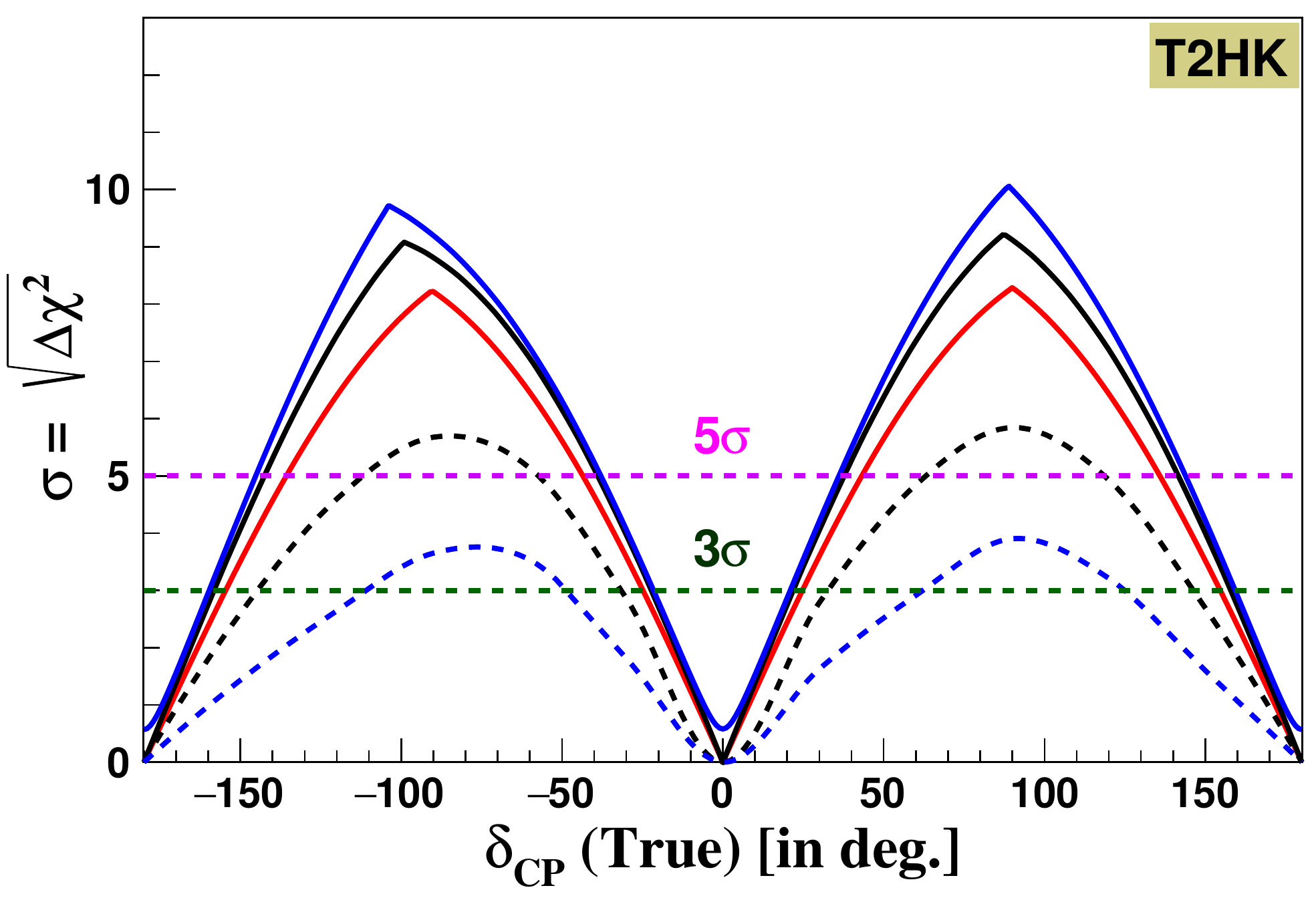} 
	\includegraphics[width=0.32\linewidth, height = 5cm]{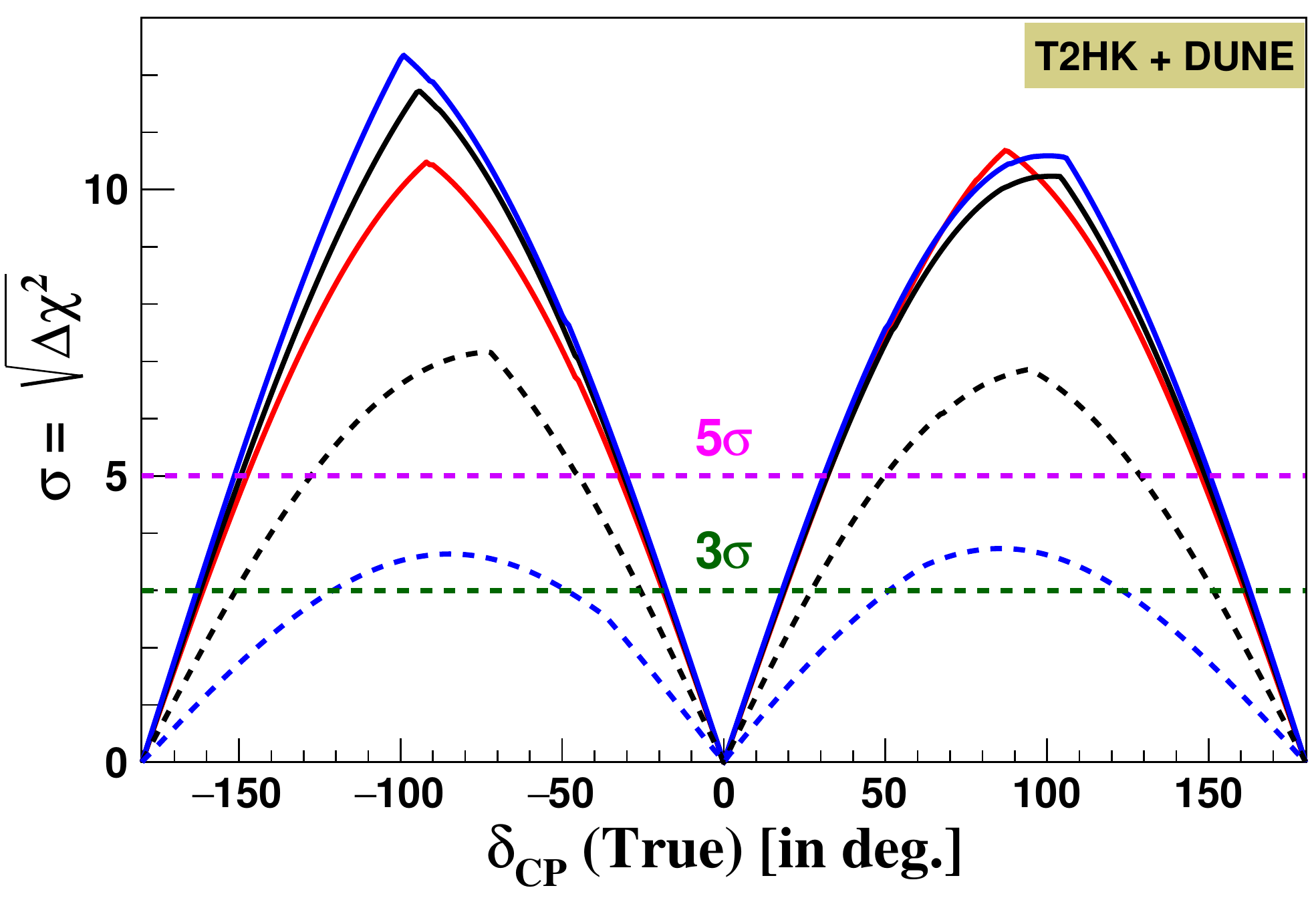} 

	\includegraphics[width=0.32\linewidth, height = 5cm ]{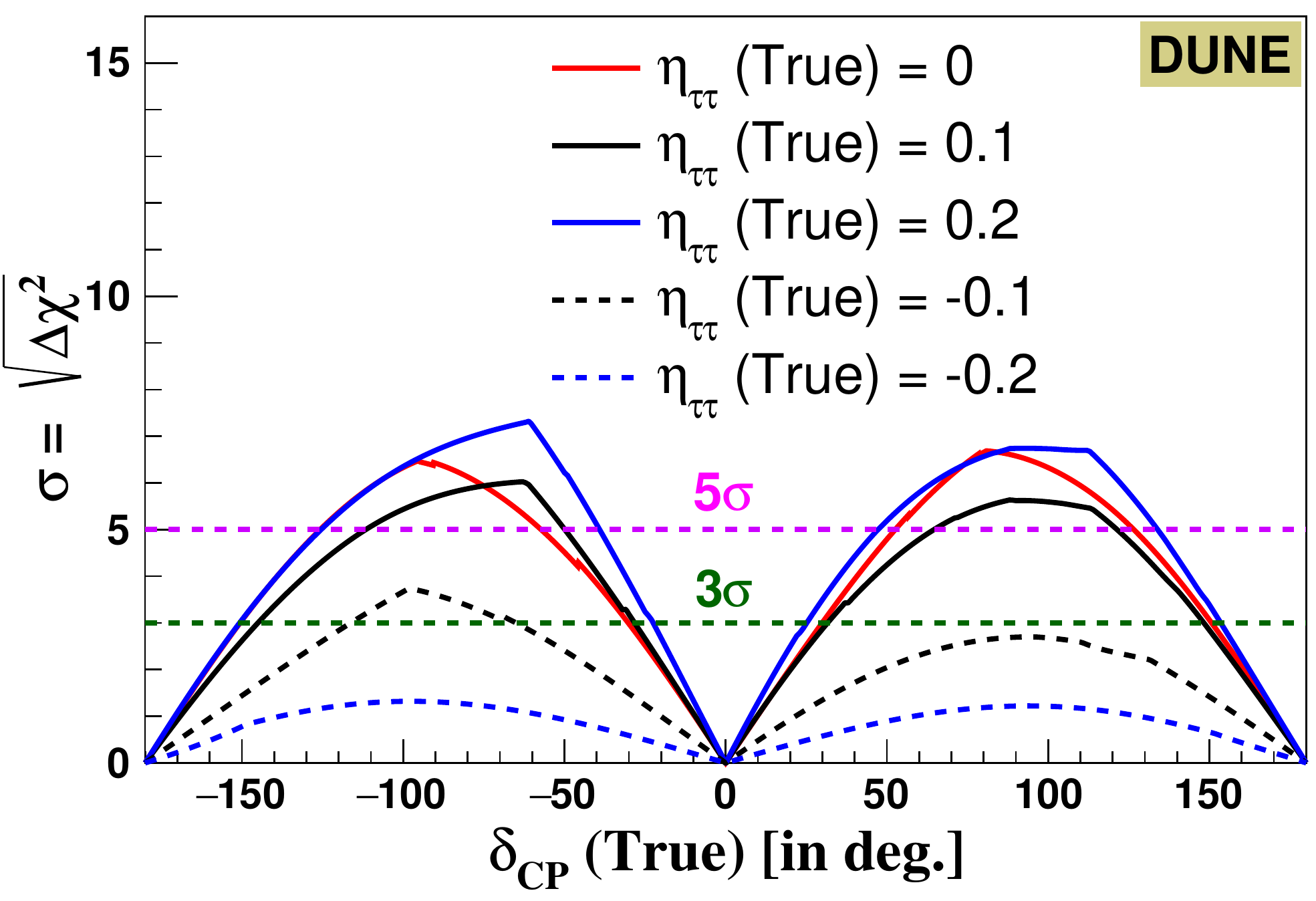} 
	\includegraphics[width=0.32\linewidth, height = 5cm ]{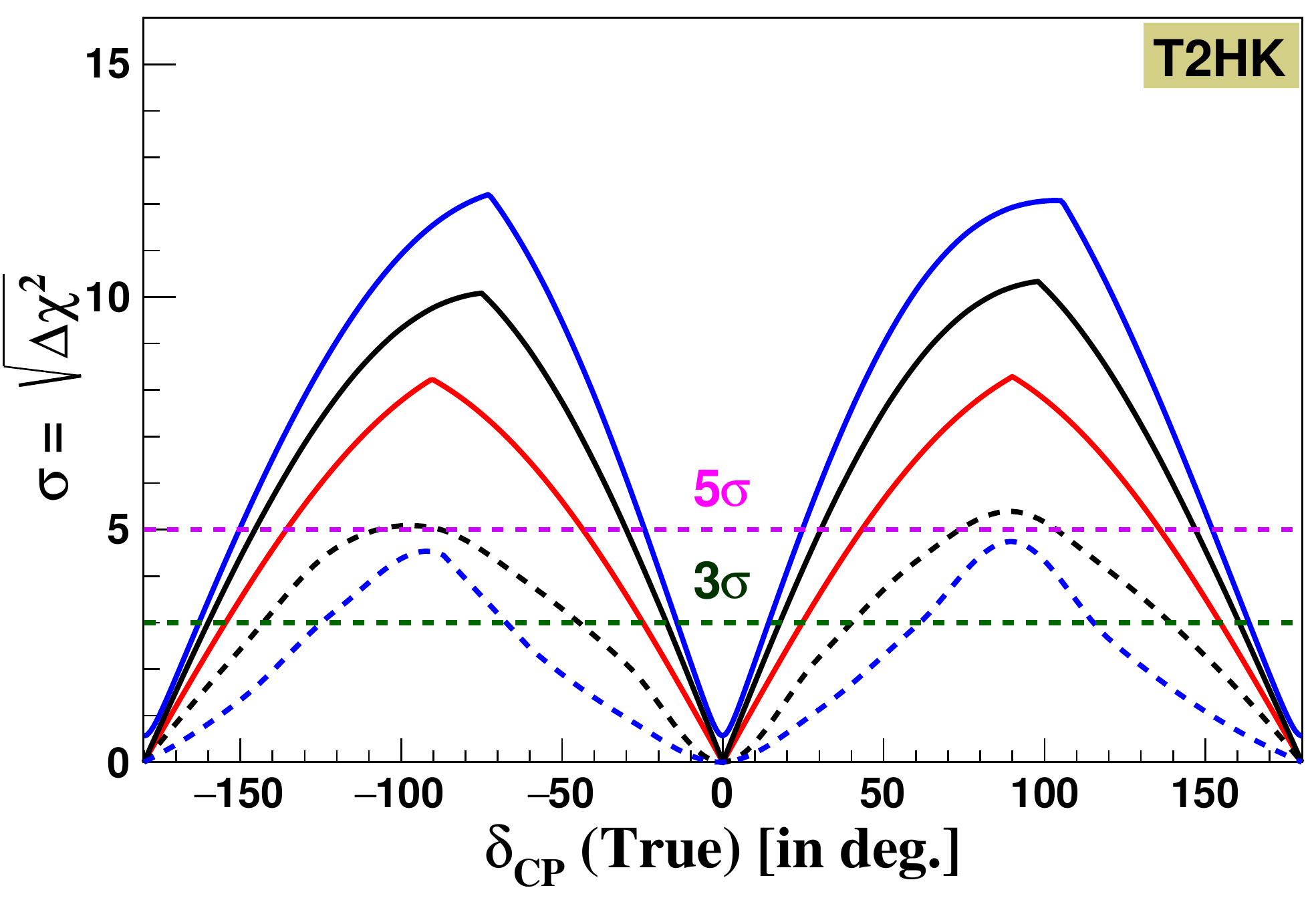} 
	\includegraphics[width=0.32\linewidth, height = 5cm ]{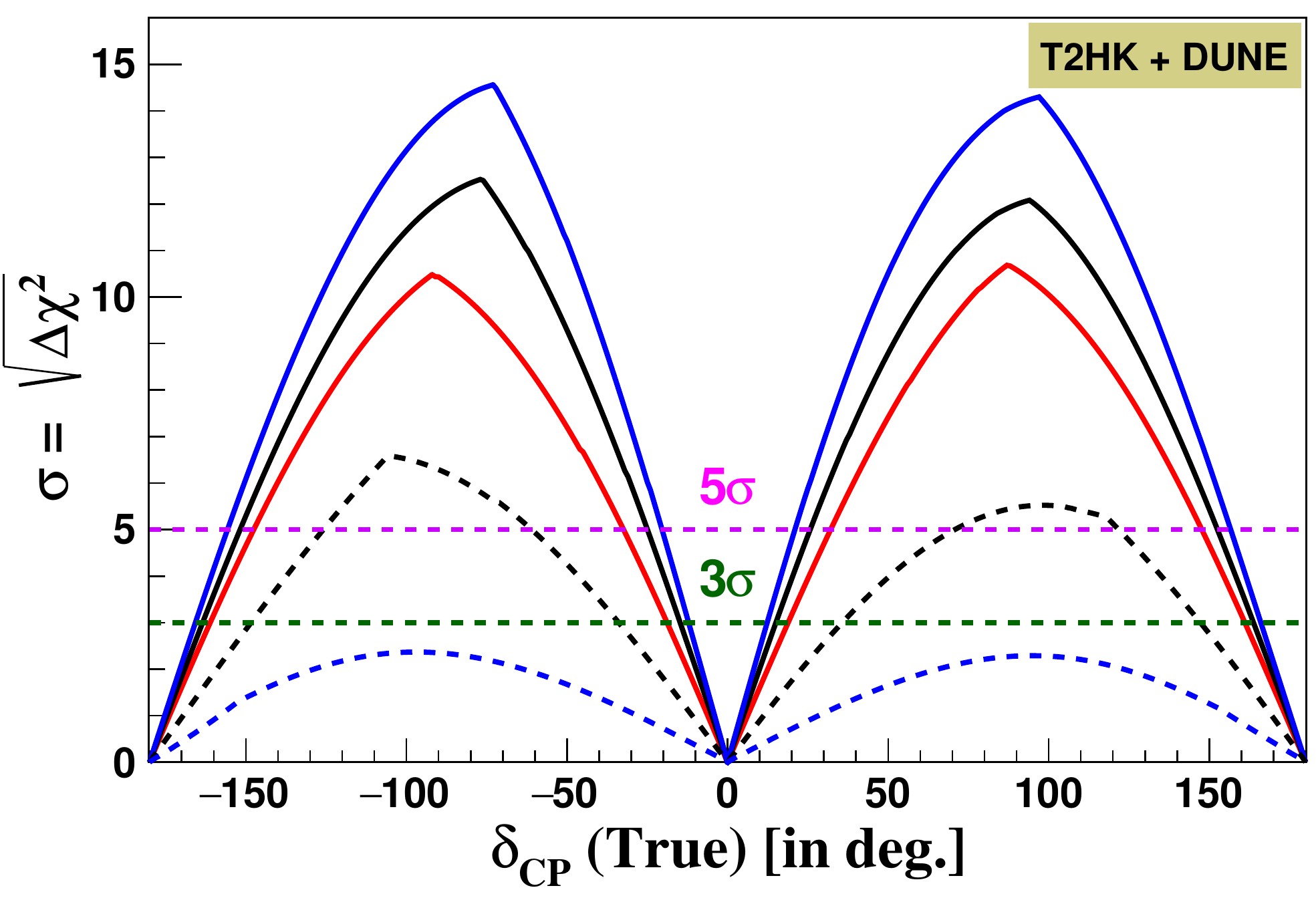} 
	\caption{The CPV sensitivity of DUNE (left--column), T2HK (middle-column) and DUNE + T2HK (right-column) in presence of scalar NSI. The plots for $\eta_{ee}$, $\eta_{\mu\mu}$ and $\eta_{\tau\tau}$ are included in the top-row, middle-row and bottom--row respectively. The solid-red curve is for the no scalar NSI case whereas solid (dashed) black and blue curves are for positive (negative) $\eta_{\tau\tau}$.}
	\label{fig:cpv_3}
\end{figure}

In figure \ref{fig:cpv_3}, we show the effects of scalar NSI on the CPV sensitivity for DUNE (left--column), T2HK (middle--column) and DUNE + T2HK (right--column). Both the experiment's (DUNE and T2HK) sensitivity towards CPV gets significantly affected by inclusion of scalar NSI. We have plotted here the statistical significance $\sigma$ (=$\sqrt{\Delta \chi^2_{CPV}}$) as a function of true $\delta_{CP}$. The plots for $\eta_{ee}$, $\eta_{\mu\mu}$ and $\eta_{\tau\tau}$ are shown on the top--row,  middle--row and bottom--row respectively. For the $\chi^2$ study, we have marginalized over the NSI parameters. In all the plots, the solid--red curve represents the no scalar NSI case i.e. $\eta_{\alpha\beta}$ = 0. The solid (dashed) black and blue curves are for chosen positive (negative) values of $\eta_{\alpha\beta}$.  The observations from figure \ref{fig:cpv_3} are listed below.

\begin{itemize}
    \item A positive (negative) $\eta_{ee}$ mostly enhances (suppresses) the CPV sensitivities at DUNE and T2HK. At $\eta_{ee}$ = 0.1 and $\delta_{CP}^{true}$ $\in$ [0, 90$^\circ$], we see that the sensitivities without and with scalar NSI almost overlap. The combined study of DUNE + T2HK improves the sensitivities (without and with NSI) for all cases including the overlapped region particularly due to the collection of data in broader range of degenerate spaces.
    \item  A positive $\eta_{\mu\mu}$ deteriorates the CPV sensitivities in the upper half plane of $\delta_{CP}$ i.e. [0, $\pi$] at DUNE, while we observe a mild fluctuation for the rest of $\delta_{CP}$. At T2HK, we see enhancements for positive $\eta_{\mu\mu}$.  For a negative $\eta_{\mu\mu}$, we observe significant suppression in the sensitivities for DUNE and T2HK. We find that, combining DUNE and T2HK improves the overall sensitivities (without and with NSI).
    \item At DUNE, for a positive $\eta_{\tau\tau}$, we see marginal fluctuations as compared to the no scalar NSI case. At T2HK, a positive $\eta_{\tau\tau}$ enhances the sensitivity. The analysis with DUNE+T2HK enhances the sensitivities (without and with NSI).
\end{itemize}

\begin{figure}[!h]
	\centering
	\includegraphics[width=0.32\linewidth, height = 5cm]{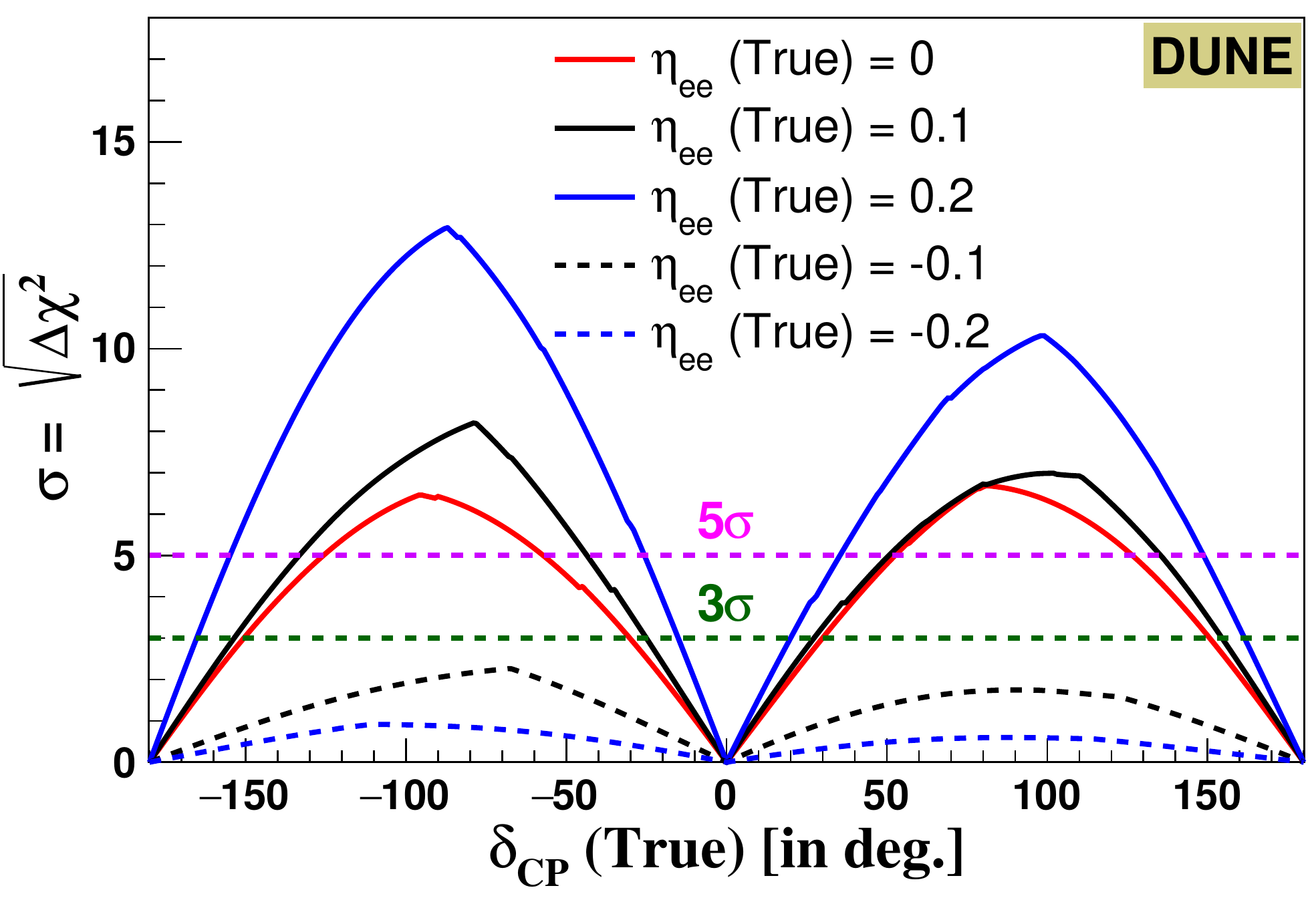} 
	\includegraphics[width=0.32\linewidth, height = 5cm]{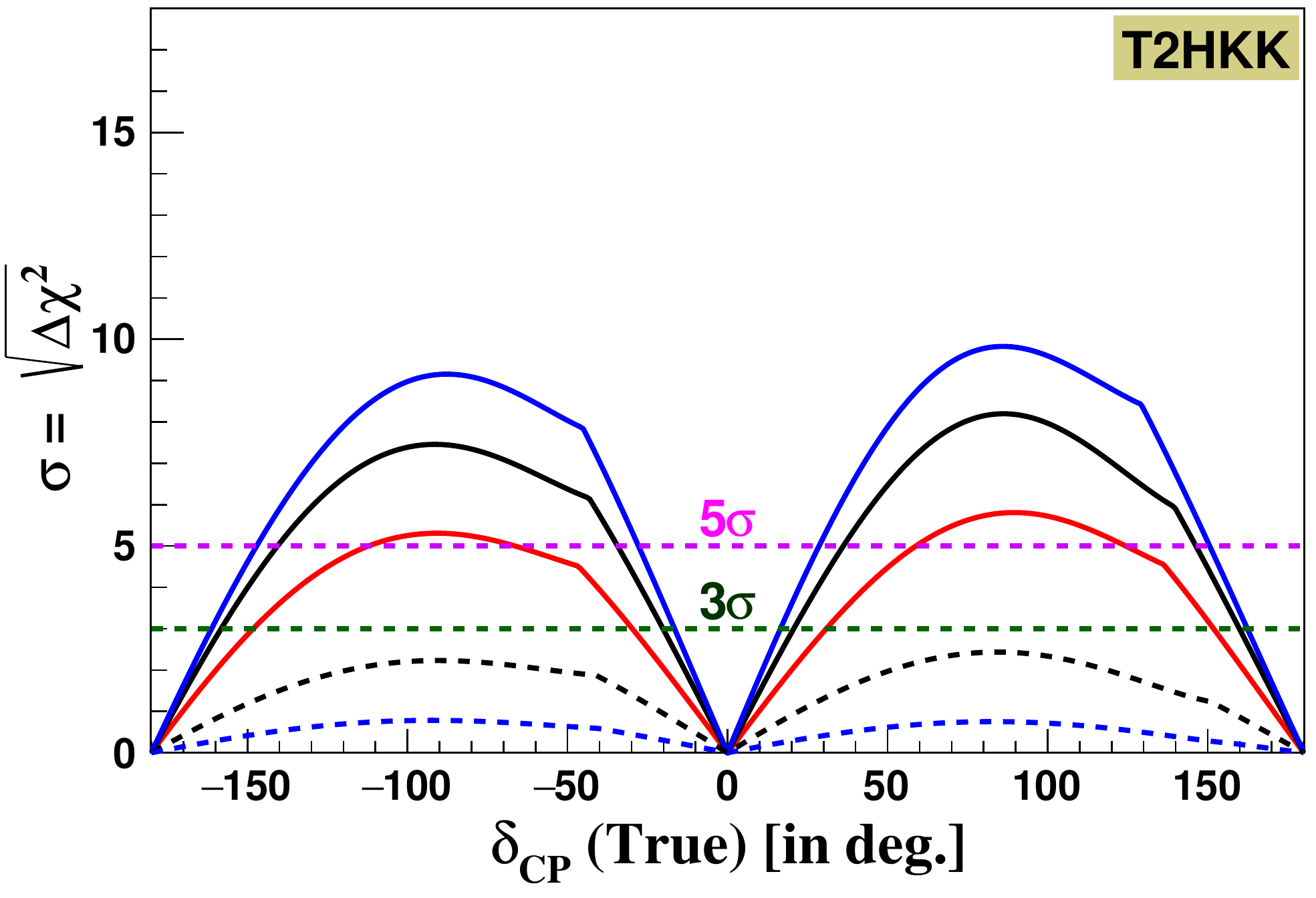} 
	\includegraphics[width=0.32\linewidth, height = 5cm]{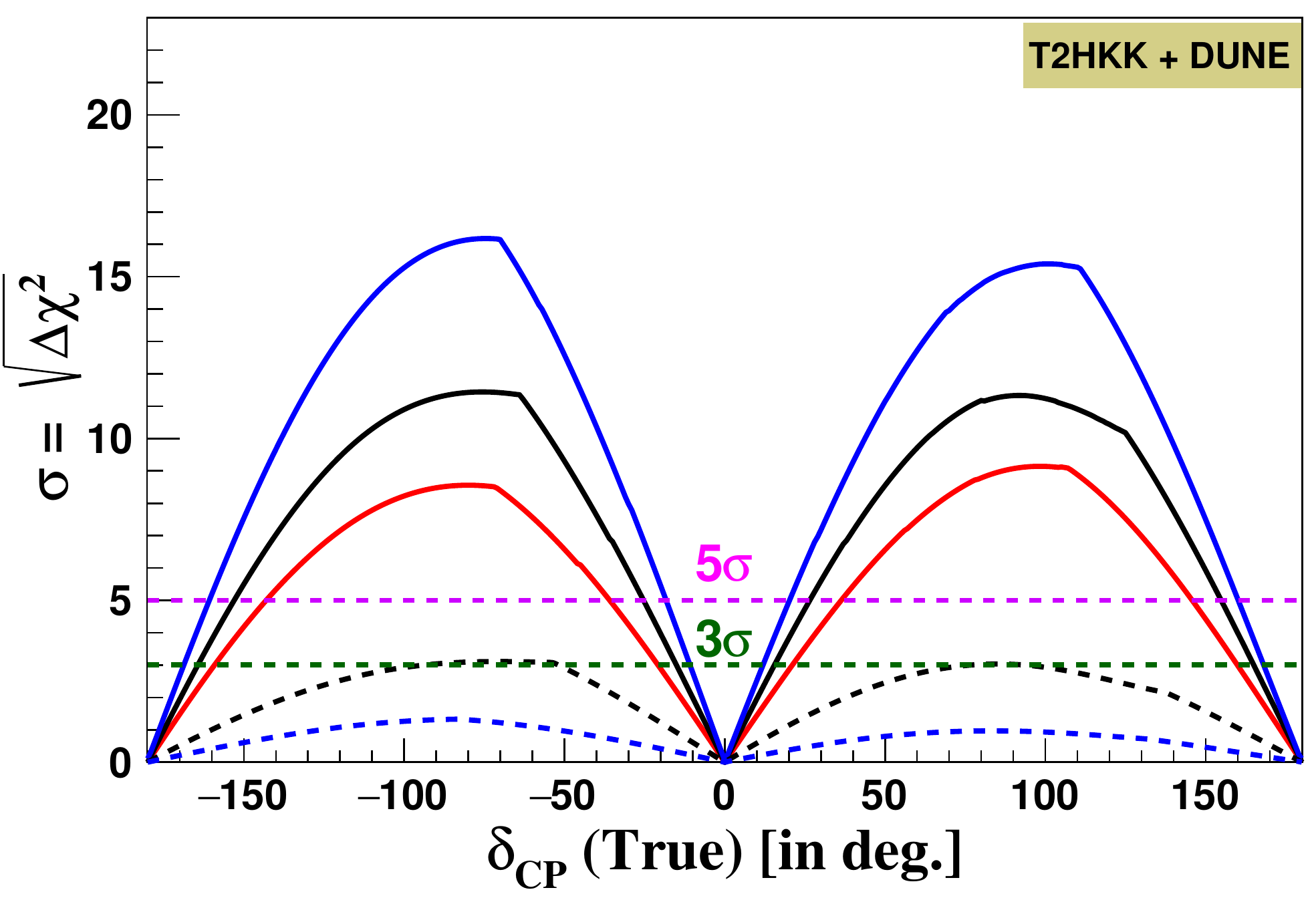} 

	\includegraphics[width=0.32\linewidth, height = 5cm]{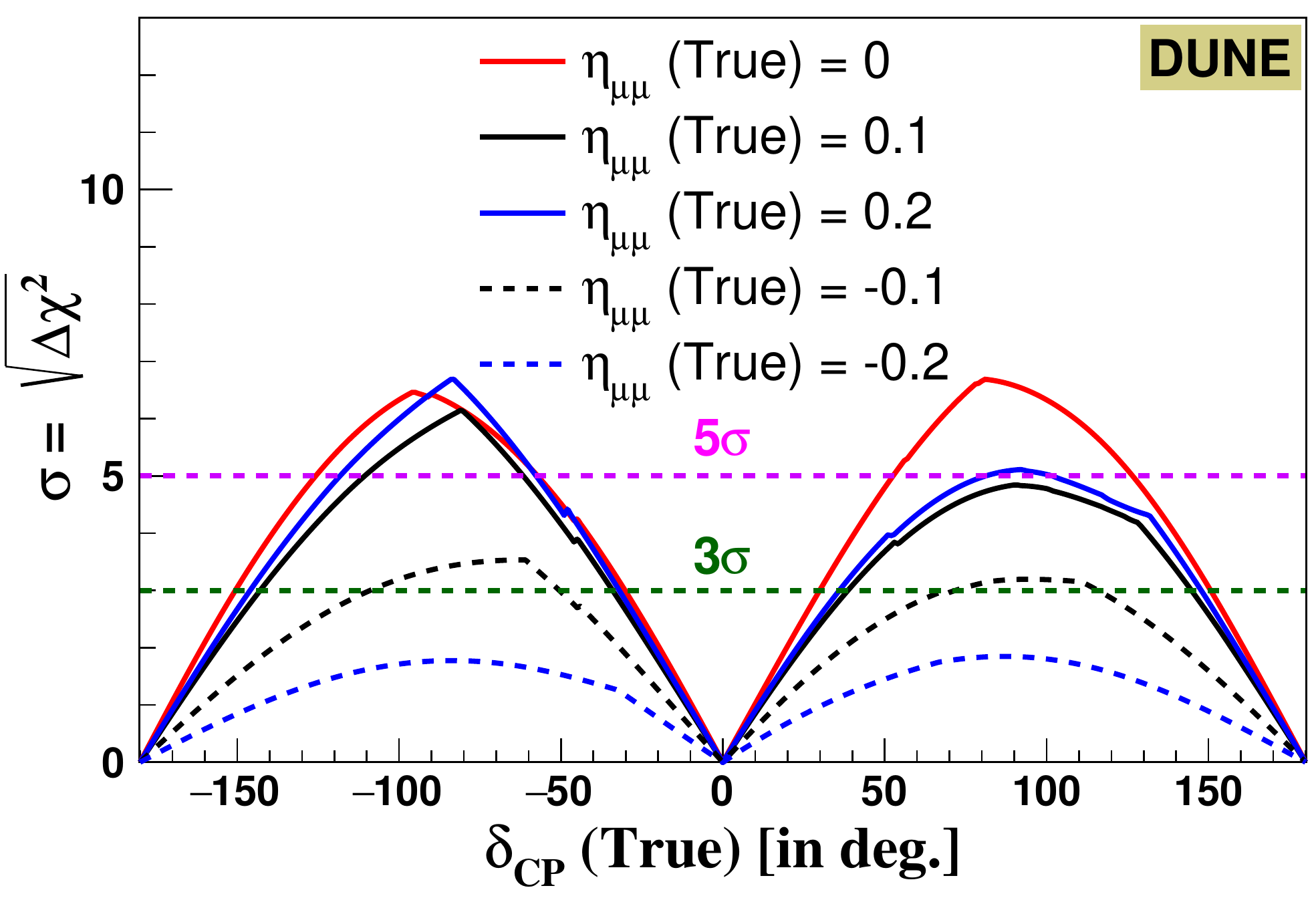} 
	\includegraphics[width=0.32\linewidth, height = 5cm]{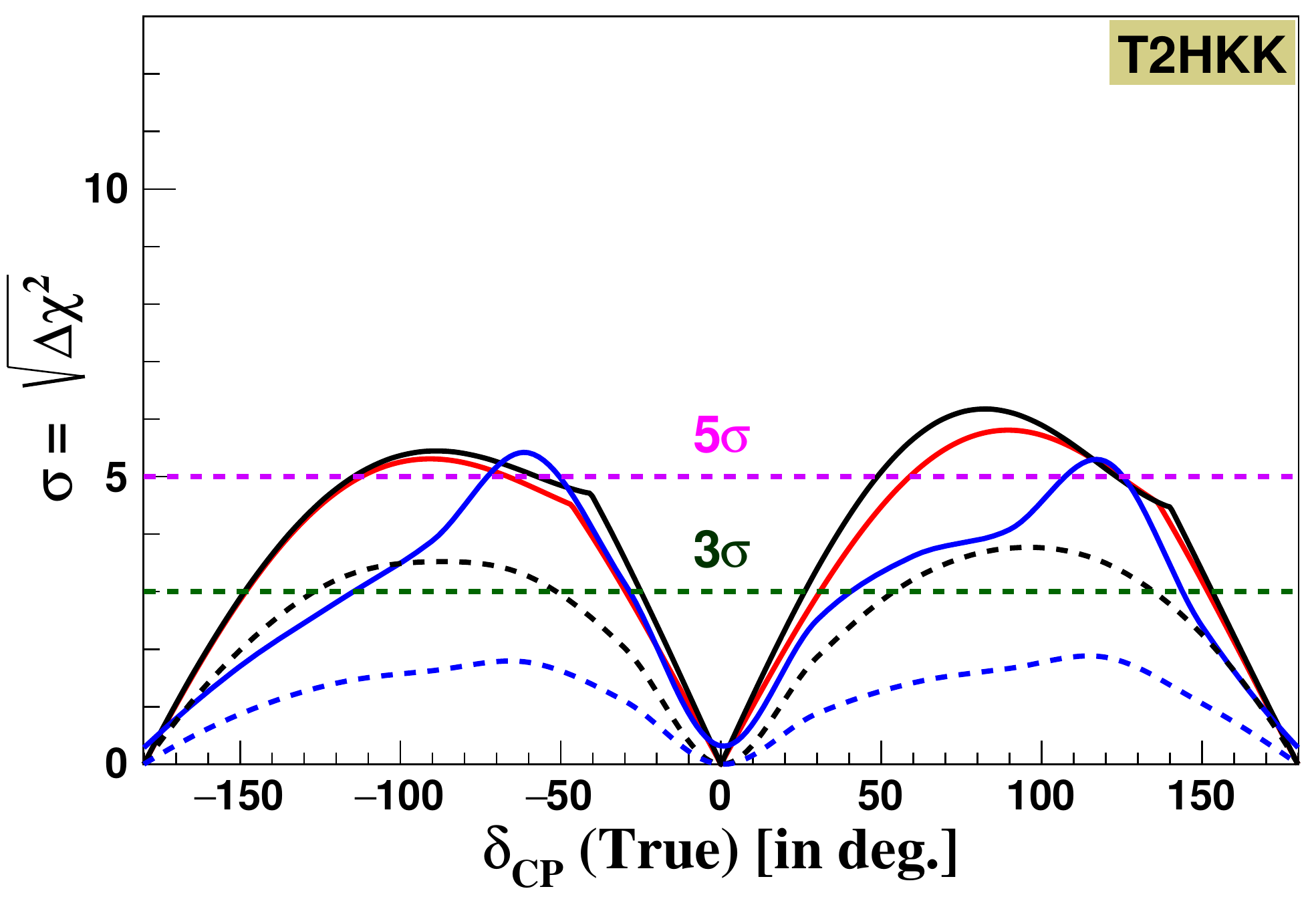} 
	\includegraphics[width=0.32\linewidth, height = 5cm]{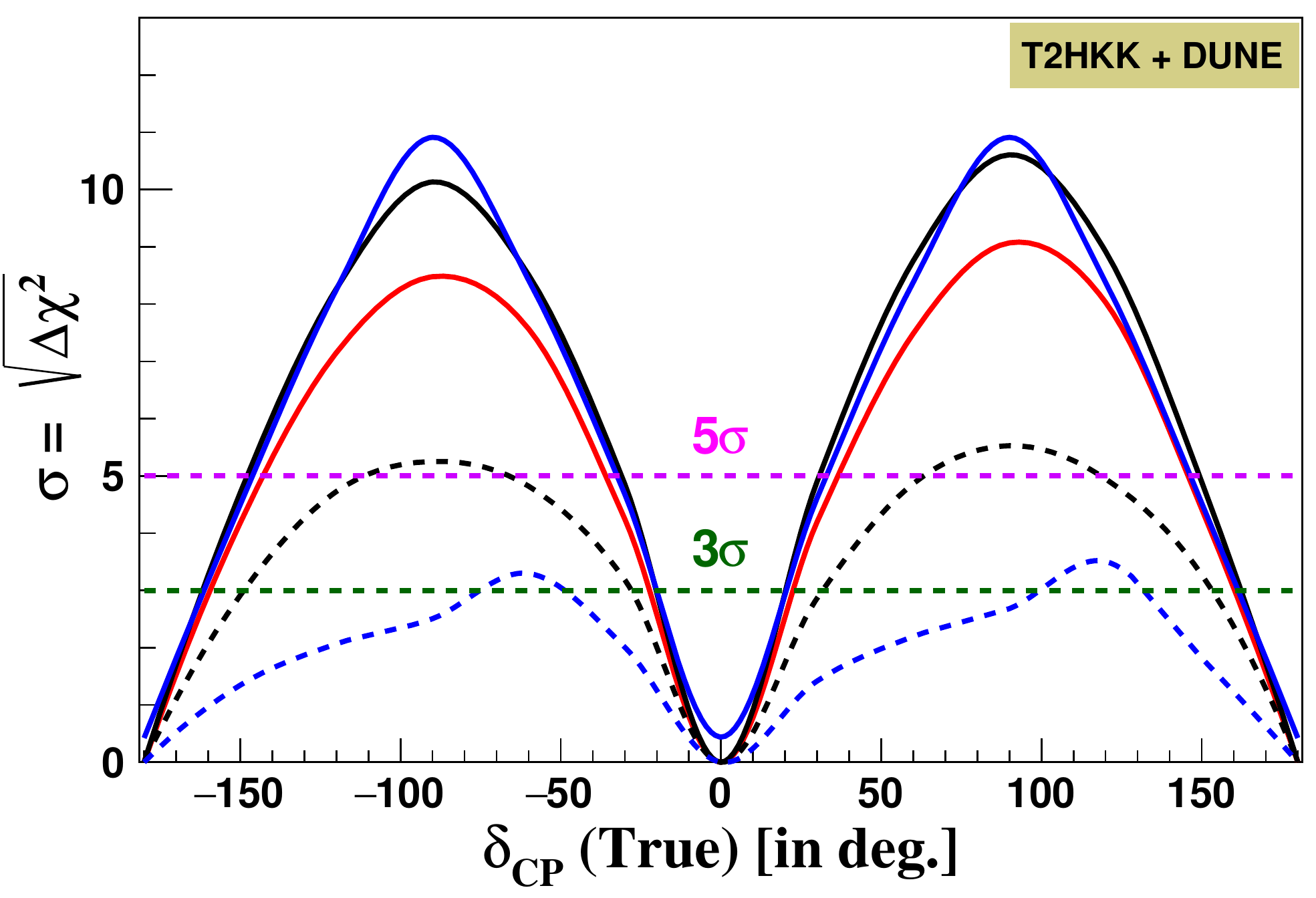} 

	\includegraphics[width=0.32\linewidth, height = 5cm]{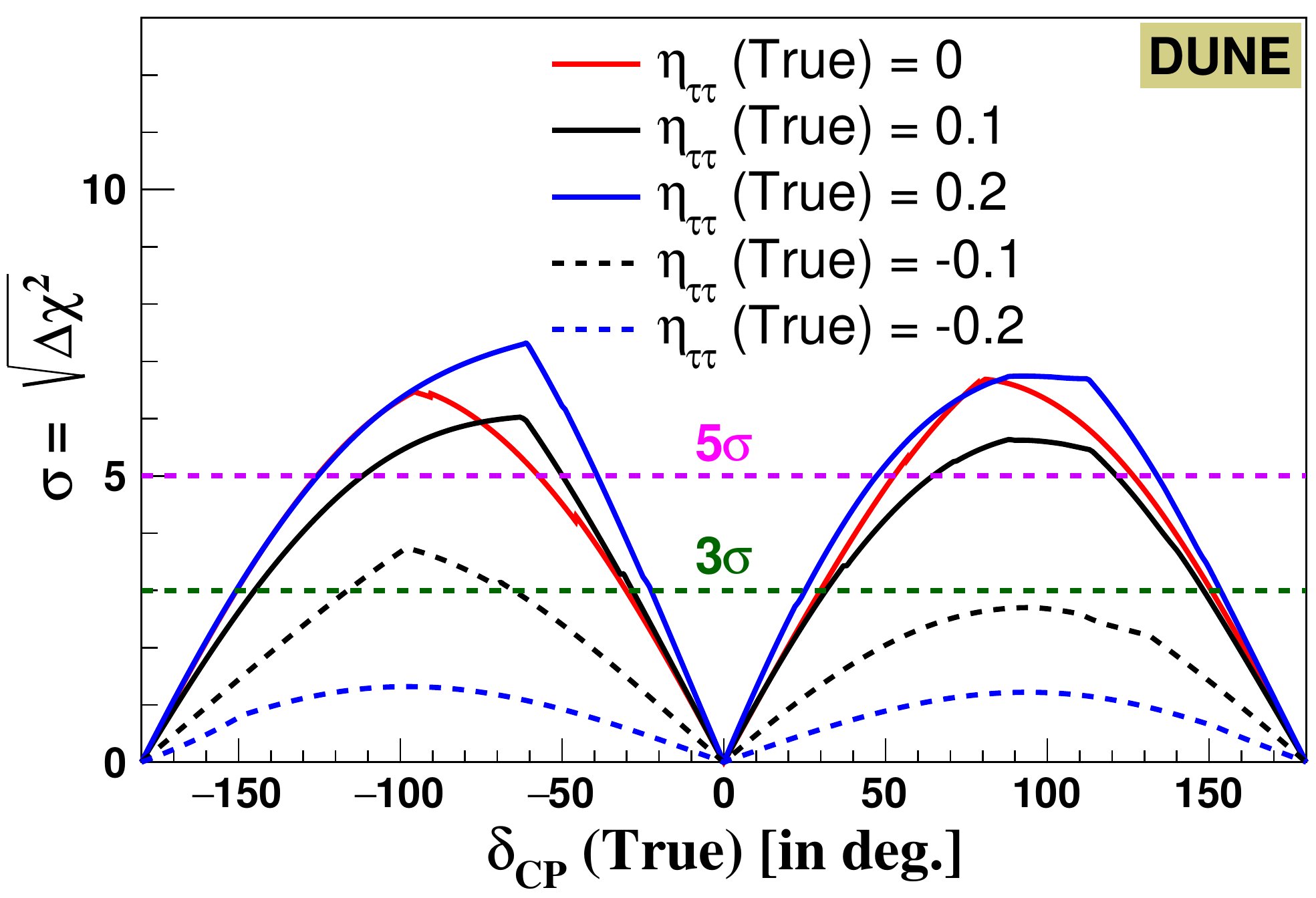} 
	\includegraphics[width=0.32\linewidth, height = 5cm]{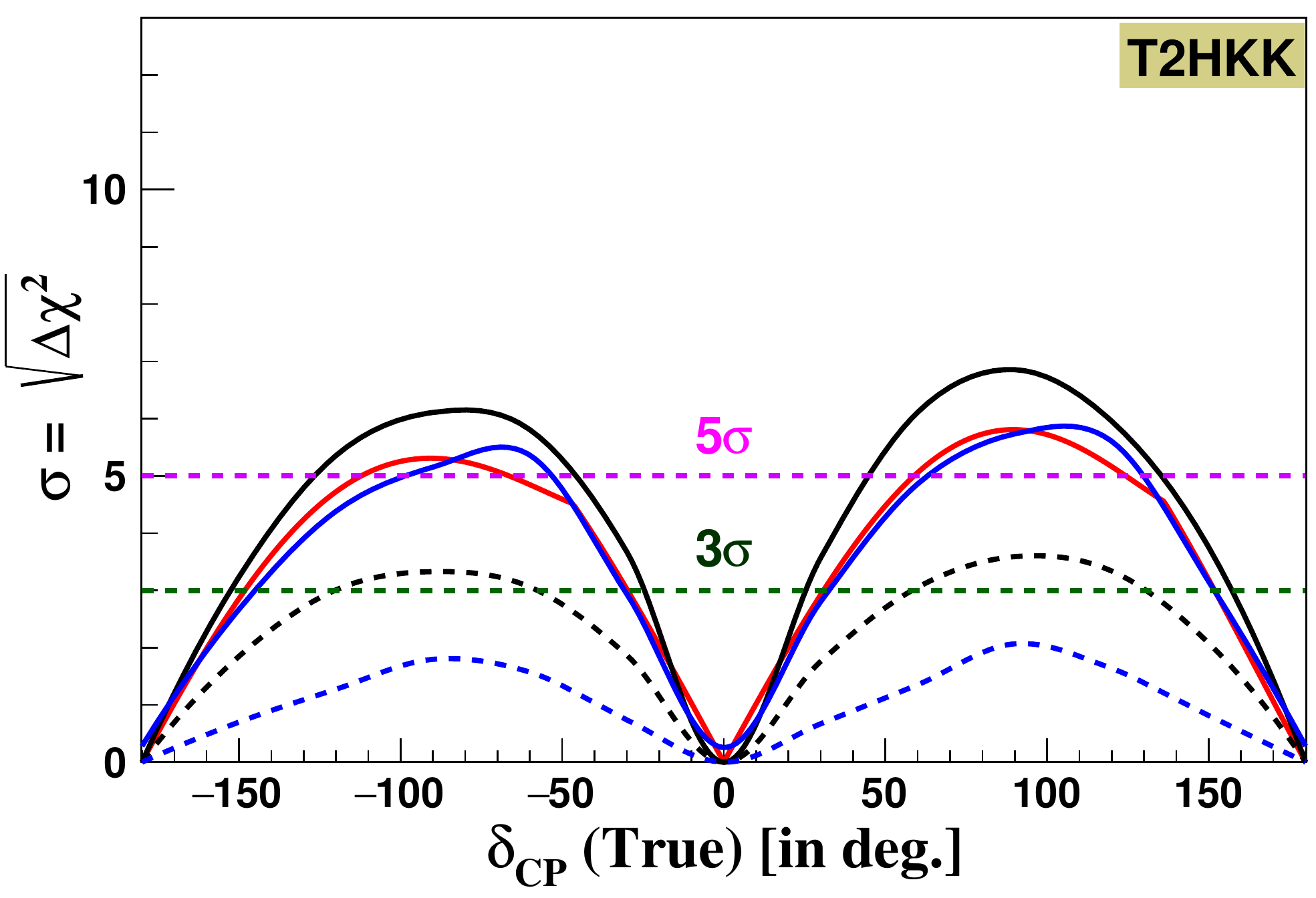} 
	\includegraphics[width=0.32\linewidth, height = 5cm]{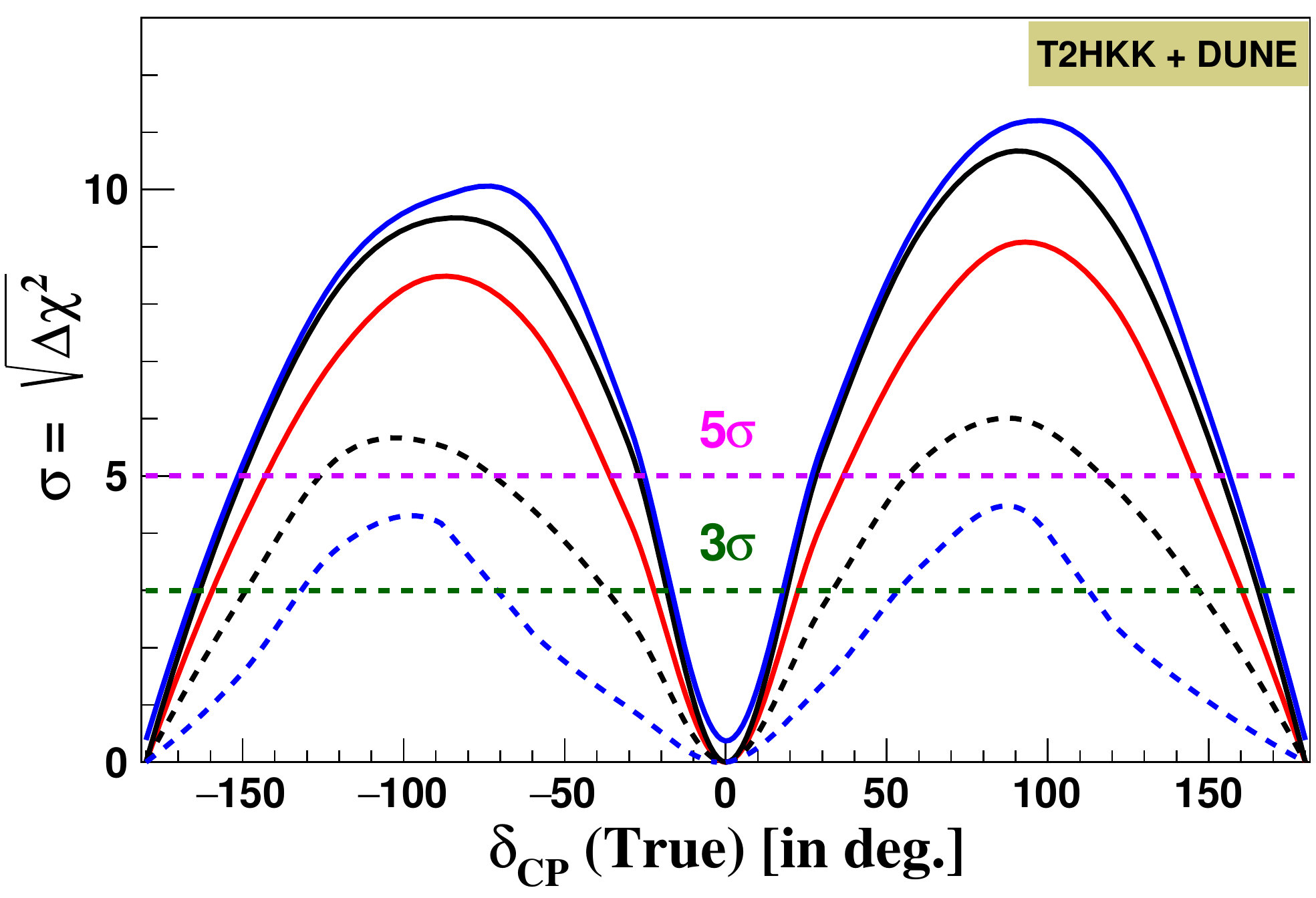} 
	\caption{The CPV sensitivity of DUNE (left-column), T2HKK (middle--column) and DUNE + T2HKK (right--column) in presence of scalar NSI. The plots for $\eta_{ee}$, $\eta_{\mu\mu}$ and $\eta_{\tau\tau}$ are included in the top-row, middle-row and bottom--row respectively. The solid-red curve is for the no scalar NSI case whereas solid (dashed) black and blue curves are for positive (negative) $\eta_{\tau\tau}$.}
	\label{fig:cpv_7}
\end{figure}

In figure \ref{fig:cpv_7}, we show the effects of scalar NSI on the CPV sensitivities at DUNE (left--panel), T2HKK (middle--panel) and DUNE + T2HKK (right--panel) respectively. It may be observed that the impact of $\eta_{\alpha\beta}$ on CPV sensitivities is significant for both DUNE and T2HKK. In the joint analysis we observe a further improved sensitivity, since by combining the two experiments larger statistics can be collected from a wider range of degenerate parameter spaces. We have marginalized over the NSI parameters as well as $\theta_{23}$ in the allowed range [$40^\circ$, $50^\circ$]. The solid red--line represents standard case, whereas other coloured solid (dashed) lines are for positive (negative) values of $\eta_{\alpha\beta}$. The effects of $\eta_{ee}$, $\eta_{\mu\mu}$ and $\eta_{\tau\tau}$ are shown in top--panel, middle--panel and bottom--panel respectively. The dashed green and the dashed magenta line show the 3$\sigma$ and 5$\sigma$ CL respectively. We see that,
\begin{itemize}
    \item A  positive (negative) $\eta_{ee}$ enhances (suppresses) the CPV sensitivities mostly in DUNE and T2HKK. In the region $\delta_{CP}^{true}$ $\in$ [0, 90$^\circ$] the sensitivities overlap for the no Scalar NSI case and for $\eta_{ee}$ = 0.1. This implies that in that range DUNE alone will not be able to distinguish a fake sensitivity coming from scalar NSI. The joint analysis of DUNE + T2HKK can lift this degeneracy and can improve the overall sensitivities (without and with NSI).
    \item A negative $\eta_{\mu\mu}$ deteriorates the CPV sensitivity while a positive $\eta_{\mu\mu}$ can create various degeneracies in CP-measurement. For example, at T2HKK the standard CPV sensitivities overlap with NSI sensitivities for $\eta_{\mu\mu}$ = 0.1 in $\delta_{CP}^{true}$ $\in$ [-180$^\circ$, -120$^\circ$] and $\delta_{CP}^{true}$ $\in$ [110$^\circ$, 180$^\circ$]. This degeneracy can be removed by the joint analysis of DUNE + T2HKK. 
    \item A negative $\eta_{\tau\tau}$ suppresses the CPV sensitivities while a positive $\eta_{tau\tau}$ mostly improves the sensitivities. The without and with scalar NSI sensitivities overlap in various region of $\delta_{CP}^{true}$ for a positive $\eta_{\tau\tau}$. This makes the experiments' capability indistinguishable to the effects from standard and non-standard interactions. The combined sensitivity of DUNE+T2HKK can lift this degeneracy with overall improvement in the CPV sensitivities (without and with NSI).
\end{itemize}

\section{Summary and concluding remarks }
 \label{sec:summary}
With the magnificent development in the field of neutrino physics and in combination with the state-of-the-art experimental set--up, the neutrino oscillation parameters are aimed at being measured with utmost accuracy. The highly ambitious upcoming flagship neutrino experiments are aiming at measuring the neutrino mixing parameters as precisely as possible. Currently, the least constrained parameters in neutrino physics are $\delta_{CP}$ and the octant of mixing angle, $\theta_{23}$. 

In this work, we have primarily explored the impact of scalar NSI on the CP--measurement sensitivities of three upcoming LBL experiments (DUNE, T2HK and T2HKK) in a model--independent way. We also look into the advantages in the sensitivity measurements from combined analyses with DUNE + T2HK and DUNE +T2HKK. If nature permits scalar NSI, we see that, the impact of scalar NSI on the CPV sensitivity may be significant. For chosen negative values of NSI parameters, we observe a deterioration in the CP measurement sensitivities. We also notice an overlapping of standard and non-standard CPV sensitivities for certain positive $\eta_{\alpha\beta}$ at DUNE and T2HKK. This makes the experiments insensitive towards the fake CP effects coming from scalar NSI in those regions. However, this can be removed by a joint sensitivity analysis of DUNE+T2HK and/or DUNE+T2HKK, mainly due to an enhanced parameter space. We observe that, T2HK shows a better constraining capability towards NSI parameters as compared to DUNE or T2HKK due to its giant detector size ($\sim$ 374 kton fiducial mass). A synergy between two experiments ( DUNE+T2HK or DUNE+T2HKK ) helps in collecting tremendous statistics over an enhanced parameter space. and as a result, the overall sensitivities get improved for all non-zero NSI parameters. It may be noted that, for a positive (negative) $\eta_{\alpha\beta}$, an analysis combining all the three experiments shows a significant improvement (deterioration) in CPV sensitivities. We see that, the element $\eta_{ee}$ has the highest sensitivity towards CPV for all the considered NSI parameters.

It is crucial to identify these subdominant effects of neutrinos and its impact on the physics reach of various neutrino experiments. This study was primarily on understanding the impact of scalar NSI for three upcoming LBL experiments. We are also working on the possible exploration of impact on NSI at other physics sensitivities of different neutrino experiments. A combined efforts of all the solar, atmospheric, reactor etc experiments are needed to understand the impact of NSI. It is equally important to put some stringer constrain on the the effects of scalar NSI for accurate interpretation of data from various neutrino experiments.     

\section*{Acknowledgments}
We acknowledge the support of the Research and Innovation grant 2021 (DoRD/RIG/10-73/1592-A) funded by Tezpur University. AM and MMD would also acknowledge the support of the DST SERB grant CRG/2021/002961. AM thanks Dr. Pritam Das for the useful suggestions and discussions during the work. The authors also acknowledge the support of the DST FIST grant SR/FST/PSI-211/2016(C) of the Department of Physics, Tezpur University. The authors thank the referee for the critical reading of the manuscript.

\bibliographystyle{JHEP}
\bibliography{scalar_NSI_synergy}
\end{document}